\documentclass{article}

\usepackage{PRIMEarxiv}

\usepackage[utf8]{inputenc} 
\usepackage[T1]{fontenc}    
\usepackage{hyperref}       
\usepackage{url}            
\usepackage{booktabs}       
\usepackage{amsfonts}       
\usepackage{nicefrac}       
\usepackage{microtype}      
\usepackage{lipsum}
\usepackage{fancyhdr}       
\usepackage{graphicx}       
\graphicspath{{media/}}     
\usepackage{tabularx}
\usepackage{seqsplit}

\usepackage{longtable}
\usepackage{array} 
\usepackage{helvet}
\usepackage{xcolor} 
\usepackage{caption} 
\usepackage{array}
\usepackage{enumitem}

\usepackage{soul}

\newcommand{\redhighlight}[1]{{\sethlcolor{red}\hl{#1}}}

\usepackage[numbers]{natbib}
\title{X-LoRA: Mixture of Low-Rank Adapter Experts, a Flexible Framework for Large Language Models with Applications in Protein Mechanics and Molecular Design

}

\author{
  Eric L. Buehler\thanks{Academy for Science and Design (ASD), 9 Townsend W, Nashua, NH 03063, USA} \\
   \And
  Markus J. Buehler\thanks{Massachusetts Institute of Technology (MIT), 77 Mass. Ave 1-165, Cambridge, MA 02139, USA} \\
  \\
  \texttt{mbuehler@MIT.EDU} \\
}

\begin{document}
\maketitle

\begin{abstract}
We report a mixture of expert strategy to create fine-tuned large language models using a deep layer-wise token-level approach based on low-rank adaptation (LoRA). Starting with a set of pre-trained LoRA adapters, our gating strategy uses the hidden states to dynamically mix adapted layers, allowing the resulting X-LoRA model to draw upon different capabilities and create never-before-used deep layer-wise combinations to solve tasks. The design is inspired by the biological principles of universality and diversity, where neural network building blocks are reused in different hierarchical manifestations. Hence, the X-LoRA model can be easily implemented for any existing large language model (LLM) without a need for modifications of the underlying structure. We develop a tailored X-LoRA model that offers scientific capabilities including forward/inverse analysis tasks and enhanced reasoning capability, focused on biomaterial analysis, protein mechanics and design. The impact of this work include access to readily expandable and adaptable models with strong domain knowledge and the capability to integrate across areas of knowledge. Featuring experts in biology, mathematics, reasoning, bio-inspired materials, mechanics and materials, chemistry, protein biophysics, mechanics and quantum-mechanics based molecular properties, we conduct a series of physics-focused case studies. We examine knowledge recall, protein mechanics forward/inverse tasks, protein design, adversarial agentic modeling including ontological knowledge graph construction, as well as molecular design. The model is capable not only of making quantitative predictions of nanomechanical properties of proteins or quantum mechanical molecular properties, but also reasons over the results and correctly predicts likely mechanisms that explain distinct molecular behaviors. 
\end{abstract}


\section{Introduction}
Large language models (LLMs)~\cite{Vaswani2017AttentionNeedc,Touvron2023LlamaModels,OpenAI2023GPT-4Report,Chowdhery2022PaLM:Pathways,Jiang2023Mistral7B,Gunasekar2023TextbooksNeed} have gained significant popularity, including in the development of special-purpose models that are experts in certain types of tasks, reasoning, or scientific domains~\cite{Bubeck2023SparksGPT-4,Buehler2023MechGPTModalities,Nejjar2023LLMsAnalysis,Buehler2023GenerativeDesign,Luu2023BioinspiredLLM:Materials,Luu2023GenerativeSolvents,Buehler2023MeLMProblemsb,Ge2023OpenAGI:Experts}.%
However, training such models can be costly, especially when diverse sets of capabilities are needed. Methods like low-rank adapters (LoRA)\cite{hu2021lora} have been proposed as a more efficient alternative, but adaptations are usually focused on narrower fields of knowledge.  The underlying concept in LoRA modeling is the use of low-rank matrices that are added to the original full-scale matrix, and selecting those low-rank matrices as the only trainable component of the model. Since only the adapter layers are trainable, these models typically preserve the knowledge captured during pre-training of the base model while making the model more applicable to specific tasks, in addition to being efficient in terms of computational needs. Since training of LoRA adapters is efficient, it is possible to develop a large set of special-purpose models using this strategy.

While LoRA adapters provide an efficient way to develop specialized model capabilities, it remains an open question how to integrate multiple of such systems into improved models that offer an integrated, and even improved set of capabilities. In fact, mixing LLMs has become an interesting area of research~\cite{Kim2023SOLARUp-Scaling}. Other experimental works, such as using spherically interpolated model parameters (SLERP) and other concepts~\cite{Arcee-ai/mergekit:Models.} have shown interesting results and a promise of mixing models to achieve new capabilities. Mixed models created in these ways are usually computed {\it a priori} according to a particular algorithm and while they have shown promising performance in some benchmarks, more research is necessary into rational development of mixed models and how certain features emerge. 
We hypothesize that dynamical mixing of parameters can be a more general approach as it allows for the mixed model to explore new combinations during inference.

\begin{figure}
    \centering
    \includegraphics[width=.9\linewidth]{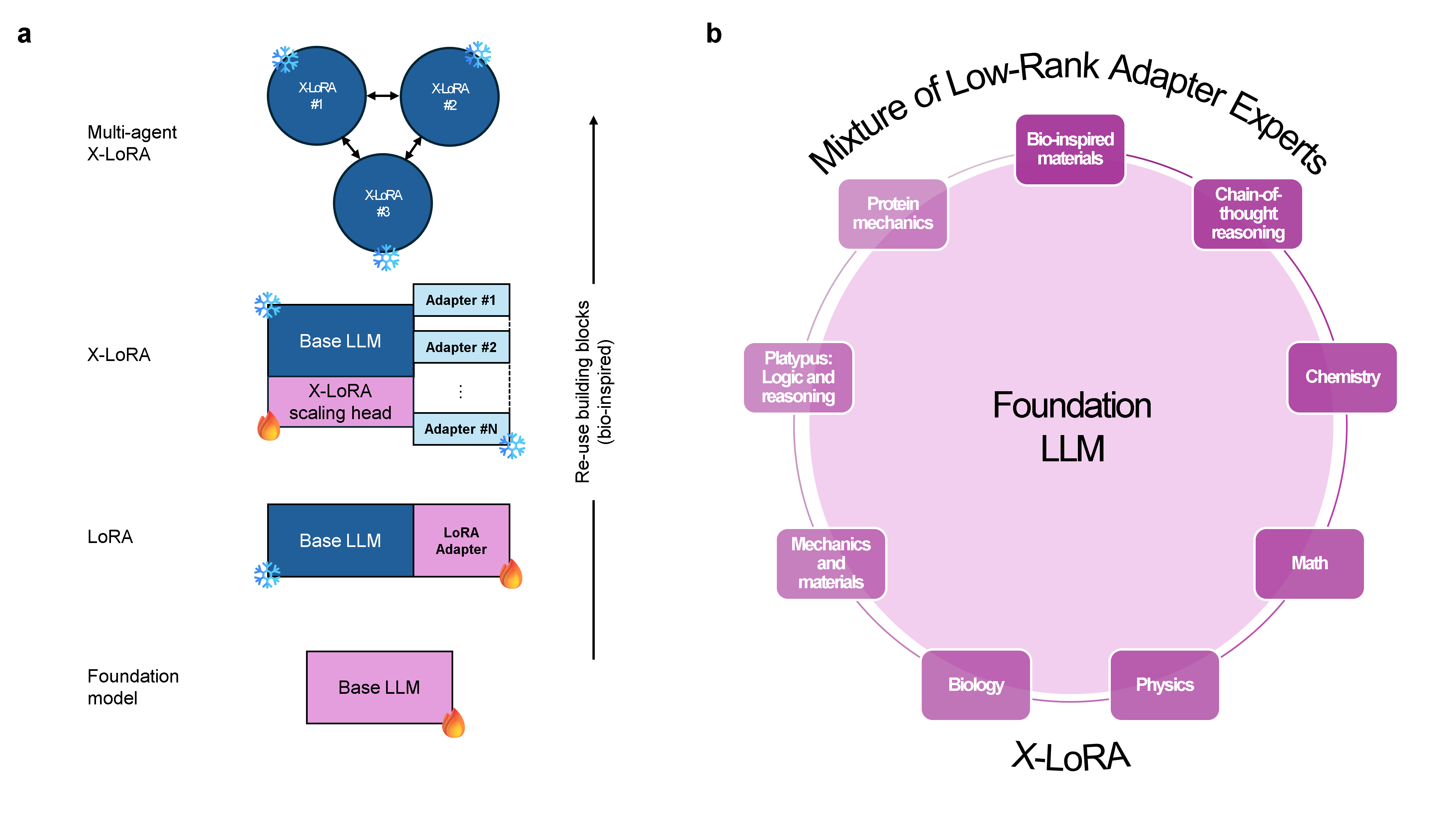}
    \caption{Multi-hierarchical design principle of adapted and agentic neural network architectures, following a bio-inspired paradigm that bases its foundation in re-use of existing building blocks. (a), in the schematic, blue color indicates a frozen (non-trainable) component, and reddish color indicates trainable components. The visual reflection of the re-use and adaptation through smaller trainable components is visible. Potentially, trainable components could be added also at the multi-agent level, albeit this is not yet implemented in this work. The training conducted as part of this work focuses on the LoRA adapter and X-LoRA levels of the model as a pre-trained foundation model is used as the basis. (b), Overview of the set of adapters used to construct the X-LoRA model, featuring experts in bioinspired materials, chain-of-thought (CoT) and reasoning, chemistry, mathematics, physics, mechanics and materials, logic and reasoning, and protein mechanics.}
    \label{fig:hierarchical_concept}
\end{figure}

Related approaches have used mixture of experts (MoE) strategies~\cite{Jacobs1991AdaptiveExperts,Hampshire1992TheRecognition,Jordan1994HierarchicalAlgorithm}, including recently released Mixtral LLM~\cite{Jiang2024MixtralExperts}. However, these can be computationally expensive with significant demand even during inference. Here, we propose a computationally efficient mixture-of-expert approach using a set of distinct LoRA adapters that can easily be trained individually. Our approach implements a high-granularity and flexible method, inspired by a biological paradigm that re-uses neural network building block in the construction of a hierarchy of interacting models, from the foundation model to agentic systems (Fig.~\ref{fig:hierarchical_concept}). To assert its impact quantitatively, our work specifically focuses on using this strategy for science-focused LLMs, here specifically applied to protein mechanics problems in the broader space of bio-inspired materials analysis and design.

We propose an algorithm to dynamically mix adapters at a token-level state, and with fine granularity to enable mixing of adapters at the levels of individual layers. The approach, referred to as X-LoRA, offers access to diverse sets of capabilities that are critical for scientific uses of multimodal language modeling~\cite{Hu2023DeepScience,Buehler2023AMetamaterials,Buehler2022ProteinNetworks,}.

\subsection{Fundamental concepts of X-LoRA}

We first briefly review the concepts used in the development of LoRA adapters. The basic strategy behind low-rank adaptation (LoRA)~\cite{hu2021lora}, hypothesizes that updates to the weights have a low “intrinsic dimension” and takes advantage of this by freezing the original weights $W_0 \in \mathbb{R}^{d \times k}$ and constraining the updating to the low-rank decomposition 
\begin{equation}
W_0 + \Delta W_0 = W_0 + BA 
\end{equation}
where $B \in \mathbb{R}^{d \times r}$  and $A \in \mathbb{R}^{r \times d}$ with the rank $r \ll {\rm min}(d,k)$. During training with LoRA~\cite{hu2021lora}, the pretrained weights $W_0$ are frozen and only the $A$ and $B$ matrices retain trainable parameters. The forward pass of a LoRA layer may be expressed as the following equation:
\begin{equation}
    h = W_0x + \Delta W_0x = W_0x + BAx
\end{equation}
We note that in practice, LoRA adapters have a fixed scalar weight $\alpha$. This is applied to the decomposition matrices, yielding
\begin{equation}
    h = W_0x + BAx \cdot \alpha
\end{equation}

Building on this, X-LoRA is based on the idea to use a set of LoRA adapters, each of which has unique sets of capabilities, as has been shown in earlier science-focused LLM work (e.g., \cite{Buehler2023GenerativeDesign}). Unlike mixing or integrating models statically, we create a dynamic gating approach that scales individual LoRA adapter with individual token and layer granularity to facilitate mixing deep inside the model. In other words, considering a LoRA adapter $i$, the original $\alpha_i$ parameter is multiplied by a scaling value denoted by $\lambda_i$ to create the new scalings value for the adapter $\alpha^*_i$:
\begin{equation}
    \alpha^*_i = \alpha_i \cdot \lambda_i
\end{equation}
The scaling value $\lambda_i$ is predicted by a X-LoRA scaling head that utilizes the model's hidden states, forming the trainable component in the X-LoRA model. Since the scaling head can take advantage of complex encodings it can be implemented efficiently as a simple feed-forward neural network with one or few layers (in the implementation of X-LoRA users can specify the details of how this neural network is constructed).

The scaling calculation is conducted for each token in the sequence. Hence, it offers a high level of granularity as each of the LoRA adapters that operate at specific layers in the LLM is scaled. 
Henceforth, we refer to this approach of gating the various components of the adapters in this way as scaling. Further details on the approach are provided in the Materials and Methods section.

\subsection{Paper outline}

The plan of this paper is as follows. First, we discuss the approach and training strategy including relevant datasets that results in the development of an X-LoRA model with special capabilities in the physical sciences, especially focused on biomaterials including proteins. We then present a series of experiments in which we apply an X-LoRA to a variety of tasks, such as question answering, conversational and agentic modeling, protein design and analysis, and others. We conduct a detailed analysis of the scaling patterns and validate the approach through comparison with molecular modeling and other physical data and methods. 

\section{Results and discussion}

The process of developing the X-LoRA model consists of the following steps:
\begin{enumerate}
    \item Train foundational base LLM (done in earlier work)
    \item Individually train a set of adapters to develop expertise in a series of areas (these reflect distinct or overlapping areas of skills and knowledge)
    \item Train the integrated X-LoRA model
\end{enumerate}

Our experiments start with training a series of LoRA adapters. We develop a set of nine adapters, fined-tuned with distinct expertise, based on the Zephyr-7B-$\beta$ model~\cite{HuggingFaceH4/zephyr-7b-betaFace} that was built on top of the Mistral-7B model\cite{Jiang2023Mistral7B}:
\begin{enumerate}
    \item Bioinspired materials 
    \item Chain-of-thought (CoT) and reasoning 
    \item Chemistry
    \item Mathematics
    \item Physics
    \item Biology
    \item Mechanics and materials
    \item Platypus: Logic and reasoning 
    \item Protein mechanics tasks (generative sequence-to-property and inverse capabilities)
\end{enumerate}
These areas of focus and expertise were chosen since we ultimately seek to have a STEM focused, materials science emphasized LLM. However, we postulate that our approach can be generally applied to all sorts of areas, especially to create modalities where connections between distinct areas of knowledge can be developed~\cite{Buehler2023MechGPTModalities,Buehler2023MeLMProblemsb}. 

After the LoRA adapters are created, we train the X-LoRA model. To do this, we use a subset of the combined set of the training samples used to train the individual LoRA adapters (details see Materials and Methods, including detailed delineations on the dataset, implementation and so on).

We use the X-LoRA system as an experimental entity to assess its performance in a few different settings. We conduct a series of experiments, including complex question answering, the use of the generative model for protein design, analysis and reasoning, as well as creating more sophisticated interaction strategies using adversarial agentic modeling. Throughout the analysis we compare performance of the base model with the X-LoRA model, and track how the scaling head utilizes different layers of the different experts across generation or development of answers. 

\subsection{Question answering and observed X-LoRA layer-wise scaling weights}

We first examine differences between the X-LoRA model and the base foundation model. We ask both models this question:
\begin{quote}
\small\texttt{What material behavior (stiffening vs softening) causes supersonic crack speeds in mode II cracks? Think step by step.}
\end{quote}
The response of the base model is:
\begin{quote}
\small\texttt{Supersonic crack speeds, also known as Rayleigh waves, occur in mode II cracks due to the stiffening material behavior of the material near the crack tip. In mode II fracture, the loading is shear-dominated and the crack propagates perpendicular to the direction of applied stress.
Here's a step-by-step explanation:
\newline\newline
As the crack tip approaches the material, the stresses in the vicinity increase rapidly. This results in a region of high stress concentrations called the stress singularity.
\newline\newline
In this stress singularity, the material experiences an elastic deformation that leads to an increase in stiffness, i.e., stiffening behavior. This stiffening effect is due to the fact that the material near the crack tip becomes more constrained as the crack approaches [...]}
\end{quote}
Using our own domain knowledge~\cite{Buehler2008AtomisticFailure}, this answer is not only incorrect but also long-winded. In contrast, the X-LoRA model responds as follows:
\begin{quote}
\small\texttt{The material behavior that causes supersonic crack speeds in mode II cracks is stiffening.}
\end{quote}
This answer is correct, clear and succinct~\cite{Buehler2008AtomisticFailure}. 

The difference between a base model and the X-LoRA model is even more clear for the specialized protein analysis and design tasks. The task given is:
\begin{quote}
\small\texttt{CalculateForceHistory< M S K S N T Y R M L V L L E D D T K I N K E D E K F L K G K P G K M H E F V D E L I L P F N V D E L D E L N T W F D K F D A E I C I P Y V E I L K E S G M K > }
\end{quote}
The response of the base foundation model is:
\begin{quote}
\small\texttt{I'm sorry but the input provided is not a valid mathematical expression. It appears to be a sequence of characters that do not form a calculable function. Please provide a valid mathematical expression for me to calculate.
\newline\newline
However, I can tell you that the sequence of characters you provided is actually a series of function and variable names that may be part of a larger program or script in a specific programming language. Without context, it's impossible to determine the intended calculation.
\newline\newline
If you provide more information, such as the programming language and the context in which this expression appears, I may be able to provide more insight into its meaning and possible calculations. }
\end{quote}
The X-LoRA model responds as follows, correctly carrying out the design task of generating the force-deformation curve associated with this protein:
\begin{quote}
\small\texttt{[0.013,0.066,0.078,0.045,0.072,0.126,0.127,0.107,0.119, 
0.074,0.106,0.098,0.103,0.128,
0.077, 0.069,0.131,0.108,0.144,0.174, 
0.166,0.183,0.185,0.217,0.175,0.230,0.187,0.224,
0.267,0.301,0.315]}
\end{quote}

\begin{figure}
    \centering
    \includegraphics[width=.55\linewidth]{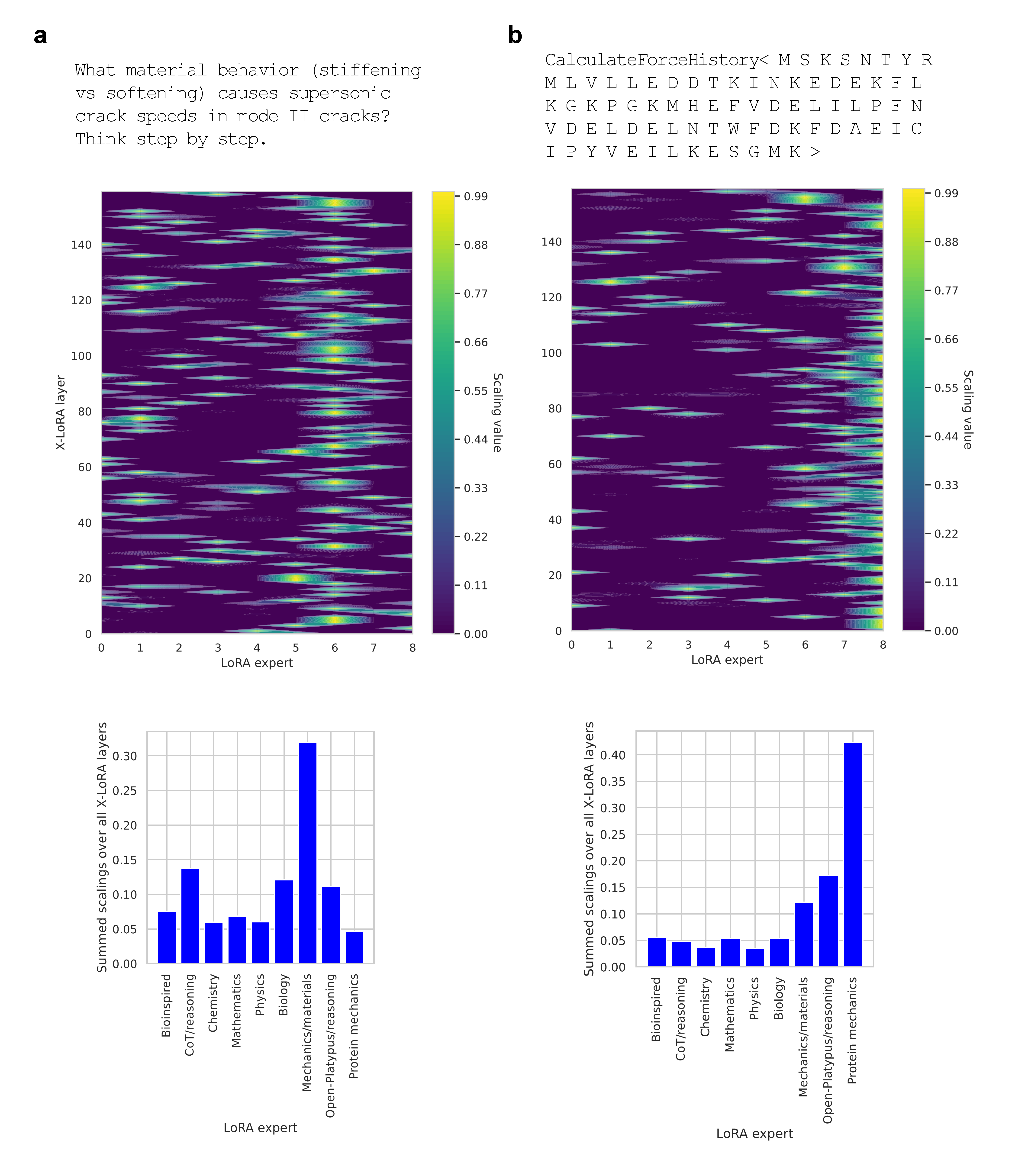}
    \caption{Results from question answering and observed X-LoRA scaling weights, for the trial questions used to compare X-LoRA with the base foundation model (panels a and b show results for two different tasks solved by the model). At the top of each sub-panel we show the question asked, followed by the X-LoRA scaling weights plotted over X-LoRA layers and LoRA experts, respectively. The lower bar plot summarizes the scalings over all layers, indicating an overall measure for which adapter is most prominently used. Panel a, featuring a question about dynamic fracture, heavily uses the Mechanics/Materials expert. In contrast, as shown in panel b, the protein analysis task results in the use of the protein mechanics adapter. It is observed that a complex pattern of scaling values is used in each case, suggesting that the X-LoRA model takes advantage of mixing different adapters heterogeneously across layers. Since the scaling weights determine which expert to use for a given input, the heatmaps show how this decision changes across different layers of the model, revealing a great level of heterogeneity.    
    }
    \label{fig:QA_weights_sample_0}
\end{figure}

Fig.~\ref{fig:QA_weights_sample_0} shows results from this question answering and observed X-LoRA scaling weights.  The two queries, one focused on mechanics and materials (Fig.~\ref{fig:QA_weights_sample_0}a) and protein mechanics (Fig.~\ref{fig:QA_weights_sample_0}b) show distinct patterns.  Importantly, we find that a complex pattern of scaling values is used in each case, suggesting that the X-LoRA model takes advantage of mixing different adapters heterogeneously across layers. We further find that the heatmap in Fig.~\ref{fig:QA_weights_sample_0}a has a broader range of activation across many experts and layers, with some areas of high activation. The heatmap in Fig.~\ref{fig:QA_weights_sample_0}b shows a more concentrated pattern of activation, with high values being less widespread. We believe that this could indicate that the gating function on the right is more selective in its activation of experts.

In Fig.~\ref{fig:QA_weights_sample_0}, the regions that resemble bright lines represent the paths of high activation, meaning the scaling function is activating certain experts more for specific layers. The patterns observed suggests there is not a uniform distribution of importance across all experts and layers; instead, it tends to be rather sparse and selective. This sparsity is characteristic of what has been observed in other mixtures-of-experts models, where only a subset of experts is chosen for any given input to specialize in different parts of the problem space.
By comparing the maps for different tasks, both the sparsity and patterns could be important for further understanding which experts the model relies on for different types of inputs or at different stages of processing within the neural network (we will address the evolution of the maps as it changes dynamically during solving complex tasks later on). 

\begin{figure*}
    \centering
    \includegraphics[width=.8\linewidth]{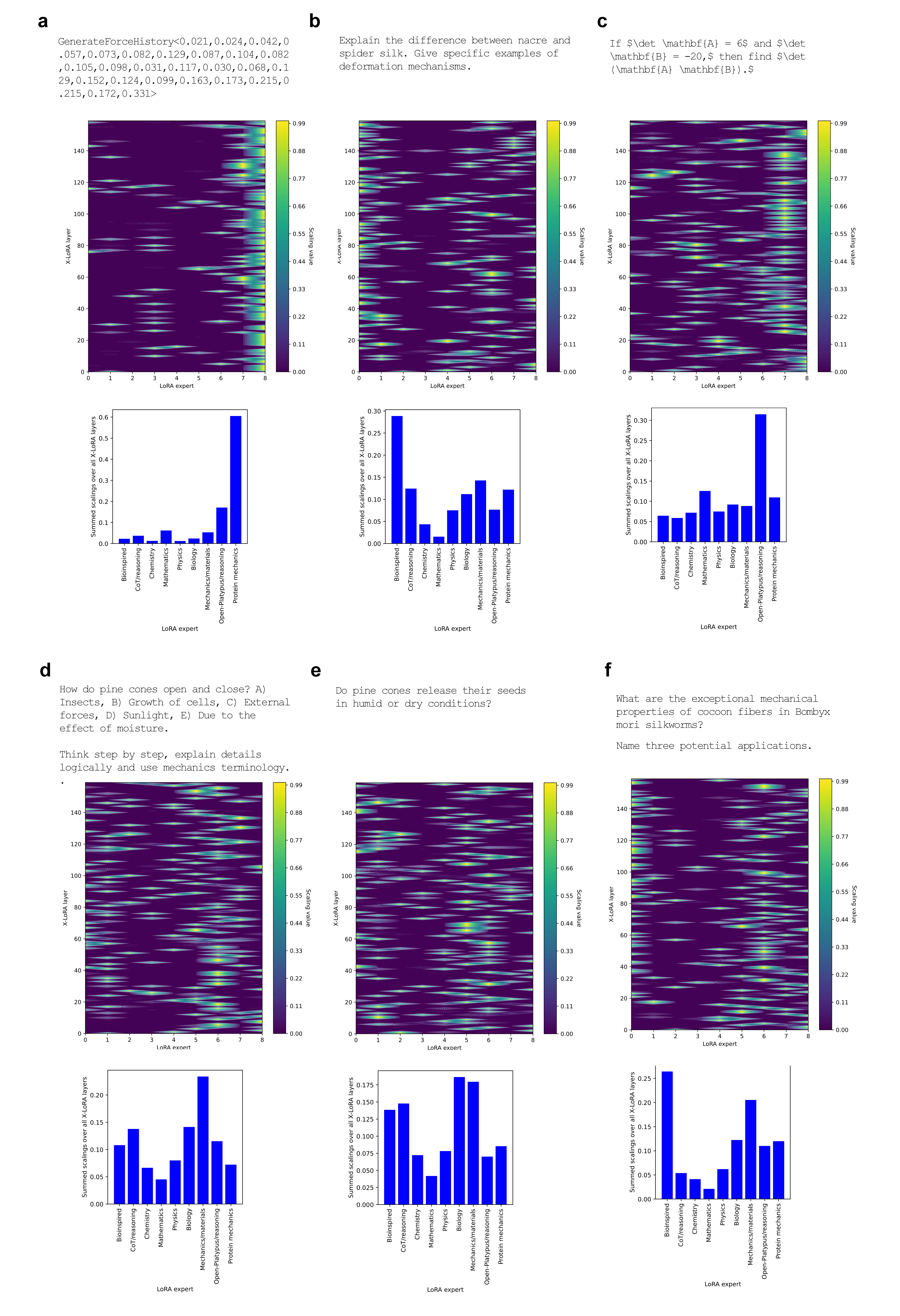}
    \caption{Results from question answering and observed X-LoRA scaling weights (panels a-f show results from different tasks solved by the model, covering a range of domain areas and types). At the top of each sub-panel we show the question asked, followed by the X-LoRA scaling weights plotted over X-LoRA layers and LoRA experts, respectively. The lower bar plot summarizes the scalings over all layers, indicating an overall measure for which adapter is most prominently used. As before, it is observed that a complex pattern of scaling values is used in each case, suggesting that the X-LoRA model takes advantage of mixing different adapters heterogeneously across layers.}
    \label{fig:QA_weights}
\end{figure*}

To further analyze the behavior, we now explore a greater variety of questions to determine how the X-LoRA model uses the various experts, given a set of distinct prompts. We are specifically interested to see whether, and if yes, how, the X-LoRA model mixes different adapters in questions that overlap across different skills or domains.  Fig.~\ref{fig:QA_weights} shows results from question answering and observed X-LoRA scaling weights, with responses to the queries listed in Table~\ref{tab:table_qa}.  In  Table~\ref{tab:table_qa} we show both, the response form X-LoRA and the response from the base model, Zephyr-7B-$\beta$. The differences between the two models are striking.
To visually show this clearly, we mark incorrect or questionable response in \redhighlight{red highlight}. The comparison shows that X-LoRA provides far higher accuracy, as well as more concise answers. 

As hypothesized, we observe complex mixing of adapters and often the activation of several dominant LoRA experts. For instance, Fig.~\ref{fig:QA_weights}d reveals that a question about the mechanics of pine cones leads to strong usage of the mechanics/materials adapter combined with the CoT/reasoning and bioinspired adapters. While the correct letter is picked by X-LoRA, the answer is not entirely correct as pine cones open in dry, not wet conditions. A follow-up question to further probe into whether release happens in dry or wet conditions results in the correct answer, however. 
In a protein-focused question Fig.~\ref{fig:QA_weights}f reveals adapter mixing in response to a question related to the mechanics of Bombyx mori silk mechanics, activating a combination of the bioinspired, mechanics/materials and biology adapters. For clearly specialized tasks like protein design (Fig.~\ref{fig:QA_weights}a) we witness a largely singular activation of one of the experts.  

A deeper analysis of the heatmap of scalings depicted in Fig.~\ref{fig:QA_weights}a shows that in addition to the obvious high relevance of the protein mechanics expert, there are several bands where certain experts are given high importance across several layers. This is particularly noticeable for the mechanics/materials and reasoning experts. These bands are not continuous and show a varying pattern, which might suggest specific conditions or features that these experts handle effectively. In fact, the reasoning expert (immediately to the left of the protein mechanics expert) stands out with several layers where they have the highest activation, especially in the upper layers (around 100 to 140). We believe that this could imply a significant role in the final decision-making or output of the model, albeit more research is necessary to solidify this. There is also a more localized but strong activation for the physics and biology experts at certain layers, which may indicate a specialization of this expert for features or representations in those regions.

\begin{figure*}
    \centering
    \includegraphics[width=.6\linewidth]{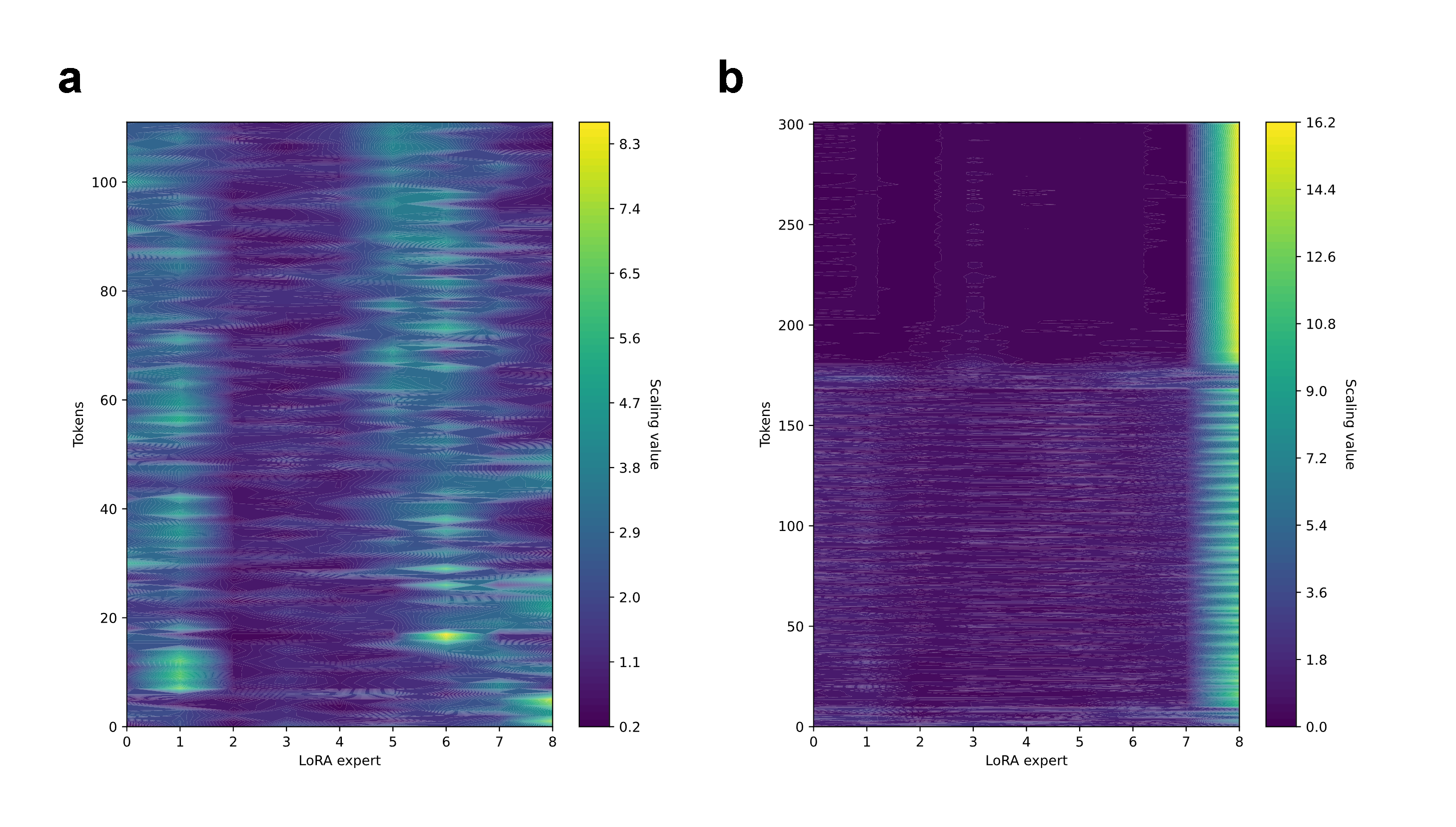}
    \caption{Summed up scalings over all layers, as a function of token history. This plot shows histories for two examples, in panel a task related to pine cones (around 110 tokens total), and in panel b the result of a series of protein mechanics tasks (around 300 tokens; first property calculations, then a generative task to design a new protein towards the end). While the use of experts is generally stable, there are noticable changes in the use of the LoRA experts during the process. The expert numbering (0 to 8) reflects the same organization as in the earlier plots, e.g. refer to Fig.~\ref{fig:QA_weights} for the labels.}
    \label{fig:QA_token_history}
\end{figure*}

Further, Fig.~\ref{fig:QA_token_history} depicts a token-level history over the summed up scalings over all layers. This analysis reveals histories for two examples, in Fig.~\ref{fig:QA_token_history}a task related to pine cone seed release, and in Fig.~\ref{fig:QA_token_history}b the result of a series of protein mechanics tasks. The protein mechanics task  involved a series of prompt-answer interactions: First two property calculations, then a generative task to design a new protein towards the end. While the use of experts is generally stable, there are noticable changes in the use of the LoRA experts during the process. Specifically, the protein mechanics expert is highly focused in the last stages of generation when a new protein is being designed. 

Table~\ref{tab:table_qa} shows diverse use cases across different domains and a comparison to the performance of the base model, on its own. We discuss a few of them in more depth.

For instance, we pose a question that involves a mix of protein science and biology with logic, asking about protein expression patterns. X-LoRA gets the answer right. The  response from the base model contains some inaccuracies and logical inconsistencies when analyzing the expressions of marker proteins in each cell type. The base model correctly identifies that fibroblasts express Protein B and do not express Protein A due to the mutual exclusivity of Protein A and Protein B. However, it does not explicitly mention the expression of Protein C by fibroblasts. Given that there's no rule against expressing Protein C alongside Protein B, and since fibroblasts don't have a restriction against Protein C, they logically express Protein C as well. For Neurons, the base model starts correctly by stating that neurons express Protein A and not Protein B due to the mutual exclusivity rule. However, the discussion about the potential for neurons to express Protein C or both Protein C and Protein B is confusing and contains a mistake. Since neurons express Protein A and not Protein B (as stated), the only other option, given the rules, is for them to express Protein C. There is no scenario where neurons could express Protein B, as mentioned in the initial analysis. Therefore, neurons express Protein A and Protein C, not Protein B. For Epithelial cells, the base model correctly notes that epithelial cells do not express Protein A but expresses both Protein B and Protein C. However, the explanation given is convoluted and includes incorrect statements about the mutual exclusivity of Protein A and Protein C, which was not part of the original observations. The simple reasoning is that epithelial cells do not express Protein A but must express the two remaining proteins, Protein B and Protein C, because the initial instructions only specify that Protein A and Protein B cannot be co-expressed. 
X-LoRA, in contrast, correctly identifies the marker proteins expressed by each cell type based on the provided observations and logical deductions. For fibroblasts, X-LoRA correctly deduces that fibroblasts cannot express Protein A because the presence of Protein A excludes the expression of Protein B. Since fibroblasts do express Protein B, and given that each cell type expresses a unique combination of three marker proteins (which was an incorrect assumption added), as the original statement didn't specify the number of proteins each cell type expresses, but rather that they express a unique combination), the conclusion that fibroblasts also express Protein C is correct based on the information provided (except the part about the number of proteins, which seems to be a misunderstanding). For neurons, the model correctly identified as expressing Protein A and Protein C. The reasoning is sound, noting that Protein B cannot be expressed alongside Protein A, and since neurons express Protein A and one other marker protein (not Protein B), they must express Protein C. Finally, for epithelial cells the X-LoRA model accurately concludes that epithelial cells express Protein B and Protein C. This is because epithelial cells do not express Protein A, and by process of elimination, given that they express two marker proteins, these must be Protein B and Protein C.

In the question about pine cone release in humid versus dry conditions, the base model gives a convoluted answer that is factually incorrect. X-LoRA gives a precise answer with clearl reasoning. There are other related examples, such as in the question about the exceptional mechanical properties of Bombyx mori silk, where the base model incorrectly asserts that silk fibers conduct electricity. The model also fails in solving several other domain specific tasks, such as to the question about the role of hyperelasticity in maximum fracture speeds where X-LoRA provides a concise and accurate answer, whereas the base model fails to answer. In a question about strong adhesion of gecko feet, X-LoRA correctly identifies the importance of setae and nano-sized effects, whereas the base models makes incorrect claims about the role of special purpose glue proteins in the process. These and other examples discussed in the table and elsewhere in the paper provide strong evidence for the enhanced capabilities of X-LoRA over the base model.

\begingroup 
\tiny 

\begin{longtable}{|p{0.2\linewidth}|p{0.35\linewidth}|p{0.35\linewidth}|}

\caption{Summary of question-answer pairs, in the domains of biological materials, mechanics of materials, protein properties, logic and reasoning, and related areas, focusing on challenging questions. Several sample questions are posed and answered by X-LoRA and the Zephyr-7B-$\beta$ base model for comparison. Associated X-LoRA scaling values for some of the queries are presented in Figure~\ref{fig:QA_weights} to show how the model is able to mix different adapters and associated capabilities depending on the context. We mark incorrect or questionable response in \redhighlight{red highlight}. The comparison shows that X-LoRA provides far higher overall accuracy, as well as more concise answers. }
\label{tab:mylongtable} \\
\hline
\textbf{Question}  &  \textbf{X-LoRA model response} &  \textbf{Zephyr-7B-$\beta$ base model response} \\
\hline
\endfirsthead

\multicolumn{3}{c}%
{{\tablename\ \thetable{} -- continued from previous page}} \\
\hline
\textbf{Question}  &  \textbf{X-LoRA Answer} &  \textbf{Zephyr-7B-$\beta$ base model Answer} \\
\hline
\endhead

\hline
\multicolumn{3}{r}{{Continued on next page}} \\
\hline

\endfoot

\hline
\endlastfoot

    \hline
    \hline
    GenerateForceHistory
\seqsplit{<0.021,0.024,0.042,0.057,0.073,0.082,0.129,0.087,0.104,
0.082,0.105,0.098,0.031,0.117,0.030,0.068,0.129,0.152,
0.124,0.099,0.163,0.173,0.215,0.215,0.172,0.331>} & \seqsplit{[ M S K S N T Y R M L V L L E D D T K I N K E D E K F L K G K P G K M H E F V D E L I L P F N V D E L D E L N T W F D K F D A E I C I P Y V E I L K E S G M K ]} & \redhighlight{The provided sequence of numbers represents the history of force measurements over time. Here's a possible interpretation:}
\begin{itemize}[leftmargin=*,align=left]
\item \redhighlight{The first seven values (0.021, 0.024, ..., 0.082) represent the force readings at different points in time. These values appear to be relatively small and fluctuating around a low average.}
\item  \redhighlight{The next three values (0.129, 0.087, 0.104) show an increase in force, followed by a decrease and then another increase. This could indicate some kind of cyclical behavior or trend.}
\item  \redhighlight{After this, there are a few more fluctuations with some larger values (0.105, 0.098, ..., 0.173). It's unclear whether these represent significant changes or just noise in the measurement process.}
\item  \redhighlight{Finally, the last four values (0.215, 0.215, 0.172, 0.331) show a dramatic increase in force, possibly indicating a significant event or anomaly. However, it's not clear from this data alone what caused this spike. More context and additional measurements would be needed to fully understand what happened here.}
\end{itemize}

\\ \hline
Explain the difference between nacre and spider silk. Give specific examples of deformation mechanisms.  & Nacre, also known as mother-of-pearl, is a biomineralized material found in the shells of mollusks such as oysters and pearl mussels. It consists of layers of aragonite platelets held together by an organic matrix of proteins and polysaccharides. Nacre's strength and toughness come from its hierarchical structure, which includes nanoscale aragonite crystals arranged in a brick-and-mortar arrangement with organic interlayers. This structure allows for efficient load transfer and energy dissipation during deformation. 
\newline
\newline
Spider silk, on the other hand, is a protein fiber produced by spiders for various purposes, such as capturing prey or building webs. Spider silk is composed of fibroin proteins that form long, thin fibers with high tensile strength and elasticity. The mechanical properties of spider silk are due to its semicrystalline structure, which includes beta-sheet crystals and amorphous regions. Spider silk exhibits excellent toughness and can undergo significant deformation before breaking.
\newline
\newline
In terms of deformation mechanisms, both nacre and spider silk exhibit plastic deformation, meaning they can be permanently deformed without breaking. However, the mechanisms behind this deformation differ. In nacre, plastic deformation occurs through interfacial debonding between the aragonite platelets and the organic matrix, as well as crack deflection at the interface. These mechanisms allow for energy dissipation and stress redistribution, resulting in improved toughness.
\newline
\newline
In contrast, spider silk deforms plastically through the sliding of beta-sheet crystals past each other, which can occur along the direction of the fiber or transverse to it. This sliding mechanism allows for large strains without fracture, contributing to the exceptional toughness of spider silk. Additionally, the presence of amorphous regions in spider silk facilitates molecular slippage during deformation, further enhancing its plasticity.
\newline
\newline
Overall, while both nacre and spider silk exhibit plastic deformation, the underlying mechanisms are distinct, reflecting their different structures and compositions. &
Nacre, also known as mother-of-pearl, is a biomaterial found in the shells of mollusks such as oysters, clams, and mussels. Spider silk, on the other hand, is a protein fiber produced by spiders for their webs and other purposes. Although both materials have remarkable mechanical properties, they differ significantly in terms of structure and deformation mechanisms.
\newline
\newline
Structure: Nacre consists of thin platelets (called tabular crystals) made of calcium carbonate (CaCO3), arranged in layers with organic material (primarily proteins) sandwiched in between. This layered structure gives nacre its characteristic iridescence. In contrast, \redhighlight{spider silk is composed of fibroin}, a protein that forms fibers with a unique hierarchical structure. \redhighlight{The fibers are made up of nanoscale fibrils, which are further organized into larger fibers.}
\newline
\newline
Deformation Mechanisms: The deformation behavior of nacre and spider silk is quite different due to their distinct structures. \redhighlight{When subjected to tension, nacre exhibits a distinctive brittle fracture behavior. As the load increases, the tabular crystals separate from each other, causing cracks to propagate through the material. However, the organic matrix between the plates prevents complete failure of the material, resulting in a relatively high toughness. An example of this behavior can be seen in the brittleness of pearl buttons when exposed to water or humidity, where the tabular crystals delaminate and detach from the button surface.}
\newline
\newline
Spider silk, on the other hand, displays highly elastic behavior under tension. \redhighlight{The fibroin protein molecules in spider silk form extended beta-sheets, which provide the material with exceptional strength and stiffness.} As the material is stretched, these sheets slide past one another, allowing the fiber to extend without breaking. This sliding mechanism results in a large strain-to-failure, making spider silk several times stronger than steel of similar weight. An example of this behavior can be seen in the ability of spider webs to stretch and recoil upon impact, enabling spiders to capture prey.
\newline
\newline
In summary, while both nacre and spider silk exhibit impressive mechanical properties, their differences in structure and deformation mechanisms result in unique behaviors under tension. Nacre's layered structure allows it to absorb energy during deformation, while spider silk's hierarchical structure enables it to stretch and recoil without breaking. These properties make both materials promising candidates for developing novel bioinspired materials with tailored mechanical properties for various applications.
\\ \hline
If $\det \mathbf{A}$=$ 6$ and $\det \mathbf{B} $=$ -20,$ then find $\det (\mathbf{A} \mathbf{B}).$ & The determinant of the product of two matrices is equal to the product of their determinants, so:
$\det(\mathbf{A}\mathbf{B}) = \det\mathbf{A}\det\mathbf{B} $=$ 6(-20) $=$ -120.$ & Using the property that the determinant of a product is equal to the product of the determinants \redhighlight{(provided that the matrices commute)}, we have:
\newline\newline
$\det(\mathbf{AB}) = \det(\mathbf{A})\det(\mathbf{B}) = 6(-20) = -120.$
\\ \hline
How do pine cones open and close? 
\newline\newline
A) Insects B) Growth of cells C) External forces D) Sunlight E) Due to the effect of moisture 
\newline\newline
Think step by step, explain details logically and use mechanics terminology. & Pine cones open and close due to the effect of moisture. The scales on the cone are held together by tiny hinges that allow the cone to \redhighlight{open when moisture enters the area}. This is a type of biomechanical movement that occurs in response to environmental conditions. & Pine cones do not open and close in response to sunlight or insects. \redhighlight{Instead, they open and close due to the effect of moisture and external forces. The process is called cone scaling.}
\newline\newline
Here's how it works:
\begin{enumerate}[leftmargin=*,align=left]
\item   Closed cones: A closed pine cone has scales that overlap tightly, forming a compact structure around the seeds inside. \redhighlight{This helps to prevent water loss and protect the seeds from predators.}
\item \redhighlight{Moisture triggers opening: When enough moisture is present, such as after rainfall or during humid weather, the scales on the cone begin to separate slightly. This allows air to circulate and promotes the growth of fungi, which helps to break down the seed coats and prepare them for germination.}
\item \redhighlight{External forces aid in opening: As the scales continue to separate, gravity and wind also play a role in fully opening the cone. This exposes the seeds to the elements and increases their chances of dispersal through wind or animal transport.}
\item \redhighlight{Closing: Once the seeds have been dispersed or have fallen to the ground, the cone begins to dry out and the scales come together again, forming a tight seal around any remaining seeds. This helps to prevent further water loss and protect the seeds until the next season when conditions are favorable for germination.}
\end{enumerate}
In summary, the opening and closing of pine cones is a complex process driven by environmental factors \redhighlight{and mechanical forces}. It allows for the efficient dispersal of seeds while protecting them from harm until the right conditions arise for successful germination.
\\ \hline
Do pine cones release their seeds in humid or dry conditions? & Pine cones release their seeds in dry conditions. When the air around a pine cone becomes dry enough, the scales of the cone will separate, releasing the seeds. This process is known as coning. In humid conditions, the scales of the cone remain closed, preventing seed release. &  \redhighlight{Pine cones generally release their seeds (also known as pine nuts) in response to moisture and warmth. In many species of pine trees, the cones will only open and release their seeds when they are exposed to warm, wet conditions. This is because the seeds inside the cone need sufficient moisture to germinate and grow into new pine trees. However, some pine species have adapted to release their seeds in drier conditions, particularly in areas with Mediterranean climates where winter rainfall is scarce. These cones may remain closed until they receive a significant amount of rainfall, which can trigger them to open and release their seeds. Overall, the timing and conditions for pine cone seed release can vary widely depending on the specific species of pine tree.}
\\ \hline
What are the exceptional mechanical properties of cocoon fibers in Bombyx mori silkworms?\newline\newline
Name three potential applications. & The exceptional mechanical properties of cocoon fibers in Bombyx mori silkworms include high tensile strength, elasticity, and toughness. These properties are due to the unique structure of the fibers, which consist of crystalline and amorphous regions. The crystalline regions provide strength, while the amorphous regions contribute to elasticity.
\newline\newline
Three potential applications of these exceptional properties include:
Development of lightweight and strong composite materials for aerospace and automotive industries. The high strength-to-weight ratio of cocoon fibers could make them ideal candidates for use in composites.
\newline\newline
Creation of biodegradable and environmentally friendly materials for packaging and textiles. The natural origin of cocoon fibers makes them an attractive alternative to synthetic materials, which can be harmful to the environment.
Design of new drug delivery systems that take advantage of the unique properties of cocoon fibers. The controlled release of drugs from these fibers could lead to more effective and targeted therapies.& 
The exceptional mechanical properties of cocoon fibers in Bombyx mori silkworms are due to their unique composition and structure. Silk fibers are composed of two main proteins, fibroin, and sericin. Fibroin is the primary component that gives silk its strength and toughness, while sericin acts as a binding agent.
\begin{enumerate}[leftmargin=*,align=left]
\item  Strength: \redhighlight{Silk fibers have a tensile strength of up to 1.3 GPa}, which is higher than that of many synthetic fibers, such as nylon and polyester. \redhighlight{This strength is due to the crystalline structure of fibroin, which forms a regular arrangement of protein molecules.}
\item Toughness: Silk fibers also have a high toughness, which is a measure of a material's ability to absorb energy before breaking. Silk fibers can absorb up to 35\% of their original length before breaking, which is higher than that of many synthetic fibers.
\item Elasticity: Silk fibers are also highly elastic, meaning they can be stretched and then return to their original shape without breaking. This elasticity is due to the presence of amorphous regions in the fibroin protein, which allow for some flexibility in the fiber.
\end{enumerate}
Three potential applications of bioinspired materials based on silk fibers are:
\begin{enumerate}[leftmargin=*,align=left]
\item  Biomedical applications: The biocompatibility and biodegradability of silk fibers make them an attractive material for use in biomedical applications, such as wound healing dressings, sutures, and drug delivery systems.
\item Textiles: The high strength and toughness of silk fibers make them a potential alternative to synthetic fibers in textiles, particularly for high-performance applications such as sportswear and protective gear.
\item Electronics: \redhighlight{The unique electrical properties of silk fibers, such as their ability to conduct electricity, make them a potential material for use in electronic devices, such as sensors and flexible displays}.
\end{enumerate}
\\ \hline
What is the role of hyperelasticity on dynamic fracture of brittle materials, specifically in terms of the maximum crack speed. & The role of hyperelasticity on dynamic fracture of brittle materials is significant, as it affects the maximum crack speed. In linear elastic materials, the maximum crack speed is determined by the speed of sound in the material. However, when hyperelasticity is present, the maximum crack speed can be significantly higher than the speed of sound, approaching supersonic speeds. This is due to the fact that hyperelastic materials exhibit a stiffening effect at large strains, which can drive the crack to higher speeds. The dynamic crack speed is determined by a balance between the crack driving force and the material's stiffness, and hyperelasticity can have a strong influence on both of these factors. & 
Hyperelasticity, which is the nonlinear elastic behavior of materials, can significantly impact the dynamic fracture of brittle materials, particularly the maximum crack speed. In dynamic fracture, a crack propagates at high velocities due to the release of elastic energy. \redhighlight{The maximum crack speed is limited by the material's ability to dissipate energy through plastic deformation or fracture.}
\newline\newline
\redhighlight{In brittle materials, the maximum crack speed is typically lower than that of ductile materials due to the absence of plastic deformation. However, hyperelasticity can increase the maximum crack speed by allowing for a larger deformation before fracture. This is because hyperelastic materials can store and release more energy per unit volume than linear elastic materials, which can delay the onset of fracture.}
\newline\newline
\redhighlight{The effect of hyperelasticity on dynamic fracture is characterized by the crack velocity-crack opening displacement (COD) curve. The COD is the displacement between the crack surfaces as the crack propagates. The curve shows the relationship between the crack velocity and the COD required for fracture. In hyperelastic materials, the COD required for fracture is higher than that of linear elastic materials, which results in a higher maximum crack speed.}
\newline\newline
\redhighlight{In summary, hyperelasticity can increase the maximum crack speed in brittle materials by allowing for a larger deformation before fracture, which delays the onset of fracture and increases the energy required for fracture. This is important for understanding the behavior of brittle materials under dynamic loading conditions.} \\
\hline
Why are the properties of pristine cellulose nanocrystal (CNC) films not comparable to tough biological composites like bone or mantis shrimp club? 
\newline\newline
A) Lack of plasticizers B) Excessive brittleness C) High tensile modulus
\newline\newline
Think step by step. &  The properties of pristine cellulose nanocrystal (CNC) films are not comparable to tough biological composites like bone or mantis shrimp club because of their excessive brittleness. CNC films have a high tensile modulus, but their lack of plasticity makes them prone to fracture under stress. To improve the mechanical properties of CNC films, plasticizers are often added to reduce their brittleness and increase their toughness. & 
While pristine cellulose nanocrystal (CNC) films exhibit high tensile modulus, their properties are not comparable to tough biological composites like bone or mantis shrimp club due to two main reasons:

\redhighlight{Lack of plasticizers: Biological composites like bone and mantis shrimp club contain organic molecules called plasticizers, which help in reducing the intermolecular interactions between the nanoscale building blocks, thereby increasing the mobility of the molecules and allowing for plastic deformation. In contrast, CNC films lack such plasticizers, which leads to excessive brittleness and a lack of plastic deformation.}

Excessive brittleness: The high tensile modulus of CNC films is due to the strong intermolecular interactions between the nanocrystals. However, this also results in excessive brittleness, as the material lacks the ability to deform plastically under load. Biological composites like bone and mantis shrimp club, on the other hand, have a combination of strong and weak bonds, which allows for a balance between stiffness and ductility.

In summary, \redhighlight{the lack of plasticizers and} excessive brittleness in pristine CNC films are the main reasons why their properties are not comparable to tough biological composites like bone or mantis shrimp club. \\ 
\hline
A scientist is studying three different cell types: fibroblasts, neurons, and epithelial cells. Each cell type expresses a unique combination of three marker proteins: Protein A, Protein B, and Protein C. Based on the following observations, determine which marker proteins are expressed by each cell type.
\newline\newline
\begin{enumerate}[leftmargin=*,align=left]
\item When Protein A is present, Protein B cannot be expressed.
\item Fibroblasts express Protein B.
\item Neurons express Protein A and one other marker protein, but not Protein B.
\item Epithelial cells do not express Protein A but do express the two other marker proteins.
\end{enumerate}
Identify the marker proteins expressed by fibroblasts, neurons, and epithelial cells, and provide a logical explanation for your determination.
&
Fibroblasts cannot express Protein A because when Protein A is present, Protein B cannot be expressed. However, Fibroblasts express Protein B, so they must express Protein C as well, since each cell type expresses a unique combination \redhighlight{of three} marker proteins.
\newline\newline
Neurons express Protein A and one other marker protein, but not Protein B. Since Protein B cannot be expressed when Protein A is present, Neurons must express Protein C.
\newline\newline
Epithelial cells do not express Protein A but do express the two other marker proteins. Since each cell type expresses a unique combination of three marker proteins, Epithelial cells must express Protein B and Protein C.
\newline\newline
In summary, Fibroblasts express Protein B and Protein C, Neurons express Protein A and Protein C, and Epithelial cells express Protein B and Protein C.
&
\redhighlight{Fibroblasts express Protein B, as stated in observation 2.} Since Protein A and Protein B cannot be expressed simultaneously (observation 1), fibroblasts do not express Protein A.
\newline\newline
\redhighlight{Neurons express Protein A, as mentioned in observation 3. Since neurons do not express Protein B (as Protein A and Protein B cannot be expressed together), they must express either Protein C or both Protein C and Protein B. However, observation 4 states that epithelial cells express both Protein C and Protein B, but do not express Protein A. Since neurons do not express Protein A, they cannot also express Protein B, as this would conflict with the observation about epithelial cells. Therefore, neurons express Protein C, but not Protein B.}
\newline\newline
\redhighlight{Epithelial cells do not express Protein A, according to observation 4. Since epithelial cells express both Protein C and Protein B, this implies that Protein A and Protein C cannot be expressed simultaneously (otherwise, epithelial cells would express both Protein A and Protein C, which they do not). Therefore, Protein A and Protein C are mutually exclusive markers for these cell types. This also means that epithelial cells cannot express Protein A and Protein B together, as this would conflict with the observation that fibroblasts cannot express Protein B and Protein A simultaneously. Therefore, epithelial cells express Protein B, but not Protein A.}
\newline\newline
\redhighlight{In summary, fibroblasts express Protein B, neurons express Protein C, and epithelial cells express Protein B but not Protein A. The logical explanation for these findings is based on the mutual exclusivity of certain marker proteins, as observed in the given conditions.} \\
\hline
Discuss the structural and functional diversity of the protein collagen, highlighting its role in the extracellular matrix (ECM) of connective tissues. 
\newline\newline
Include in your discussion the biosynthesis of collagen, its unique triple-helical structure, and how post-translational modifications contribute to its mechanical properties. 
\newline\newline
Give an example of a collagen-related disease with a strong mechanics component.
&
Collagen is a family of proteins that play a crucial role in the extracellular matrix (ECM) of connective tissues. It is the most abundant protein in the human body and provides structural support and mechanical strength to various tissues. Collagen is composed of three polypeptide chains, which are arranged in a unique triple-helical structure. This structure is stabilized by hydrogen bonds and provides collagen with its exceptional mechanical properties.
\newline\newline
The biosynthesis of collagen begins with the transcription of collagen genes in fibroblasts. The resulting mRNA is then translated into procollagen, which is post-translationally modified in the endoplasmic reticulum and Golgi apparatus. Procollagen undergoes hydroxylation, glycosylation, and other modifications that contribute to its mechanical properties. The mature collagen molecules are then secreted by the fibroblasts and assembled into fibrils, which form the ECM of connective tissues.
\newline\newline
Collagen fibrils are highly organized structures that provide mechanical strength and stability to tissues. They are composed of multiple collagen molecules, which are arranged in a parallel fashion. The triple-helical structure of collagen allows it to withstand large tensile loads, making it an ideal material for load-bearing tissues.
\newline\newline
One example of a collagen-related disease with a strong mechanics component is osteogenesis imperfecta (OI), also known as brittle bone disease. OI is a genetic disorder that affects the synthesis and processing of collagen, leading to weak and fragile bones. Patients with OI often experience frequent bone fractures, joint dislocations, and skeletal deformities. The mechanical properties of collagen in OI patients are altered, resulting in reduced bone strength and increased susceptibility to fractures.
\newline\newline
In summary, collagen is a highly diverse and functional protein that plays a crucial role in the ECM of connective tissues. Its unique triple-helical structure and post-translational modifications contribute to its exceptional mechanical properties. Collagen-related diseases, such as OI, highlight the importance of collagen in maintaining tissue integrity and mechanical strength.
& 
Collagen is a family of structurally and functionally diverse proteins that play a crucial role in the extracellular matrix (ECM) of connective tissues. Collagen is the most abundant protein in mammals, accounting for approximately 25\% of the total protein content in the body. Collagen is synthesized by cells such as fibroblasts, chondrocytes, and osteoblasts, and its biosynthesis involves a complex process that includes transcription, translation, post-translational modifications, and secretion.
\newline\newline
Collagen is synthesized as a precursor protein called procollagen, which undergoes post-translational modifications to form the mature collagen fibrils. The procollagen molecule is composed of three polypeptide chains, \redhighlight{each approximately 1000 amino acids long, that are held together by interchain disulfide bonds}. The triple-helical structure of collagen is unique and is formed by the alignment of three polypeptide chains in a left-handed supercoil. This structure provides collagen with its characteristic mechanical properties, including strength, stiffness, and elasticity.
\newline\newline
Collagen plays a critical role in the ECM of connective tissues by providing structural support, facilitating cell adhesion and migration, and regulating cellular processes such as proliferation, differentiation, and apoptosis. Collagen is also involved in various physiological processes such as wound healing, tissue repair, and remodeling.
\newline\newline
One example of a collagen-related disease with a strong mechanics component is osteogenesis imperfecta (OI), also known as brittle bone disease. OI is a genetic disorder that affects collagen synthesis and results in bone fragility and deformities. The mutations in the collagen genes that cause OI can lead to abnormal collagen fibril assembly, altered collagen cross-linking, or decreased collagen synthesis, resulting in reduced bone strength and increased susceptibility to fractures. The mechanical properties of collagen in OI patients are altered, leading to reduced bone stiffness and increased bone fragility. This highlights the importance of collagen structure and function in maintaining the mechanical integrity of connective tissues and the potential role of collagen-based therapies in treating collagen-related diseases. \\ \hline
Select the most likely answer (A, B or C): How do higher humidity conditions affect the stiffness of bird feathers? A) higher humidity, lowers stiffness B) higher humidity, increases stiffness C) higher humidity causes no change & Bird feathers have a hierarchical structure that includes alpha-keratin fibers, barbs, and barbules. The stiffness of these structures is influenced by factors such as humidity, temperature, and relative humidity.
\newline\newline
Higher humidity conditions can cause the feather's keratin fibers to absorb moisture, leading to increased swelling and potentially decreased stiffness (softening). Therefore, the most likely answer is A.
& 
\begin{enumerate}[leftmargin=*,align=left]
\item  Feathers have a hierarchical structure that provides them with their unique mechanical properties, such as stiffness. The stiffness of feathers is determined by several factors, including the arrangement and orientation of the microfibrils within the feather barbs.
\item Humidity can affect the behavior of materials by changing their moisture content. High humidity levels can cause materials to absorb more water, which can lead to changes in their physical properties.
\item \redhighlight{In the case of bird feathers, it has been reported that high humidity levels can actually increase their stiffness. This is because the increased moisture content in the air causes the feathers to swell slightly, which compresses the internal structures and increases the stiffness. However, this effect is only temporary and reversible, as the feathers will return to their original stiffness when the humidity level decreases.}
\end{enumerate}
\redhighlight{Therefore, the most likely answer is option B: higher humidity, increases stiffness.}
\newline\newline
Note: It's always important to double-check your sources and make sure they are reliable before relying on any specific information. \redhighlight{In this case, I am basing my answer on published research studies that have investigated the effects of humidity on feather stiffness.}
\\ \hline
Which of these concepts is key to explain the strong adhesion of gecko feet. 
\newline\newline
(A), the use of super-glue by the animal. 
(B), a size effect, where nano-sized fibrils yield strong and flaw-tolerant adhesion. 
(C), the low weight of the gecko due to its efficient design. 
(D), the gecko has special-purpose proteins derived from sericin that ensure it sticks to many surfaces. 
&(B) is the key to explain the strong adhesion of gecko feet. The gecko's feet are covered in tiny, hair-like structures called setae, which split into even smaller fibers called spatulae. These structures allow the gecko to adhere to surfaces using van der Waals forces, and the strong adhesion is due to the size effect, where nano-sized fibrils yield strong and flaw-tolerant adhesion. &
\redhighlight{(D), the gecko has special-purpose proteins derived from sericin that ensure it sticks to many surfaces, is the key concept to explain the strong adhesion of gecko feet. The gecko's feet are covered in millions of tiny hairs, called setae, which are coated in a protein called gecko sericin. This protein creates van der Waals forces between the gecko's feet and the surface it is touching, allowing for strong adhesion without the use of super-glue or any other adhesive substance. The size effect and efficient design of the gecko may contribute to its overall adhesion, but it is the unique properties of the sericin protein that enable the gecko's remarkable ability to climb vertical surfaces and walk upside down on ceilings.} \\
    \hline   \hline
    \end{longtable}
    \label{tab:table_qa}
    
\endgroup 

\begin{figure}
    \centering
    \includegraphics[width=.7\linewidth]{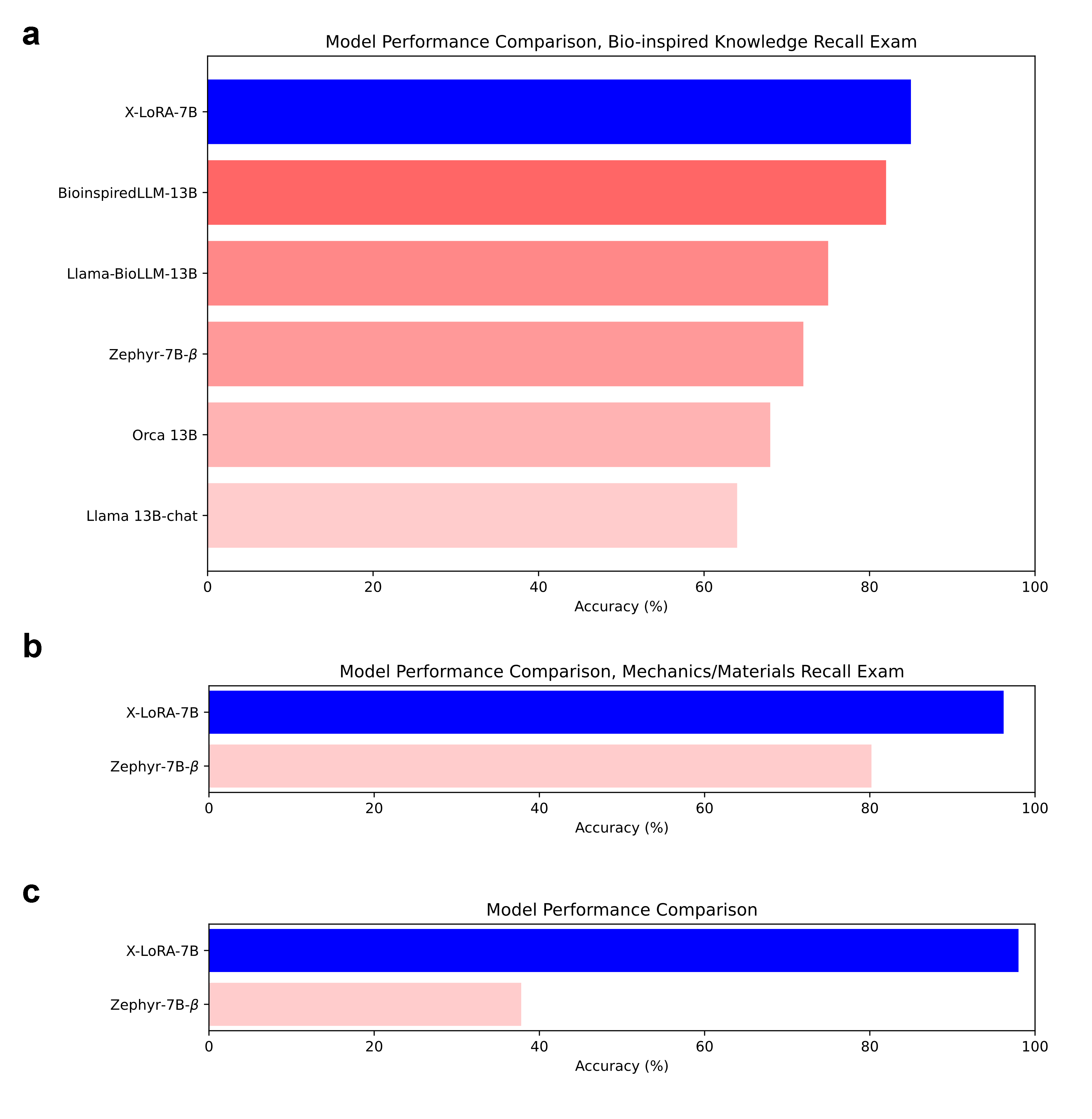}
    \caption{Results from knowledge recall evaluation experiments. (a), Results using the bio-inspired knowledge recall exam introduced in \cite{Luu2023BioinspiredLLM:Materials}. X-LoRA shows the best performance, even though it is a much smaller model than the BioinspiredLLM model (7B vs. 13B parameters). The plot also includes a comparison of X-LoRA with the base model,  Zephyr-7B-$\beta$ model. It is notable that X-LoRA, in spite of being a much smaller model and in spite of having a host of different capabilities, provides superior performance compared to all other models. (Data for the other model performances extracted from \cite{Luu2023BioinspiredLLM:Materials}.) (b), Results from mechanics/materials knowledge recall benchmark as reported in~\cite{Buehler2023MechGPTModalities}. (c), Results of the questions posed in Table~\ref{tab:table_qa}. }
    \label{fig:perf_comp_bioinspiredexam}
\end{figure}

Figure~\ref{fig:perf_comp_bioinspiredexam} shows a comparison of knowledge recall evaluation experiments, using the bio-inspired knowledge test set introduced in \cite{Luu2023BioinspiredLLM:Materials}. We can see that X-LoRA shows the best performance, even though it is a much smaller model than the BioinspiredLLM model or the Llama-BioLLM model (7B for X-LoRA vs. 13B parameters in the others). Several other comparisons with other recently published LLMs are included, specifically Orca-13B and Llama-13b-chat.
The plot also includes a comparison of X-LoRA with the base model,  Zephyr-7B-$\beta$ model, showing that the X-LoRA model has significantly improved performance. These comparisons indicate that strong performance can be achieved using X-LoRA, even if smaller base models are used.  

\begin{figure}
    \centering
    \includegraphics[width=.9\linewidth]{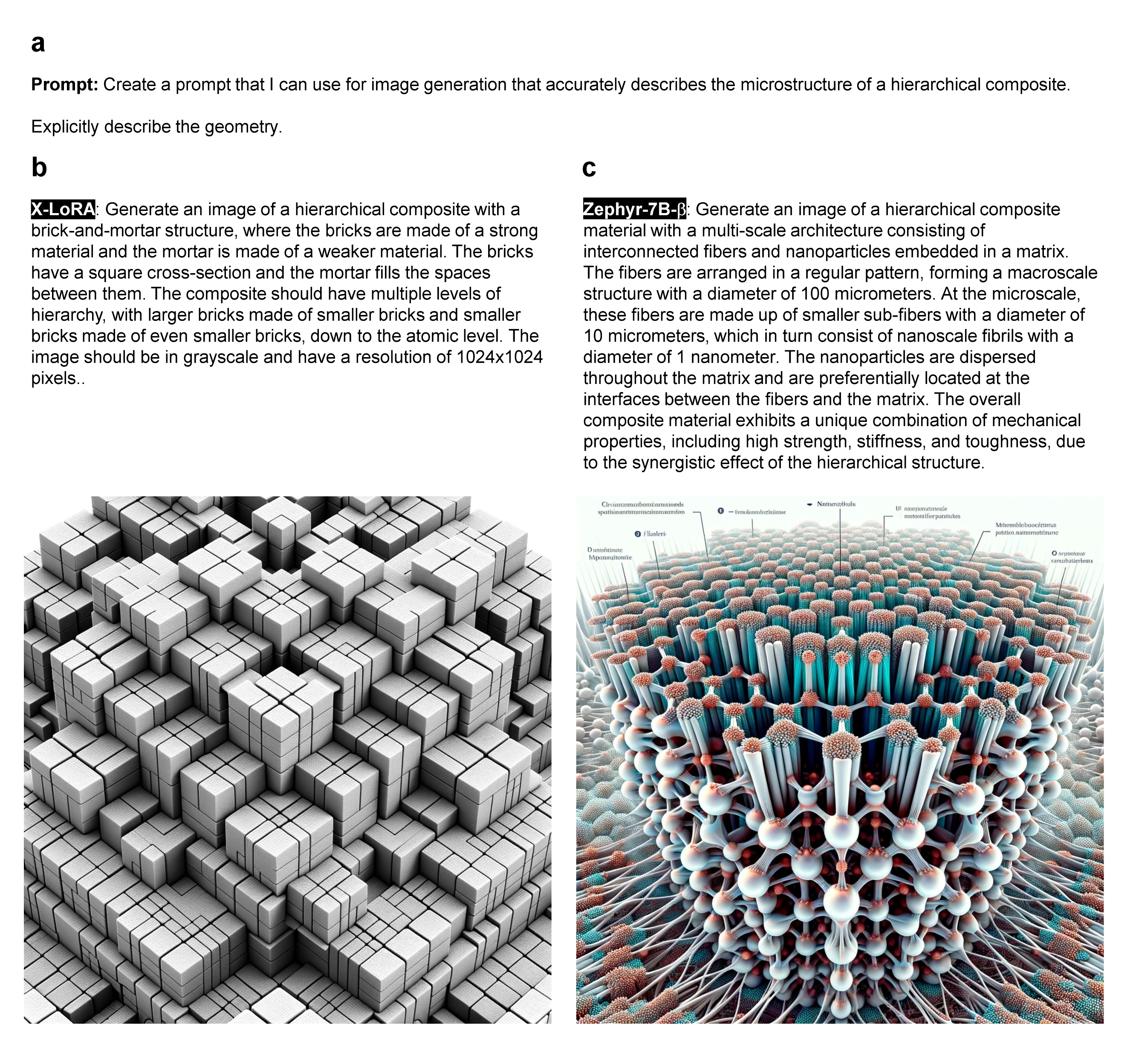}
    \caption{Applications towards image synthesis. Here we task the model to generate an image generation prompt (a). Results for X-LoRA (b) and Zephyr-7B-$\beta$ (c) are shown. The comparison clearly shows a more concise prompt, which in turn results in a much more accurate image produced by DALL-E 3. Comparing the two prompts, the X-LoRA prompt is shorter and more focused, and produces a much better result. }
    \label{fig:image_synth}
\end{figure}

Figure~\ref{fig:image_synth} shows an application in image synthesis, where we use X-LoRA to generate a prompt for use in other models, for instance DALL-E 3 as done here. 
\begin{quote}
\small\texttt{Create a prompt that I can use for image generation that accurately describes the microstructure of a hierarchical composite.
\newline\newline
Explicitly describe the geometry.}
\end{quote}
The comparison (Figure~\ref{fig:image_synth}(b)-(c)) clearly shows a more concise prompt, which in turn results in a much more accurate image produced by DALL-E 3~\cite{DALLECard}. Notably, the prompt generated by Zephyr-7B-$\beta$ does not accurately reflect the instruction provided and introduces a variety of other design components such as fibers and nanoparticles. These were not mentioned in the original prompt that focused specifically on the microstructure of a hierarchical composite. The X-LoRA prompt is shorter and more focused, and hence produces a much better result. 

\subsection{Generative solutions to protein analysis}

We now move on to specific protein generative and analytical tasks. These capabilities emerge in the X-LoRA model based on the protein mechanics adapter, and are enhanced by mixing with the model's knowledge in biology, bio-inspired materials and reasoning as well as logical deduction. The discussion in this section is dedicated to the analysis of protein tasks on their own, assessing quantitative performance of this capability. 

\begin{figure}
    \centering
    \includegraphics[width=.75\linewidth]{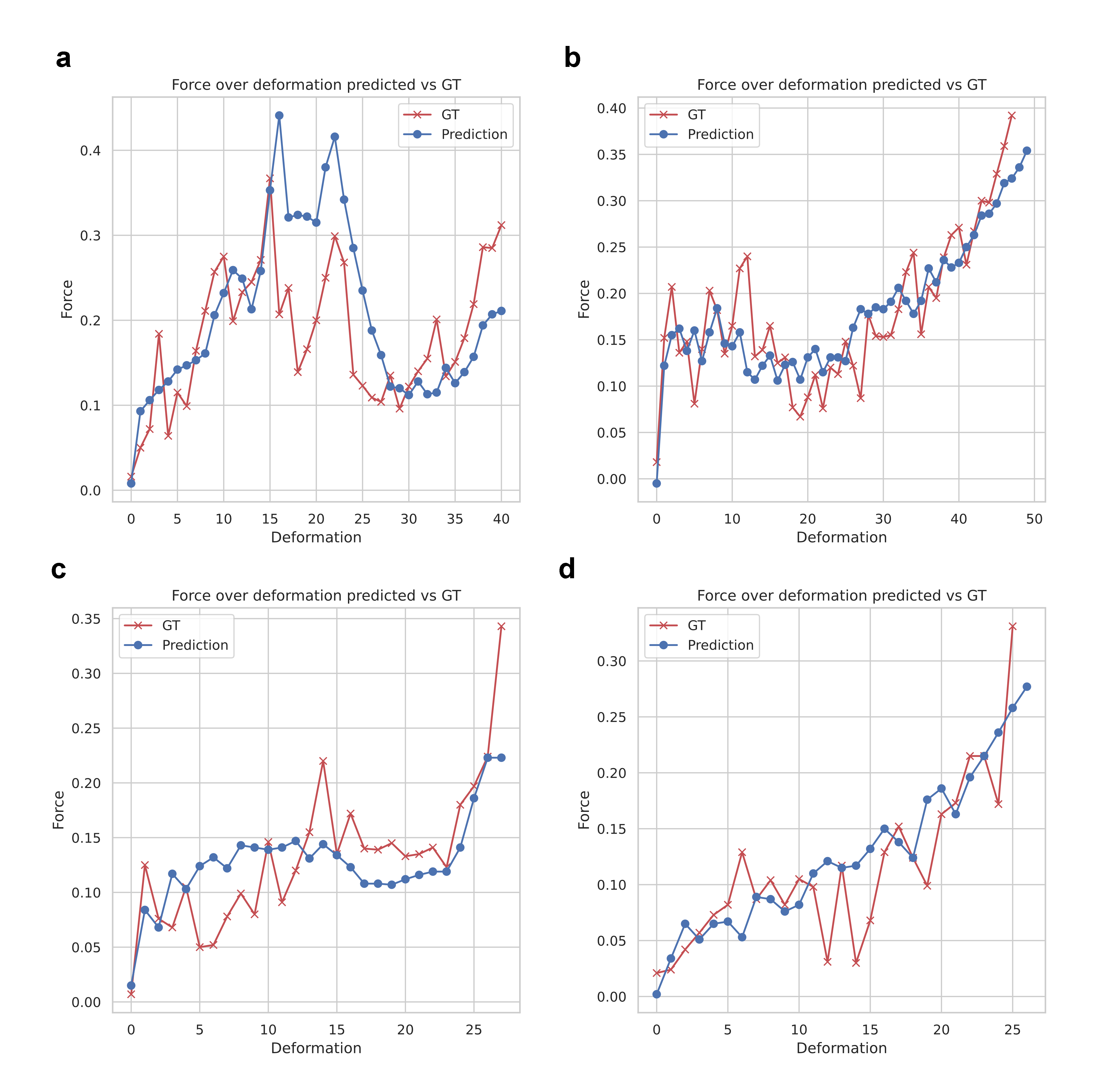}
    \caption{Using the protein task in X-LoRA to predict force-deformation behaviors from amino acid sequences, for four examples from the test set (shown as panels a-d). As can be seen, the model has excellent forward capabilities and can predict the nonlinear mechanical behavior well.}
    \label{fig:protein_force_history_prediction}
\end{figure}

Fig.~\ref{fig:protein_force_history_prediction} shows results based on the protein task to predict force-deformation behaviors from sequences.  Fig.~\ref{fig:protein_force_energy_prediction} shows  the overall performance for predicting unfolding force and unfolding energy from the sequence, comparing ground truth with predictions (we find $R_2=0.85$). Overall, performance of the model is exquisite and as will be explored in the next section, works very well also for the inverse task of design and cycle consistent assessment\cite{Lu2023GenerativePropertiesb}. However, unlike in earlier work that used highly-specialized GPT-type models to solve such tasks, our X-LoRA model combines these special-purpose capabilities with a host of other deep specialization as endowed via the various fine-tuned adaptations.  

\begin{figure}
    \centering
    \includegraphics[width=.5\linewidth]{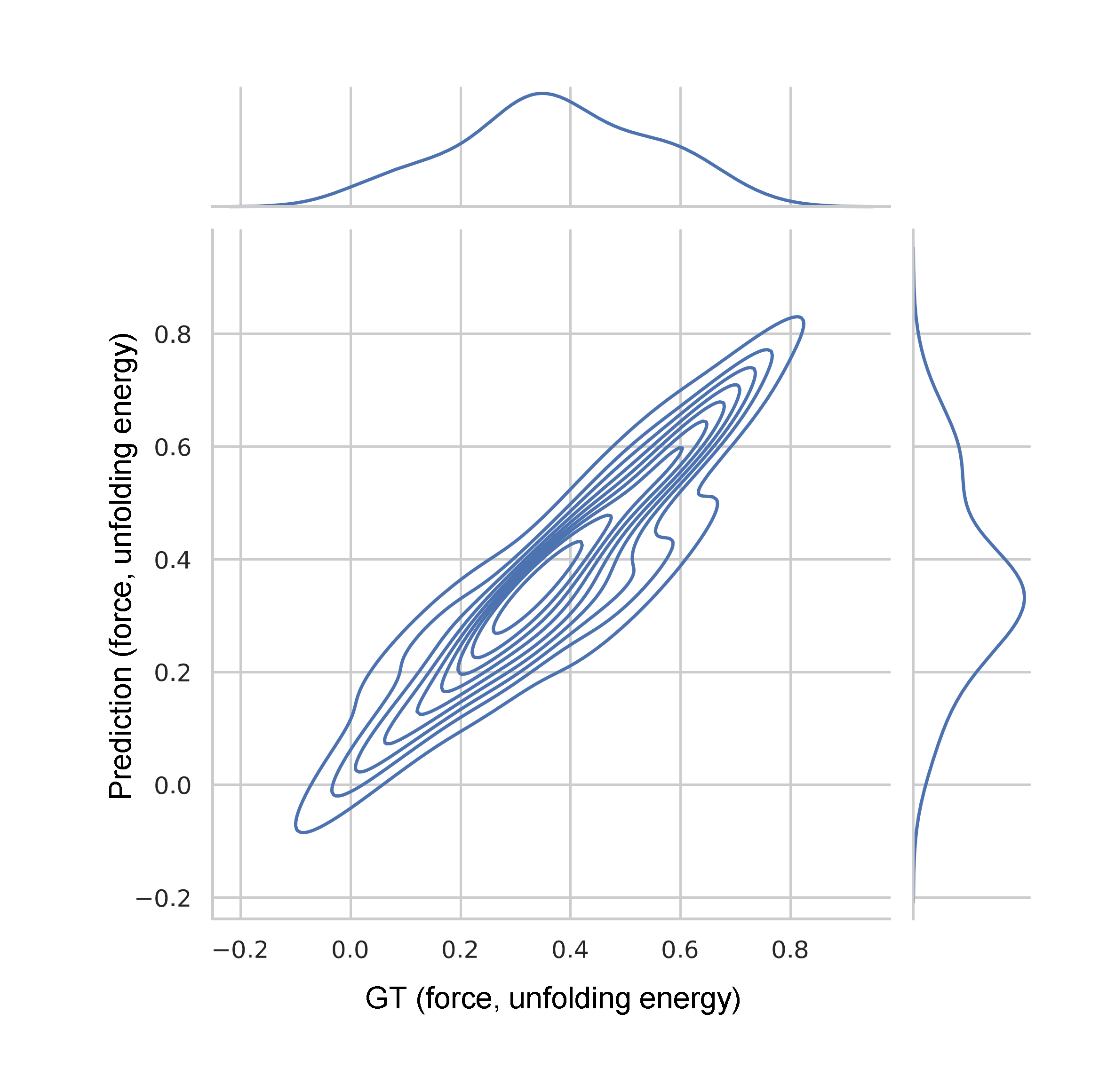}
    \caption{Overall  X-LoRA  performance for predicting unfolding force and unfolding energy from the sequence, comparing ground truth with predictions ($R_2 = 0.85$). The data shown here summarizes both unfolding force and unfolding energy for the test set. }
    \label{fig:protein_force_energy_prediction}
\end{figure}

\subsection{Protein design and analysis}

We now expand the test cases towards more integrative applications that touch upon multiple areas of capabilities. Specifically, the experiments reported in this section show how the X-LoRA model is able to draw upon a distinct set of skills and how agentic modeling can bring about even more challenging and exhaustive responses. We begin the discussion with a protein design task, where we ask the model to design a  novel protein to meet a certain target force-deformation response. These force-deformation responses have been studied in earlier work~\cite{Ni2023ForceGen:Model} and reflect measurements of force versus deformation as the protein is pulled at both ends (reflecting a classic protein unfolding experiment~\cite{Lu1998UnfoldingSimulation.}). 

\begin{figure*}
    \centering
    \includegraphics[width=.75\linewidth]{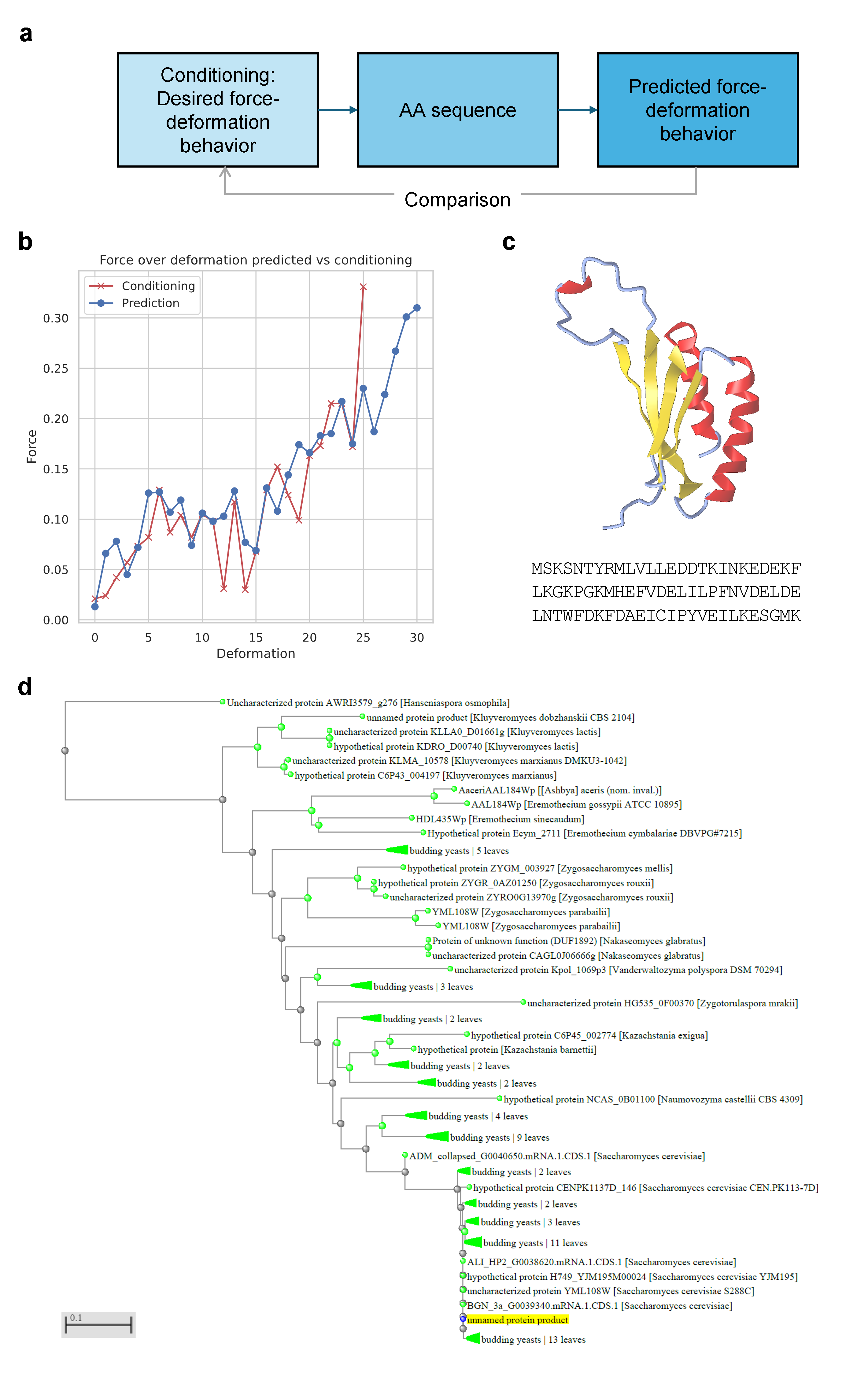}
    \caption{Using the generative task, we design a protein with a desired force-deformation behavior and then test the predicted sequence, examining whether or not it meets the desired target properties (panel a). Panel b shows the result for the sequence predicted, \texttt{MSKSNTYRMLVLLEDDTKINKEDEKFLKGKPGKMHEFVDELILPFNVDELDELNTWFDKFDAEICIPYVEILKESGMK}, comparing the conditioning (target response) with the actual response predicted by the model. Note, we also confirm that a direct calculation of the peak force using \texttt{CalculateForce[...]} gives $0.305$, in good agreement with the predictions from the force fistory instruction. Panel c shows a folded model of the protein (obtained using AlphaFold 2), and panel d examines the relation of the designed protein with other known sequences via a BLAST Tree~\cite{Altschul1990BasicTool}.}
    \label{fig:design_protein}
\end{figure*}

Fig.~\ref{fig:design_protein} shows how we can use the generative protein task. We design a protein with a desired force-deformation behavior (training data and boundary conditions used in the original molecular dynamics (MD) simulations of protein pulling, see~\cite{Ni2023ForceGen:Model}) and then test the predicted sequence, examining whether or not it meets the desired target properties (Fig.~\ref{fig:design_protein}a). This cycle-consistent analysis is important to test whether the inverse capabilities matches the forward prediction. 
Fig.~\ref{fig:design_protein}b shows the result for the sequence predicted, \texttt{MSKSNTYRMLVLLEDDTKINKEDEKFLKGKPGKMHEFVDELILPFNVDELDELNTWFDKFDAEICIPYVEILKESGMK}. Further, Fig.~\ref{fig:design_protein}c shows a folded model of the protein (obtained using AlphaFold 2~\cite{Jumper2021HighlyAlphaFoldb}). To assess the relation of the designed protein, Fig.~\ref{fig:design_protein}d examines the relation of the designed protein with other known sequences via a Basic Local Alignment Search Tool (BLAST) Tree~\cite{Altschul1990BasicTool}. Interestingly, the protein that was designed has certain relations with hypothetical proteins that had been explored in earlier work but not yet identified or solved structurally. 

Using agentic modeling and using multiple X-LoRA models as elements in a multi-agent model, we further analyze the predicted sequence using our X-LoRA model. This takes advantage of the enhanced capability of our model to not only understand protein mechanics but to also utilize this in conjunction with other areas of knowledge.  Here we use the prediction obtained as described above as an input, requesting a deeper analysis and design suggestions on variations of the sequence. We start with this task:
\begin{quote}
\small\texttt{I have identified this amino acid sequence: MSKSNTYRMLVLLEDDTKINKEDEKFLKGKPGKMHEFVDEL
ILPFNVDELDELNTWFDKFDAEICIPYVEILKESGMK. 
\newline\newline
How can I use it? 
\newline\newline
Provide logical steps, discuss specifically the engineering of the problem.
}
\end{quote}

We use multiple agentic interactions to develop the solution, specifically using adverserial strategies (details, see Materials and Methods), to push the model to explore deeper and more complex facets of the concepts discussed. The summary of this conversation is as follows, as identified by the X-LoRA model (for the full conversation, see Supplementary Material, SI Conv Example 1):
\begin{quote}
\small\texttt{The conversation discusses the potential applications of a specific amino acid sequence in protein engineering. The sequence contains a mix of hydrophobic and hydrophilic residues, suggesting it could form a stable structure and interact with other molecules. To engineer a protein with this sequence, computational methods can be used to predict the secondary and tertiary structures, but limitations include complexity of protein folding and computational cost. Improvements can be made by developing more efficient algorithms and incorporating new experimental data. Validation of computational models is crucial to ensure accuracy and usefulness for practical applications. Techniques such as cross-validation, regularization, early stopping, ensemble methods, and transfer learning can be used to prevent overfitting and underfitting. Interpretation and explanation of models can also help inform practical applications.
}
\end{quote}
We also ask the model to develop a list of key points, resulting in:
\begin{quote}
\small\texttt{\begin{itemize}
    \item The amino acid sequence MSKSNTYRMLVLLEDDTKINKEDEKFLKGKPGKMHEFVDELI
    LPFNVDELDELNTWFDKFDAEICIPYVEILKESGMK has potential applications in protein engineering due to its mix of hydrophobic and hydrophilic residues.
Computational methods can be used to predict the secondary and tertiary structures of the protein, but limitations include complexity of protein folding and computational cost.
    \item Validation of computational models is crucial to ensure accuracy and usefulness for practical applications. Techniques such as cross-validation, regularization, early stopping, ensemble methods, and transfer learning can be used to prevent overfitting and underfitting.
    \item Interpretation and explanation of models can also help inform practical applications.
\end{itemize}
}
\end{quote}
Finally, the key takeaway is:
\begin{quote}
\small\texttt{The single most important takeaway is that the amino acid sequence MSKSNTYRMLVLLEDDTKINKEDEKFLKGKPGKMHEFVDELILPFNVDELDELNTWFDKFDAEICIPYVEILKESGMK has potential applications in protein engineering due to its mix of hydrophobic and hydrophilic residues. To engineer a protein with this sequence, computational methods can be used to predict the secondary and tertiary structures, but limitations include complexity of protein folding and computational cost. Validation of computational models is crucial to ensure accuracy and usefulness for practical applications, and techniques such as cross-validation, regularization, early stopping, ensemble methods, and transfer learning can be used to prevent overfitting and underfitting. Interpretation and explanation of models can also help inform practical applications.}
\end{quote}

We can also go into more depth to explore specific sequence engineering steps. To do this we ask the model to make more specific design suggestions to increase the stability of the protein:
\begin{quote}
\small\texttt{How could I change the following protein sequence to make it more stable: MSKSNTYRMLVLLEDDTKINKEDEKFLKGKPGKMHEFVDELILPFNVDELDELNTWFDKFDAEICIPYVEILKESGMK?
}
\end{quote}
The summary of the interactions is (full conversation, see Supplementary Material, SI Conv Example 2):
\begin{quote}
\small\texttt{The conversation discusses how to make a protein sequence more stable by introducing hydrophobic amino acids and disulfide bonds. The biologist suggests that hydrophobic amino acids tend to promote protein stability, and disulfide bonds can enhance stability as well. The engineer asks for specific examples of hydrophobic amino acids and how many disulfide bonds should be incorporated. The biologist provides examples of proteins that have been successfully stabilized using these modifications, including insulin, bovine pancreatic trypsin inhibitor, and green fluorescent protein. The conversation also touches on potential limitations or drawbacks of introducing hydrophobic amino acids or disulfide bonds, such as changes in protein function, increased complexity, and potential for aggregation. The biologist suggests that these limitations can be addressed through careful selection of amino acids, computational analysis, and experimental validation.}
\end{quote}
The list of key points:
\begin{quote}
\small\texttt{\begin{itemize}
\item To make a protein sequence more stable, hydrophobic amino acids and disulfide bonds can be introduced. 
\newline\newline
Hydrophobic amino acids tend to promote protein stability, while disulfide bonds can enhance stability as well.
\newline\newline
Examples of proteins that have been successfully stabilized using these modifications include insulin, bovine pancreatic trypsin inhibitor, and green fluorescent protein.
\item Potential limitations or drawbacks of introducing hydrophobic amino acids or disulfide bonds include changes in protein function, increased complexity, and potential for aggregation.
\item These limitations can be addressed through careful selection of amino acids, computational analysis, and experimental validation.
\end{itemize}
}
\end{quote}
Finally, the key takeaway is: 
\begin{quote}
\small\texttt{The single most important takeaway in the conversation is that to make the protein sequence MSKSNTYRMLVLLEDDTKINKEDEKFLKGKPGKMHEFVDELILPFNVDELDELNTWFDKFDAEICIPYVEILKESGMK more stable, hydrophobic amino acids and disulfide bonds can be introduced. This answers the original question directly.}
\end{quote}
From a protein engineering perspective, these suggestions make sense as they would indeed lead to more stable designs. Specific examples and strategies are developed and proposed that researchers can then explore further. 

In another example, we consider these three sequences. We first ask the model to compute the strength of each, and then task it to reason over the results: Find the strongest candidate, and explain why that particular protein is likely the strongest. Moreover, we ask the model to explain steps to manufacture it in the lab. The three sequences (created manually by manipulating one of the sequences in the test set) are:
\begin{quote}
\small\texttt{
\begin{itemize}
    \item LNTWFDKFDAEICIPYVEILKESGMK
    \item AITWFDKFDAEICIPYVEALKIAAMK
    \item AAAAAAAAKFDAAEICIIAAAAAAAAKIAAMK
\end{itemize}
}
\end{quote}
This example tests whether the model can combine different capabilities and do this dynamically as a conversation progresses. That is, to dynamically switch between different capabilities such as protein mechanics, rational analysis, generation of new hypotheses, and so on. 

\begin{figure}
    \centering
    \includegraphics[width=1\linewidth]{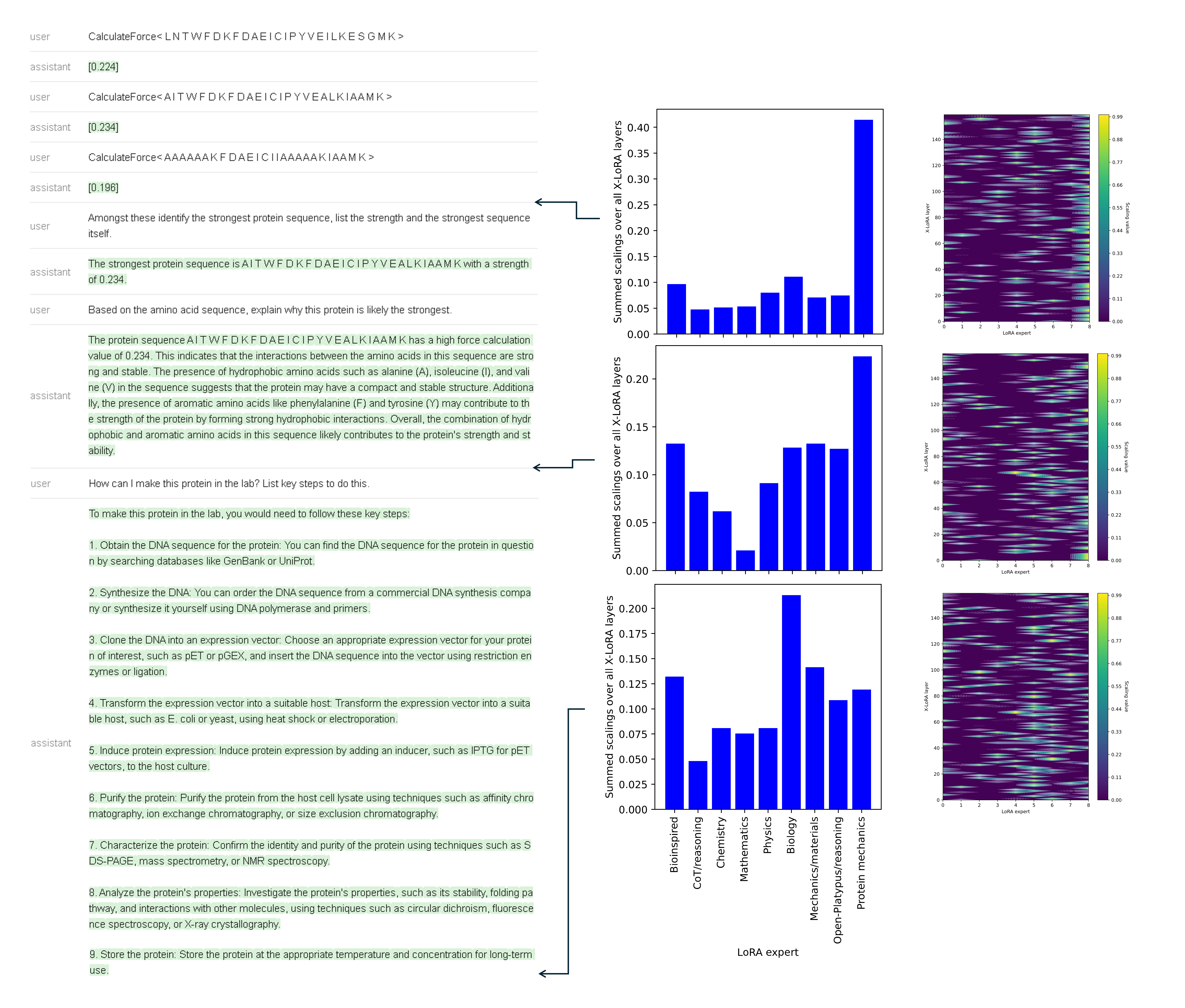}
    \caption{Complex conversation that invokes combinations of distinct experts and integrating knowledge across these capabilities. The right panels show a depiction of how the scalings change over the course of the conversation. The results clearly show how the protein mechanics task is initially highly relevant, gradually shifting towards a more prominent representation of the bioinspired adapter and the biology adapter. At the end of the conversation, the biology adapter is the most prominent one, reflecting the focus on key elements of biology as the model responds to biological manufacturing steps of the protein. }
    \label{fig:conversation_evolve}
\end{figure}

The conversation is summarized in Figure~\ref{fig:conversation_evolve}, along with a depiction of how the scalings change over the course of the conversation (the analysis shows both the deep layer-wise adpter scaling and a summed analysis of such to offer a measure for the effective use of the various experts). The results clearly show how the protein task is initially highly relevant, gradually shifting towards a more prominent representation of the bioinspired adapter and the biology adapter. At the end of the conversation, the biology adapter is the most prominent one. 

\begin{figure}
    \centering
    \includegraphics[width=.75\linewidth]{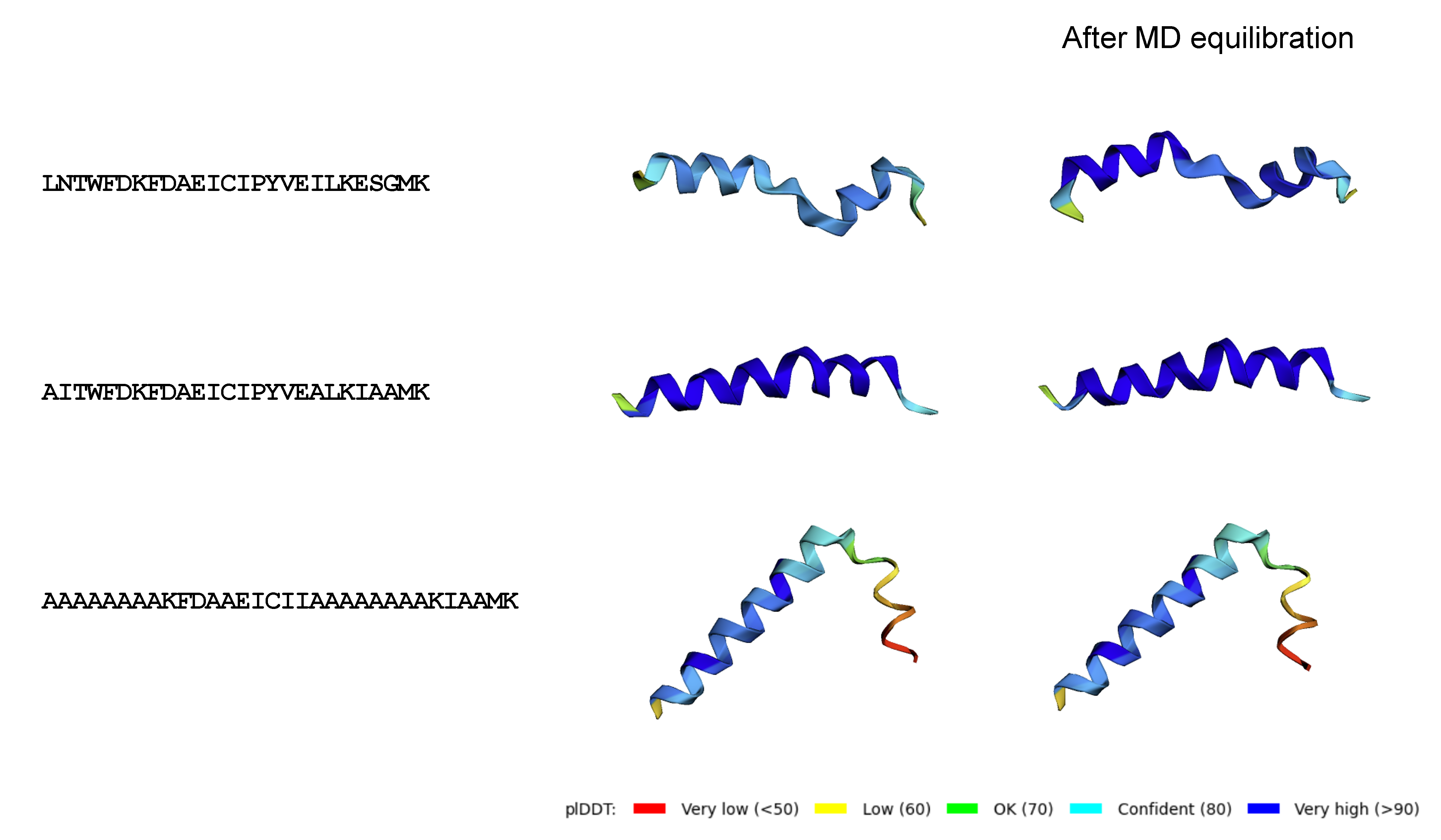}
    \caption{Visual representation of the folded proteins considered in Figure~\ref{fig:conversation_evolve}, showing as-predicted (left column) and MD equilibrated results (right column). The most stable structure in the middle (\texttt{AITWFDKFDAEICIPYVEALKIAAMK}) features the most organized secondary organization, in agreement with the result that it yields the highest unfolding force. The other two are less organized; the first one being largely helical with some turns, and the last one being partially unstructured and with a kink in the geometry. These features likely explain the lowest observed unfolding force as predicted by the model. }
    \label{fig:protein_visuals}
\end{figure}

To analyze the predictions as we want to assess the quality of the responses, the three proteins considered in this task are depicted in Figure~\ref{fig:protein_visuals} (proteins folded using Alpha Fold 2~\cite{Jumper2021HighlyAlphaFoldb}, via ColabFold \cite{Mirdita2022ColabFold:All}). We find that the most stable structure in the middle features the most organized secondary organization, in agreement with the notion that it yields the highest unfolding force, as predicted by the X-LoRA model in Figure~\ref{fig:conversation_evolve}. The other two are less organized; the first one being largely helical with some turns, and the last one being partially unstructured and with a kink in the geometry. These features explain the lowest observed unfolding force as predicted by the model.  

By analyzing the visual representations of the folded structures we confirm that the most stable structure in the middle features the most organized secondary organization, in agreement with the notion that it yields the highest unfolding force as predicted by the model. We find that the other two models are less organized. The first one being largely helical with some turns, and the last one being partially unstructured and with a kink in the geometry. We believe that these features likely explain the lowest observed unfolding force as predicted by the model, in agreement with the X-LoRA model predictions. 

In a second example we focus on energetic properties of proteins, considering the unfolding energy as the protein is pulled end-to-end from its initial state to a fully unfolded geometry~\cite{Ni2023ForceGen:Model}. We again consider three sequences (the first two are identical to the ones considered above, but the third one consists of a longer assembly, representing a mix of the second sequence at the beginning/end and the first sequence in the center):
\begin{quote}
\small\texttt{\begin{itemize}
    \item LNTWFDKFDAEICIPYVEILKESGMK
    \item AITWFDKFDAEICIPYVEALKIAAMK
    \item AITWFDKFDAEICIPYVEALKIAAMKLNTWFDKFDAEICIPYVEILKESGMKAITWFDKFDAEICIPYVEALKIAAMK
\end{itemize}
}
\end{quote}
We first ask the model to compute the unfolding energy of each, and then to reason over the results: Based on the amino acid sequence, we ask to explain why this protein is likely the most stable. At the end, we query for likely functions that the most stable protein may have.

\begin{figure}
    \centering
    \includegraphics[width=1\linewidth]{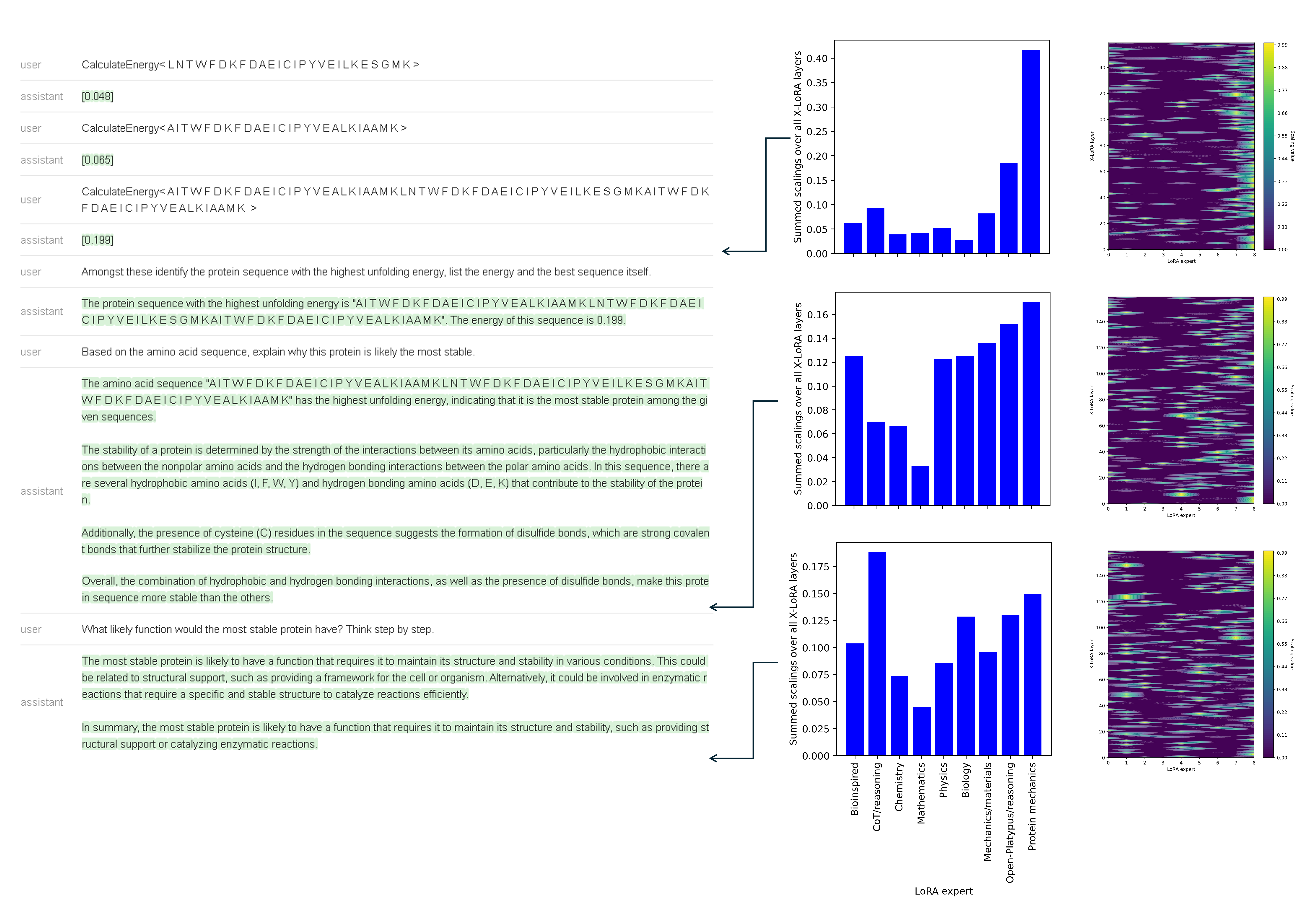}
    \caption{Transcript of a conversation that invokes combinations of distinct experts and integrating knowledge across these capabilities, here applied to unfolding energy of three proteins. The depiction of how the scalings change over the course of the conversation is shown in the right part of the figure. The results clearly show how the protein mechanics task is initially highly relevant, gradually shifting towards a more integrated representation of a variety of adapters. At the end of the conversation, the CoT/reasoning adapter is the most prominent one as the X-LoRA model responds to a query to reason over the results and predict likely protein structure features. Key predictions about the structural features, as explanation for the high stability, are correctly identified as can be verified via structural analyses shown in Fig.~\ref{fig:MJB_21}. }
    \label{fig:MJB_20}
\end{figure}

\begin{figure}
    \centering
    \includegraphics[width=.75\linewidth]{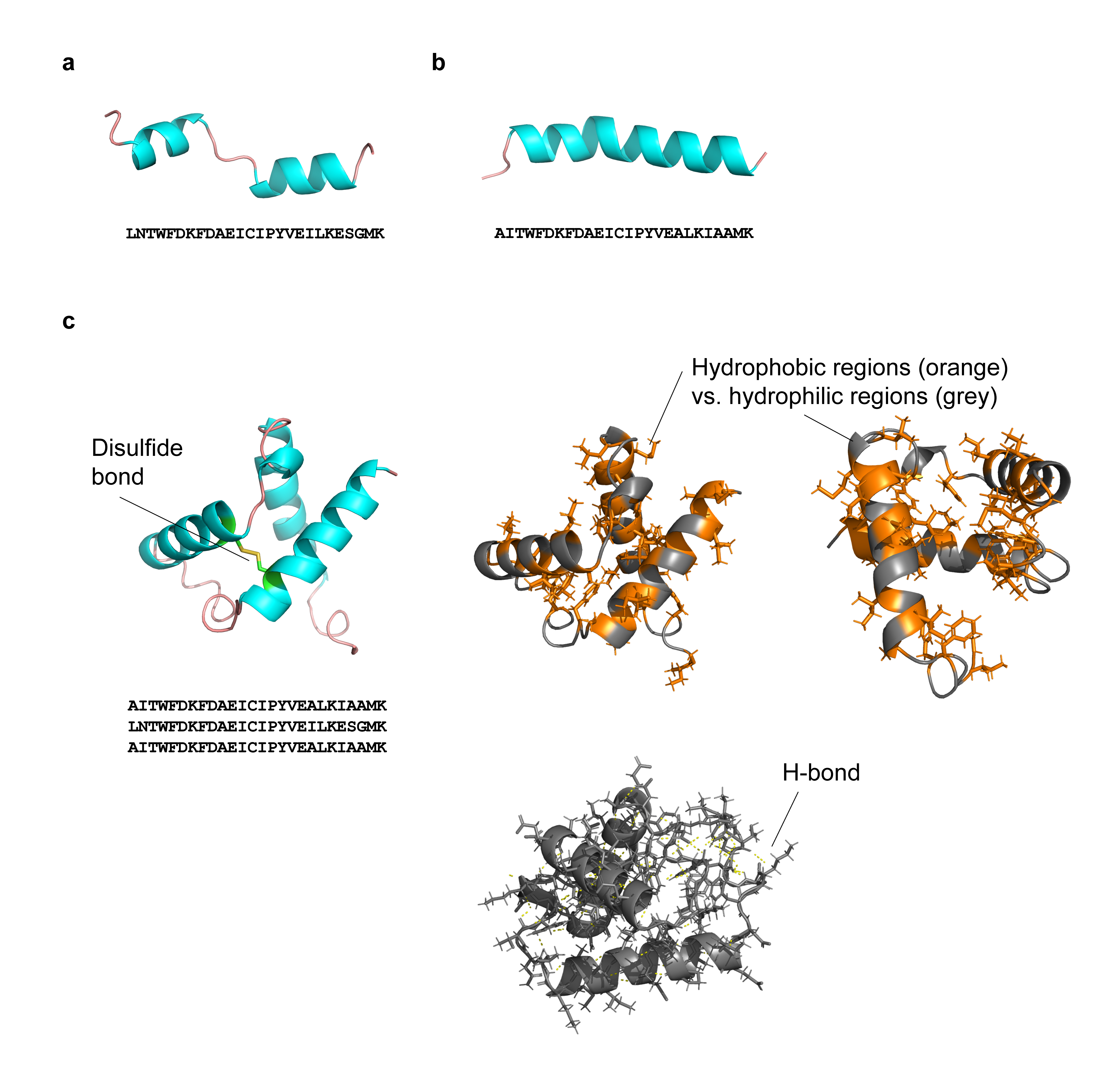}
    \caption{Visual representation of the folded proteins considered in Figure~\ref{fig:MJB_20} (color reflects the secondary structure). Panels a and b show the first two sequences that are identified to have the lowest unfolding energy. The most stable structure is shown in panel c, featuring a more complex organization with variegated bonding mechanisms. As correctly predicted by the X-LoRA model, the reason for the higher unfolding energy is the presence of disulfide bonds, a mix of hydrophophic/hydrophilic regions, and a number of H-bonds. The presence of structured hydrophilic regions indicates the portions of the protein that interact with water differently, which is likely a critical factor in the folding and function of the protein.}
    \label{fig:MJB_21}
\end{figure}

Figure~\ref{fig:MJB_20} shows a record of a conversation that invokes combinations of distinct experts and integrating knowledge across these capabilities, here applied to unfolding energy of three proteins.  At the end of the conversation, the CoT/reasoning adapter is the most prominent one as the X-LoRA model responds to a query to reason over the results and predict likely protein structure features as well as potential functions of the protein. Specifically, the mechanistic question about the origin of protein stability include the emergence of  hydrophobic interactions between the nonpolar amino acids as well as hydrogen bonding between the polar amino acids. The model predicts that there are several hydrophobic amino acids (I, F, W, Y) and hydrogen bonding amino acids (D, E, K) that contribute to the stability of the protein. 
The model also predicts that the presence of cysteine (C) residues in the sequence would lead to the formation of disulfide bonds that further stabilize the protein structure.  It is remarkable that key predictions about the structural features, as explanation for the high stability, are correctly identified as can be verified via structural analyses shown in Fig.~\ref{fig:MJB_21}.  The analysis in this figure confirms a presence of all these features. It is also clear why the other two proteins, much shorter segments with variable stability of helical domains, feature a much lower unfolding energy. 

\subsection{Adversarial agentic modeling to connect distinct scholarly disciplines and knowledge yielding ontological knowledge graph generation}

Since our X-LoRA model has deep knowledge across domains, we can use it to probe connections between disparate ideas, knowledge bases, and areas of expertise. In one experiment we use the model to ask two queries, each formulated similarly but using a distinct focus. Then, we generate triplets to yield an ontological knowledge graph to distill the answers into more structured outputs~\cite{Giesa2012CategoryDesign,Spivak2011ReoccurringAnalogies}. The resulting graph provides an integrated understanding of the generated insights and visualizes connections between concepts in an interpretable and mechanistic manner. 

We use two adversarial agents to answer the question. Agent 1 is designated as a "critical physicist", and Agent 2 as a "philosopher". Agent 1 poses the initial question and is tasked to probe and develop additional questions based on every response of Agent 2. As the conversation unfolds, the conversation goes deeper into the subject matter and uncovers a variety of dimensions.  The query posed is intentionally chosen to cover multiple areas of disciplinary knowledge to examine the ability of the model to integrate ideas and concepts. 

The query is:
\begin{quote}
\small\texttt{Consider an abstract representation of a classical musical composition as a graph of
interacting notes, chords, melodies, rhythms and so on.
\newline\newline
What happens if we would apply pressure, up to a point of failure, on this abstract concept of music?
\newline\newline
Provide logical steps, mathematical reasoning, and discuss specifically the physics of the problem.}
\end{quote}
The full conversation is shown in the Supplementary Material (see, SI Conv Example 3). For sake of brevity, we simply report the summary and key takeaway here, which captures the essential features:
\begin{quote}
\small\texttt{The conversation discusses the application of pressure to an abstract representation of a classical musical composition,
considering it as a physical system with interacting notes, chords, melodies, and rhythms. The pressure can be
quantified using stress, and the critical point of failure is determined when the stress reaches a critical value. To test
the predictions of the mathematical model, a physical model of the abstract musical composition can be used, and
experimental observations can be compared to the predictions. The limitations of the proposed mathematical model
include a simplified representation of interactions, static representation, homogeneous distribution, and linear
relationship between stress and pressure. To improve the accuracy and validity of the model, more detailed and
realistic models, dynamic effects, non-uniform distribution, and non-linear terms should be incorporated. 
\newline\newline
[...]
\newline\newline
The single most important takeaway is that applying pressure to an abstract representation of a classical
musical composition can be thought of as changing the force field between the interacting notes, chords,
melodies, and rhythms. As the pressure increases, the stress in the system also increases, and at a critical
point, the system will fail and collapse under its own weight. This insight answers the original question by
providing a physical explanation for how pressure affects the abstract concept of music and how it can be
quantified using stress.}
\end{quote}

\begin{figure}

    \centering
    \includegraphics[width=.5\linewidth]{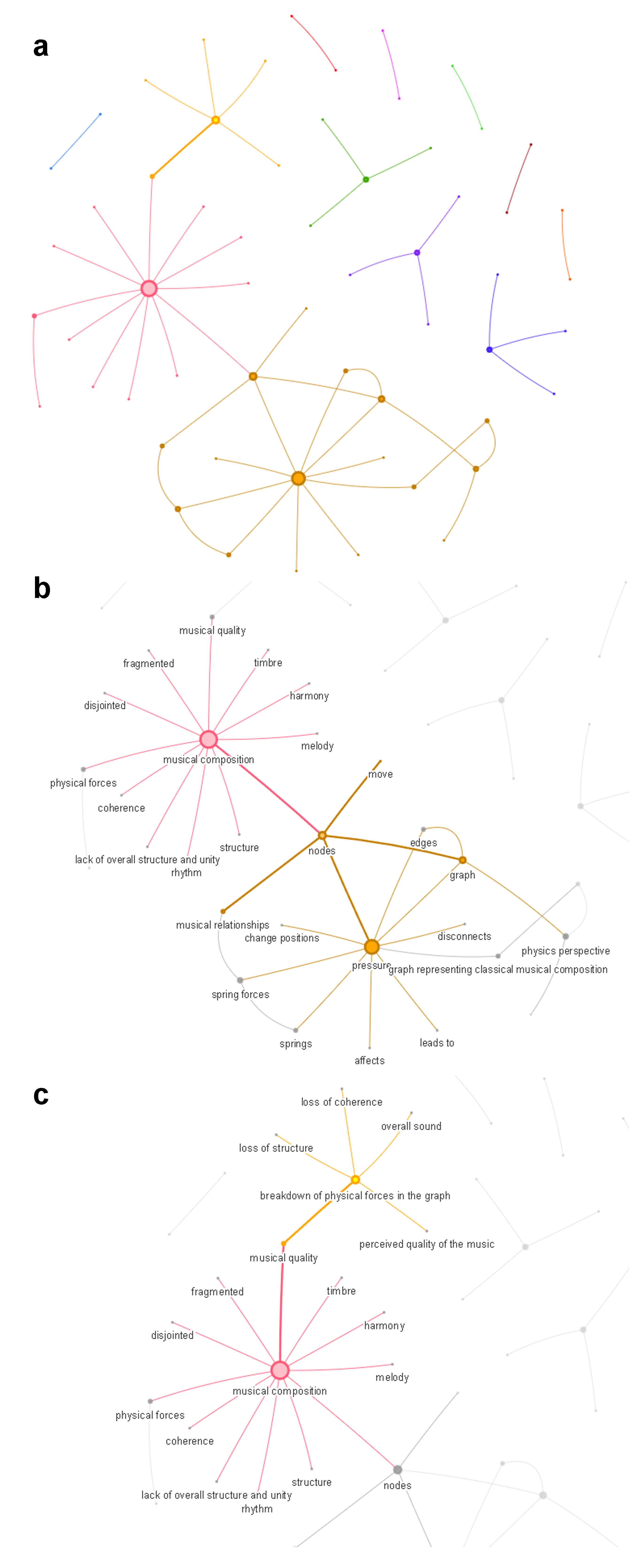}
    \caption{Ontological knowledge graph generated using agentic modeling in response to the question to explore an abstract representation of a classical musical composition as a graph of
interacting notes, chords, melodies, rhythms that is exposed to pressure. The model then responds by integrating concepts from music and art with physics, reasoning and mechanics to develop an answer. The graph representation of the produced response offers visual insights into the results and provides a human-readable analysis of the critical points made. The color reflects the 12 distinct communities identified. Panel a shows the overall graph, and panels b and c show close-up views with node labels for a deeper understanding of the result.}
    \label{fig:graph}
\end{figure}

While the generated text provides useful insights, the construction of ontological representation of the data can be beneficial for interpretable analysis~\cite{Buehler2023MechGPTModalities}. For graph generation, we concatenate the entirety of the two conversation and construct a knowledge graph. An example graph is show in Figure~\ref{fig:graph}. The resulting graph is divided into communities using the Girvan-Newman algorithm~\cite{Girvan2002CommunityNetworks}, and a total of 12 communities are identified. The figure shows both the overall graph (Figure~\ref{fig:graph}a) and detailed views with node labels (Figure~\ref{fig:graph}b-c). 

\subsection{Development of X-LoRA-Gemma with combined protein, chemical, bio-inspired and mechanics of materials capabilities}
\label{gemma-section}

To show that the proposed approach works with other base models that have distinct architectures, we trained another X-LoRA model, this time based on the Gemma-7B-it model~\cite{Google/gemma-7b-itFace}. In this model we use a set of four LoRA adapters, defined as follows:
\begin{enumerate}
    \item Bioinspired materials  
    \item Mechanics and materials
    \item Protein mechanics tasks (featuring generative sequence-to-property and inverse capabilities)
    \item Quantum-mechanics based molecular properties QM9 (featuring generative SMILES-to-property and inverse capabilities, see Table~\ref{tab:table_QM9} for a definition and summary of all properties computed in that dataset)~\cite{Ruddigkeit2012EnumerationGDB-17,Ramakrishnan2015ElectronicSpace}
\end{enumerate}
This X-LoRA-Gemma model is developed in a similar as the original X-LoRA model but aims to feature distinct sets of capabilities, in particular adding the ability to predict a set of 12 quantum mechanical properties and to design molecules to meet a set of 12 quantum mechanical properties. This new adapter is trained along with the other ones listed above that are based on the same datasets we used in the earlier experiment. Note, whereas the bulk of the paper focuses on the Zephyr-based X-LoRA model, in the scope of this section we need to distinguish the models and hence refer to the original X-LoRA model as  X-LoRA-Zephyr, and the Gemma-based model as X-LoRA-Gemma.
Figure~\ref{fig:protein_force_energy_prediction} shows the overall  X-LoRA  performance, featuring  predictions  of unfolding force and unfolding energy from the sequence ($R_2 = 0.87$), as well as performance for prediction of quantum mechanical properties of molecules based on the QM9 dataset ($R_2 = 0.96$). We observe excellent performance for both tasks, even slightly better than for the original X-LoRA model for the protein design task.

\begin{figure}
    \centering
    \includegraphics[width=1\linewidth]{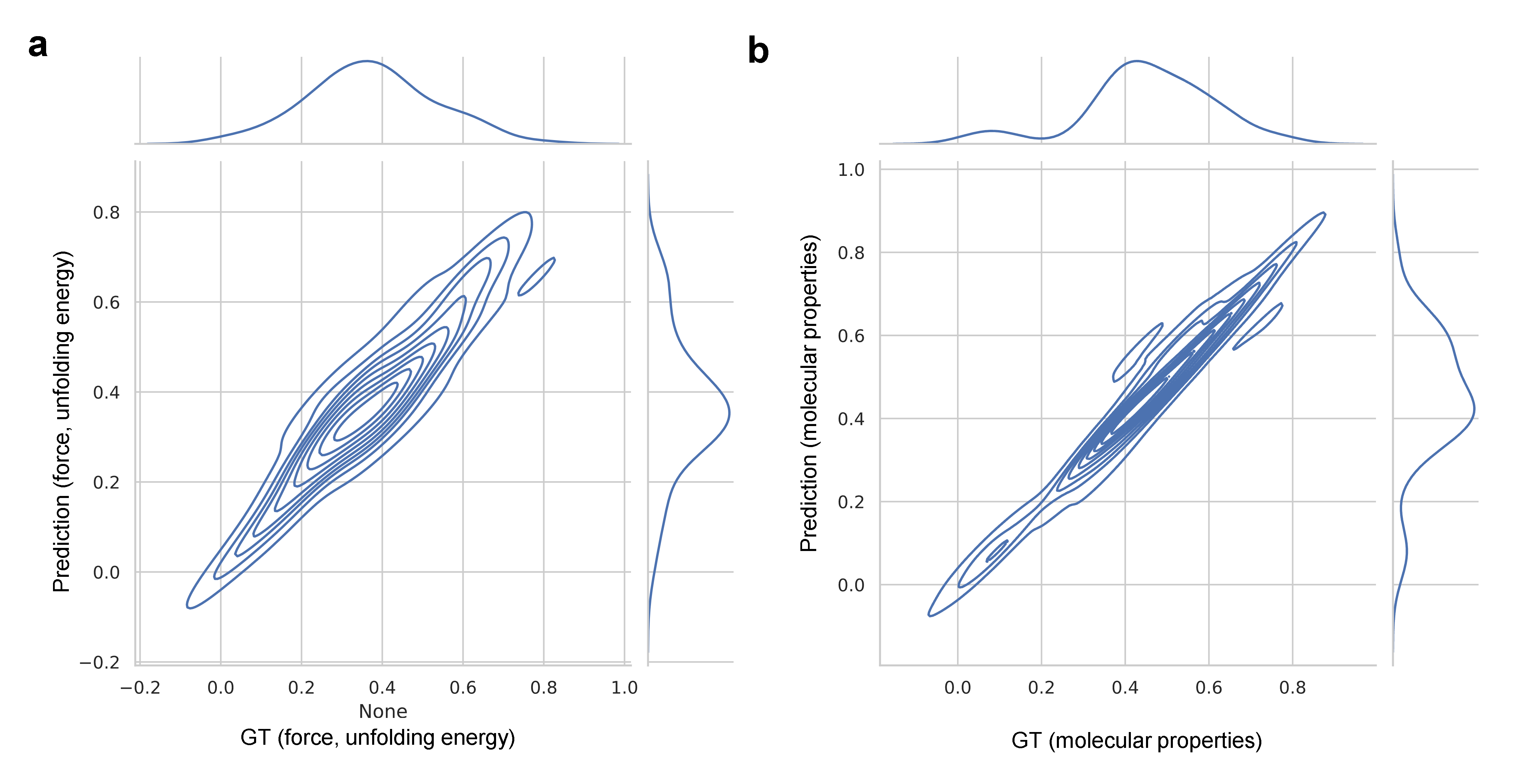}
    \caption{Overall X-LoRA-Gemma performance for predicting unfolding force and unfolding energy from the sequence comparing ground truth with predictions (panel a, $R_2 = 0.87$), as well as performance for prediction of quantum mechanical properties of molecules based on the QM9 dataset (panel b, 
    $R_2 = 0.96$). We observe excellent performance for both tasks. }
    \label{fig:protein_force_energy_prediction}
\end{figure}

\begin{figure}
    \centering
    \includegraphics[width=.75\linewidth]{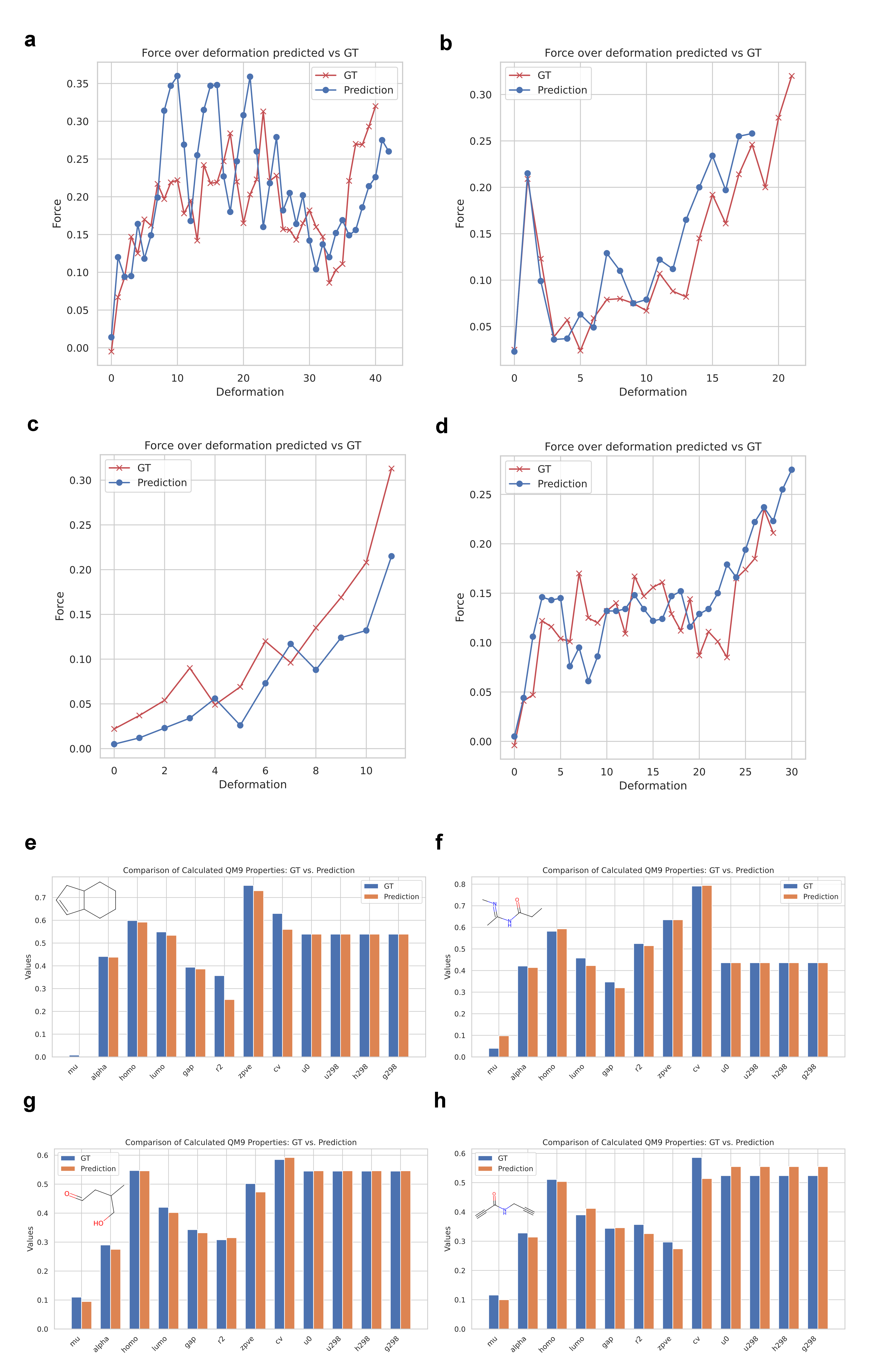}
    \caption{Predictions using the X-LoRA-Gemma model that is capable of predicting various biophysical properties of proteins (a-d) based on their amino acid sequence and quantum mechanical properties (e-h) of molecules represented in SMILES language. Panels a-d, to predict force-deformation behavior from amino acid sequences, for four examples from the test set. The model has excellent forward capabilities. Panels e-f depicts four representative examples of predicting quantum mechanical properties based on SMILES representation of molecular structures. Also here, excellent predictive capabilities are found. SMILES codes are (from e to f): \texttt{C1C=CC2CCCCC12}, \texttt{CCC(=O)NC(=NC)C}, \texttt{CC(CO)CC=O} and \texttt{CCN1CC(C1)(C\#C)O}. See Table~\ref{tab:table_QM9} for a definition of the properties predicted here. }
    \label{fig:image_Gemma_force_and_QM9}
\end{figure}

\begin{figure}
    \centering
    \includegraphics[width=.9\linewidth]{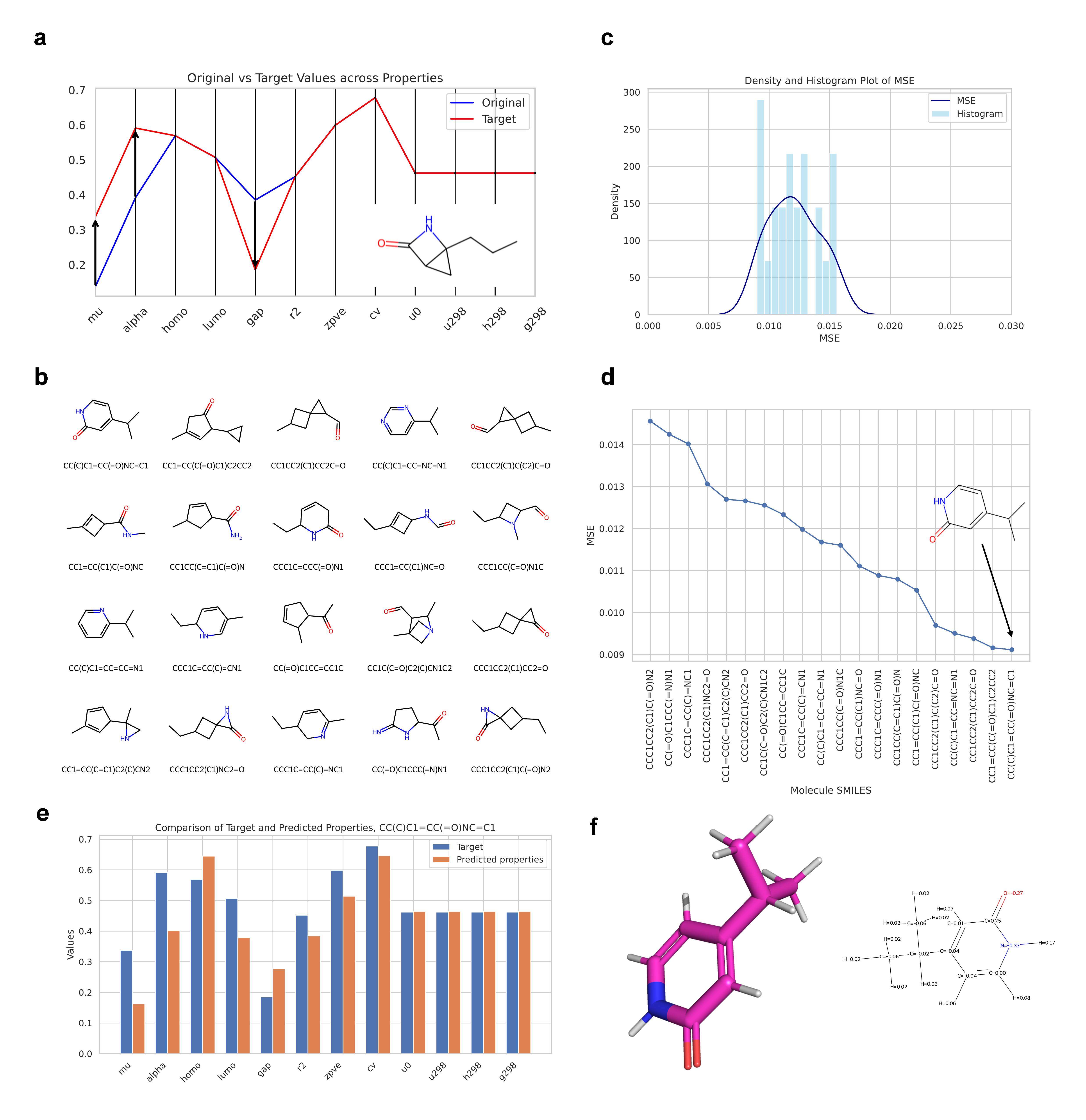}
    \caption{Design of novel molecules to meet a target objective. Here, we use \texttt{CC(C)C1=CC(=O)NC=C1} as a basis, and attempt to make the molecule more reactive (panel a shows target and design changes). Panel b shows the top 20 designs, panel c the distribution of the mean-squared error (MSE), and panel d the MSE over the top generated designs. Finally, panel e shows an assessment of how well the target matches with the predicted properties. We depict an optimized 3D molecular representation in panel f, along with a representation of atoms and Gasteiger partial charges~\cite{Gasteiger1980IterativeCharges} (a method of estimating the distribution of electric charge over the atoms in a molecule). A close analysis of the molecular structure indicates that the new molecule exhibits higher reactivity, in agreement with the design objective.}
    \label{fig:QM9_design}
\end{figure}

\begin{table}
\tiny 
\caption{Summary of QM9 molecular properties, as defined in~\cite{Ruddigkeit2012EnumerationGDB-17,Ramakrishnan2015ElectronicSpace}.}
\begin{center}
    \begin{tabular}{ |p{2cm} | p{8cm} |}
    \hline
   \textbf{Label}   &  \textbf{Definition}   \\ \hline\hline
   mu   &   Dipole moment: A measure of the separation of charge within the molecule, affecting its interaction with electric fields and other molecules. \\ \hline
   alpha   &   Polarizability: Indicates how much the electron cloud around the molecule distorts in an external electric field, influencing the molecule's optical properties and interactions. \\ \hline
   homo   &   Highest Occupied Molecular Orbital (HOMO) energy: Related to the energy of the highest occupied electron orbital, important for understanding a molecule's chemical reactivity. \\ \hline
   lumo   &   Lowest Unoccupied Molecular Orbital (LUMO) energy: Pertains to the energy of the lowest unoccupied electron orbital, also critical for reactivity and optical properties. \\ \hline
   gap   &   HOMO-LUMO gap: The energy difference between HOMO and LUMO, significant for determining a molecule's chemical stability and reactivity. \\ \hline
   r2   &   Electronic spatial extent: This is a measure of the size of the electron cloud of a molecule, related to its electronic properties. \\ \hline
   zpve   &   Zero-point vibrational energy: The energy of a molecule at its lowest vibrational state, contributing to its stability and reactivity. \\ \hline
   cv   &   Heat capacity at constant volume: Relates to the amount of heat required to change the temperature of a molecule by a certain amount, important for thermodynamics. \\ \hline
   u0   &   Internal energy at 0 K: The total energy of a molecule including electronic, vibrational, rotational, and translational contributions at absolute zero. \\ \hline
   u298   &   Internal energy at 298.15 K: Similar to u0, but measured at room temperature (approximately 25°C). \\ \hline
   h298   &   Enthalpy at 298.15 K: The total heat content of a molecule at room temperature, including internal energy and the product of pressure and volume. \\ \hline
   g298   &   Free energy at 298.15 K: Gibbs free energy of the molecule at room temperature, indicating the maximum amount of work obtainable from a thermodynamic process at constant temperature and pressure. \\ \hline
    \end{tabular}
    \label{tab:table_QM9}
\end{center}
\end{table}
We now demonstrate the capabilities of X-LoRA-Gemma in a simple multi-agent design experiment. We consider a small organic molecule, \texttt{CCCC12CC1C(=O)N2} (taken from the test set based on the QM9 dataset). This is a bicyclic structure with a ketone (=O) and an amine (N) incorporated into the ring, see Figure~\ref{fig:QM9_design}(a) for a visualization of the molecule). First we ask the X-LoRA-Gemma model:
\begin{quote}
\small\texttt{Here is a molecule in SMILES code: CCCC12CC1C(=O)N2
\newline\newline
What changes would I need to make to make this molecule more useful for polymer design. }
\end{quote}
The response is:
\begin{quote}
\small\texttt{To make this molecule more useful for polymer design, you could:
\begin{enumerate}
\item  Introduce a reactive functional group, such as an amine or hydroxyl group, to allow for further modification and crosslinking.
\item  Include a longer, more flexible carbon backbone to increase the molecule's ability to form polymers with different structures and properties.
\item  Add a side chain with a functional group, such as a carboxyl group, to allow for the formation of covalent bonds with other molecules and polymers.
\item  Incorporate a cyclic structure into the molecule, such as a lactam ring, to create a more rigid and stable polymer.
\item  Include a variety of side chains to create a range of properties and functionalities in the polymer.
\item  Use a variable number of carbons in the backbone to allow for the creation of polymers with different molecular weights and properties.
\item  Introduce a crosslinking point into the molecule to create a more stable and less reactive polymer.
\item  Include a repeating unit structure to create a polymer with a specific molecular weight and properties.
\item  Use a combination of the above strategies to create a versatile and useful polymer for various applications.
\end{enumerate}}
\end{quote}
In the following we focus on the first suggestion. 
\begin{quote}
\small\texttt{How can I increase the general reactivity of a molecule? I want to use this molecule to build a polymer material.
\newline\newline
The molecule of interest, in SMILES code, is: CCCC12CC1C(=O)N2
\newline\newline
I have access to tuning the following properties:
\newline\newline
Dipole moment, mu
Polarizability, alpha \newline
Highest Occupied Molecular Orbital energy, homo\newline
Lowest Unoccupied Molecular Orbital energy, lumo\newline
HOMO-LUMO gap: The energy difference between HOMO and LUMO, gap\newline
Electronic spatial extent\newline
Zero-point vibrational energy\newline
Heat capacity at constant volume\newline
Internal energy at 0 K\newline
Internal energy at 298.15 K\newline
Enthalpy at 298.15 K\newline
Free energy at 298.15 K
\newline\newline
What are the top THREE properties I should change, and how should I modify them (increase or decrease). }
\end{quote}
The response from X-LoRA-Gemma is:
\begin{quote}
\small\texttt{\begin{itemize}
    \item  HOMO-LUMO gap (decrease)
    \item  Dipole moment (increase)
    \item  Polarizability (increase)
\end{itemize}
By modifying these properties, you can increase the general reactivity of the molecule and make it more suitable for polymer material building.
}
\end{quote}

We use the generative and forward capabilities of X-LoRA-Gemma to design for these changes of molecular properties. 
To do the optimization we use a two-agent setup consistent of two instances of X-LoRA-Gemma, where one agent uses the generative task \texttt{GenerateMolecularProperties<...>} to generate candidates. We then use a second agent to do a calculation using \texttt{CalculateMolecularProperties<...>} and assess the error between the target and the predicted properties. This process is repeated until we reach the maximum number of iterations or until a minimum threshold for the mean-squared error is met.  
Figure~\ref{fig:QM9_design} shows the results of the entire process. Figure~\ref{fig:QM9_design}(a) depicts the target design and how it deviates from the properties of the original molecule. Figure~\ref{fig:QM9_design}(b) shows the top 20 designs generated by the algorithm, along with the distribution of the mean-squared error (Figure~\ref{fig:QM9_design}(c)) and a plot of the mean-squared error over the generated designs (Figure~\ref{fig:QM9_design}(d)). The best design is \texttt{CC(C)C1=CC(=O)NC=C1} and it meets the target relatively well, as shown in Figure~\ref{fig:QM9_design}(e). Finally, Figure~\ref{fig:QM9_design}(f) shows an optimized 3D molecular representation (we use UFF~\cite{Rappe1992UFFSimulations} to optimize the molecule's 3D structure). 

Since we wanted to assess the molecule's reactivity, we also compute the Gasteiger charges~\cite{Gasteiger1980IterativeCharges} (Figure~\ref{fig:QM9_design}(f) for a visual overview). The molecule features an aromatic ring containing a ketone (C=O), which is more electrophilic compared to carbon-carbon or carbon-hydrogen bonds, making it susceptible to nucleophilic attacks. The combination of the aromatic ring containing nitrogen and the ketone group forms a pyridone ring structure. This ring is more reactive than a simple aromatic ring because the presence of the heteroatom (nitrogen) and the ketone group introduces sites for potential chemical reactions, such as nucleophilic addition to the carbonyl carbon or electrophilic substitution at the ring. The nitrogen within the ring, being part of the heteroaromatic system, contributes to the aromaticity and reactivity of the compound by altering the electron density and distribution across the ring. This can affect how the molecule interacts with other chemical entities, making the nitrogen-containing ring a site for potential chemical reactions, particularly those involving electrophilic attack on the ring due to the electron-donating nature of the nitrogen atom.
The carbon atom in the ring, which is adjacent to both an oxygen and a nitrogen atom, having a partial positive charge (+0.25) suggests it is somewhat electron-deficient. This happens because oxygen, being more electronegative, pulls electron density towards itself, making the carbon less electron-rich. The effect is somewhat balanced by the adjacent nitrogen, which is less electronegative than oxygen but more electronegative than carbon. However, the presence of the partial positive charge indicates that the carbon is a likely site for nucleophilic attacks, where a nucleophile (an electron-rich species) would be attracted to this electron-deficient carbon.
The nitrogen atom having a partial negative charge (-0.33) indicates it is relatively electron-rich compared to its usual state. This can be due to its lone pair of electrons and the effect of the aromatic ring's electron system. In aromatic heterocycles, nitrogen's lone pair can participate in the delocalized $\pi$-electron system, but the exact contribution depends on the ring's structure and the electronegativity of adjacent atoms. The partial negative charge suggests that nitrogen is more nucleophilic, meaning it has a higher propensity to donate electrons, either forming bonds with electrophiles or engaging in protonation reactions. Further quantum mechanical calculations should be done to investigate this.
The carbon with a partial positive charge is a reactive site for nucleophilic attacks. Other molecules might react with this site through mechanisms that involve nucleophiles targeting electron-deficient carbons. 
The hydrogen atom bonded to the nitrogen atom, with a positive partial charge of 0.17, is a good candidate for forming hydrogen bonds with other molecules or atoms. Hydrogen bonds are a type of dipole-dipole interaction that occurs when a hydrogen atom, which is bonded to a highly electronegative atom (e.g.: nitrogen, oxygen, or fluorine), interacts with another electronegative atom bearing a lone pair of electrons. This makes the molecule potentially capable of acting as a hydrogen bond donor, which is an important factor in determining the molecule's solubility in water and other polar solvents, as well as its interactions in biological systems.

The presence of the ketone group adjacent to the nitrogen-containing aromatic ring offers unique reactivity that can be exploited in polymer synthesis~\cite{McQuarrieSimonPhysicalChemistry,CareySundbergAdvancedOrganicChemistry,Varghese2022BeyondApplications,Varghese2022BeyondApplications}. This molecule is a type of lactam, given the cyclic amide formed from the ketone and nitrogen in the ring. Lactams are known for their role in polymer chemistry, particularly in the synthesis of polyamides, which are a class of polymers extensively used in materials applications. The ketone could undergo various chemical reactions that are useful in polymer chemistry, such as nucleophilic addition. This could be utilized to extend the polymer chain or to introduce side chains that could modify the polymer's properties. The presence of a nitrogen atom in the aromatic ring can significantly affect the electronic properties of the polymer and can participate in hydrogen bonding, which could impact the polymer's melting point, solubility, and mechanical properties. N-heteroaromatic compounds are also known for their ability to coordinate with metal ions, which could lead to applications in areas such as catalysis or materials with electronic or photonic properties. This molecule could potentially undergo polymerization reactions through the amide (lactam) functionality; polyamides are known for their strength, elasticity, and resistance to wear and chemicals and they are used in a wide range of settings from textiles (e.g., nylon) to automotive parts~\cite{Varghese2022BeyondApplications,Varghese2022BeyondApplications}.
Alkyl substituents may increase the hydrophobic character of the molecule we anticipate that polymers derived from monomers with such alkyl groups are likely to be more hydrophobic, affecting their solubility in different solvents and their interaction with water. This could be advantageous for applications requiring water resistance or specific solubility characteristics. The presence of alkyl groups can also affect the mechanical properties of the polymer. For instance, they can act as plasticizers within the polymer matrix, increasing the flexibility of the polymer. This could lead to materials that are more pliable or have better impact resistance, depending on the length and branching of the alkyl groups.

\subsubsection{Additional benchmarking}
\begin{figure}
    \centering
    \includegraphics[width=.6\linewidth]{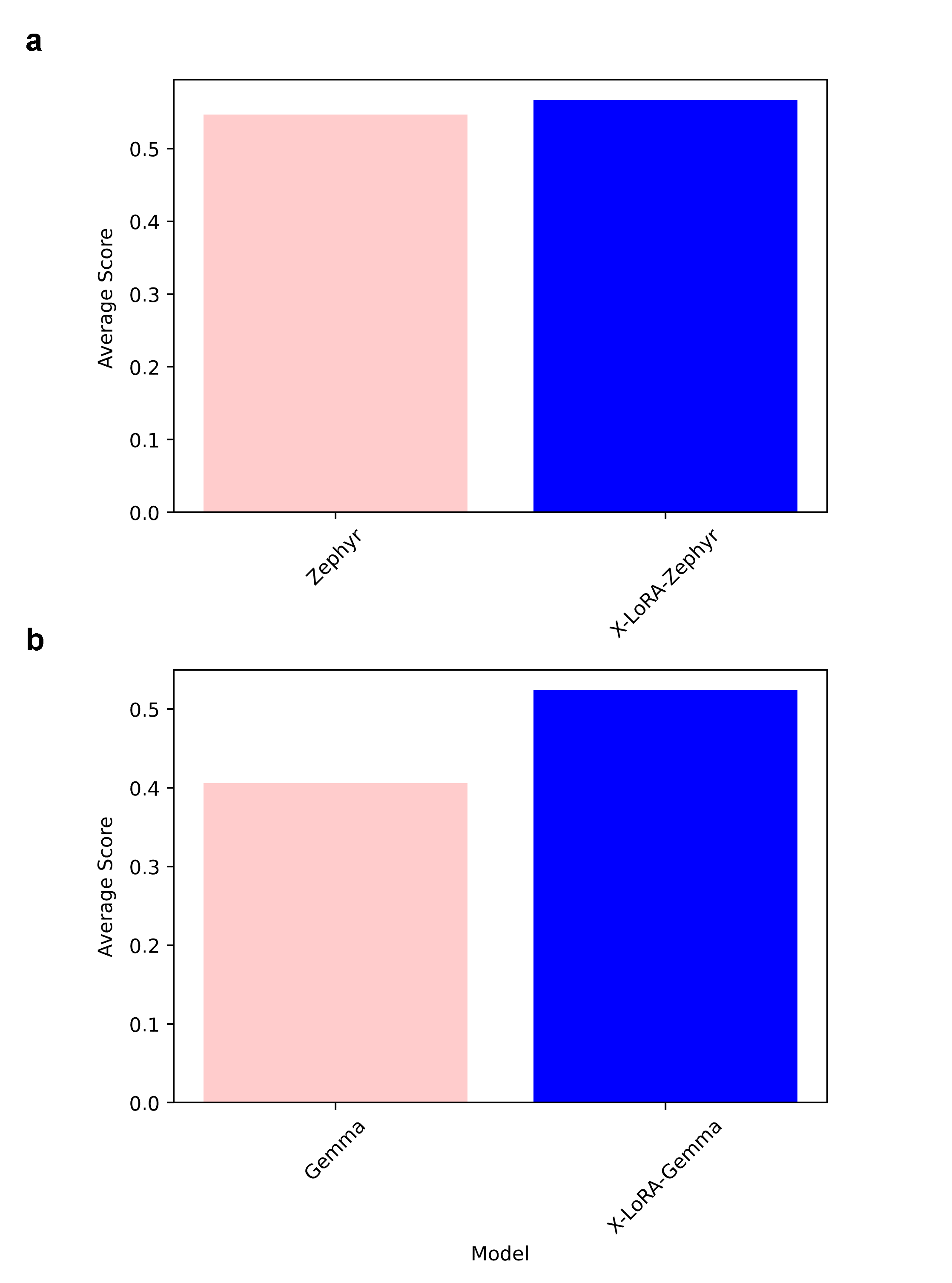}
    \caption{Performance assessment using the chemistry, physics, and biology questions in the MMLU benchmark (5-shot predictions)~\cite{Hendrycks2020MeasuringUnderstanding}, complementing the domain-specific benchmarks shown in Figures~\ref{fig:perf_comp_bioinspiredexam}, \ref{fig:protein_force_energy_prediction}, and \ref{fig:protein_force_energy_prediction}. Panel (a) shows the results for the X-LoRA-Zephyr model (referred to as X-LoRA in the other parts of the paper), and panel (b) shows results for the X-LoRA-Gemma model. The X-LoRA models outperform both base models (Zephyr-based model: 54.6\% for the base model versus 56.6\% for the X-LoRA model, and Gemma-based model: 40.6\% for the base model and 52.4\% for the X-LoRA-Gemma model). }
    \label{fig:perf_MMLU}
\end{figure}

We report additional benchmarks of the X-LoRA-Gemma model, shown in Figure~\ref{fig:perf_MMLU}, using a subset of the MMLU benchmark focused on chemistry, physics and biology (these are relatively close to the domains our model has been trained on)~\cite{Hendrycks2020MeasuringUnderstanding}. We find both X-LoRA models discussed in this paper outperform the corresponding base models. The X-LoRA models outperform both base models (Zephyr model: 54.6\% for the base model versus 56.6\% for the X-LoRA model, Gemma model: 40.6\% for the base model and 52.4\% for the X-LoRA-Gemma model). The best overall performance is observed for the X-LoRA-Zephyr model, but both are reasonably close.

\section{Conclusion}

There are many current and potential future applications of multimodal LLMs in science, along with a great need for special-purpose highly specialized expertise. In addition to general language tasks, the access of calculation or generative tasks like demonstrated here with protein mechanics is critical~\cite{Luu2023BioinspiredLLM:Materials,Buehler2023MechGPTModalities,Buehler2023MeLMProblemsb}. X-LoRA, built on open-source language models, enables us to dynamically mix expertise and knowledge across domains, as demonstrated in various examples, including Figure~\ref{fig:conversation_evolve}. This analysis showed how the X-LoRA model dynamically and autonomously evolves the use of adapters, and how it mixes certain adapters in a complex fashion across the LoRA layers. The complex mixing of LoRA layers is consistently seen across all examples studied here, such as in Figures~\ref{fig:QA_weights_sample_0} and \ref{fig:QA_weights}. This poses also an interesting problem to better understand the underlying mechanisms from a physics perspective, where viewing a LLM model as a large-scale graph forming model that exploits intricate graph structures. Thereby, the mixing of adaptation experts at deep layer scales as done here alters these mechanisms in ways to develop suitable solutions. 

The proposed algorithm augments the conventional inference strategy of a single forward pass to feature two forward passes. This trains the model to first `think' about the question and how it may reconfigure itself before responding. This implements a simple realization of `self-awareness’ whereby the model is able to adapt its own structure in order to best solve a task. As a result, the model, even though it has a relatively small parameter count of 7 billion parameters, can reason across diverse scientific domains (biological materials, math, physics, chemistry, logic, mechanics, etc.), significantly enhancing its capacity for generating innovative solutions. It may also guide the development of alternative model architectures, for instance, methods that generalize the reconfigurational strategy to parts of the model, or the entire model, of non-adapted systems. In such a scenario the scaling head would dynamically reconfigure segments of a trained model, or merge multiple models. As we have seen in the case of X-LoRA, such strategies can be powerful mechanisms to expand or integrate capabilities of different models.

We provided several performance assessments, including a comparison with much larger models (Figure~\ref{fig:perf_comp_bioinspiredexam}), where X-LoRA clearly outperforms. The detailed comparison shown in Table~\ref{tab:table_qa} showed that X-LoRA performs markedly better across a variety of tasks and application domains than the base model. Other comparisons include quantitative assessments of numerical prediction tasks, such as protein mechanics and quantum-mechanical properties of molecules, where X-LoRA shows high ac curacies with $R_2$ between 0.85 and 0.96. For comparison, these were higher than other published results, such as predicting protein unfolding force using the MeLM model~\cite{Buehler2023MeLMProblemsb} where  $R_2=0.78$ was reported. The capability to accurate solve multiple prediction tasks, along with inverse design tasks, and sophisticated reasoning capabilities were underscored in the example focused on molecular design (Section~\ref{gemma-section}). This example also illustrated that X-LoRA can be successfully implemented based on different base model architectures (Gemma vs. Zephyr/Mistral). Future work could expand this further and develop X-LoRA models for larger base models. 
 
A disadvantage of the proposed approach is that we require two forward passes, which can lead to additional computational cost: One to calculate hidden states that serve as the input to the X-LoRA scaling head, and one with $\Lambda$ applied to the adapters. However, the first pass is used for the X-LoRA scaling head and as such exploits basic capabilities of the LLM in contrast with learning the knowledge separately by training a larger scaling head. To implement effective application towards model-serving tools, separate key-value caches should be tracked for both the scaling and forward passes. To improve inference speed, we developed \texttt{Mistral.rs}, a LLM serving platform that implements X-LoRA in Rust~\cite{matsakis2014rust} and includes several methods to optimize performance. Nevertheless, the advantage of our approach is that it works for any model without a need to restructure the internal logic. We view this as an advantage, especially as it can be applied to the vast resources available in the Hugging Face ecosystem.

One advantage of the  proposed approach is that it provides a simple way to implement in any existing LLM (e.g., since the X-LoRA code has been released, it should be easily usable for any model in the Hugging Face ecosystem). Another advantage is that the scaling head exploits the innate strength of the LLM used, and learns sophisticated ways by which the adapters are best mixed for certain answers. During training of X-LoRA, it would be best to provide complex samples of question-answer or conversations that allow the model to learn the best combinatorial method. We examined several analyses of protein design and mechanics. For instance, Figure~\ref{fig:MJB_20} and Fig.~\ref{fig:MJB_21} presented a stability analysis of several protein sequences, including a comparison, reasoning over the predicted results, and a mechanistic explanation that provides a bottom-up chemical foundation of the predicted behavior. As can be see in Fig.~\ref{fig:MJB_21} this was directly confirmed via a structural analysis of the protein's folded 3D geometry.

As for other future work, additional research could explore a greater variety of adapter experts, building on the initial set of expertise in our LoRA set. The X-LoRA model exceeds the capabilities of the base model or a base model with either one adapter, and it can provide answers to complex questions while drawing on specialized skills endowed from the set of adapters. There also seem to be synergies between the various adapters, as the complex mixing results of scaling weights across the layers seems to indicate. This could be researched further, and compared with methods like SLERP that merge models with different strategies. Whereas we trained the X-LoRA scaling head with a subset of the training data, using a few hundred samples from each of the original training sets used to develop the adapters, more purpose-driven development of a training set for such a purpose can be devised. For instance, we noticed that posing multiple-choice questions tends to activate the reasoning-focused adapters since these were trained on such datasets. Additional prompting is required to help the scaling head understand a certain domain that is necessary for questions asked, for instance via the use of particular system messages or other strategies. Generally, the development of adequate training sets is an area that deserves further exploration including specific methods to invoke effecting mixing of layer-wise scaling. Our experiments of tracking these over longer conversations show that the model can clearly invoke changing scaling mechanisms to best respond to certain tasks, which is a promising feat. 

We observed fascinating patterns by which experts are activated across the layers, with interesting insights discussed in the paper. This analysis can be expanded, and may yield important insights about how and why mixing model parts or sets of layers can be advantageous. This would be particularly interesting when the X-LoRA method is applied to areas other than protein mechanics and protein design as done here. We leave this to future investigations. 

Another important avenue is the further study of agentic modeling, as done here using two X-LoRA models that interact adverserially. Future work could use more than two models to enhance interactions and add further capabilities to push models towards unexplored generative realms. This can also help with the development of methods that integrate physics-based modeling or other validation steps, using code-writing/executing agents or agents that use function calling or other processing techniques~\cite{Ghafarollahi2024ProtAgents:Learning}.

\section{Materials and Methods}

X-LoRA codes and examples are available at \url{https://github.com/EricLBuehler/xlora}. Model weights and data associated with the X-LoRA discussed in this paper, along with additional examples, are available at \url{https://huggingface.co/lamm-mit/x-lora}. The X-LoRA-Gemma model is available at \url{https://huggingface.co/lamm-mit/x-lora-gemma-7b}.  The \texttt{Mistral.rs} inference engine is available at \url{https://github.com/EricLBuehler/mistral.rs}.

\subsection{Mathematical formulation of X-LoRA}
We define the $\mathbf{scalings}$ to be a matrix containing token and layer level coefficients for LoRA adapters. The X-LoRA $\mathbf{scaling}$ $\mathbf{head}$ takes the hidden states from the base model and applies linear fully-connected layers to predict scalings, after which a softmax function is applied to the final dimension to ensure scalings across all experts add up to one. 
We use ReLU activation functions and dropout layers intercalated between linear fully-connected layers in the X-LoRA scaling head. The side of the hidden layers can be set, and chosen to be succesfuly reduce the size from the hidden dimension to the scaling dimension, or implement a widening and narrowing approach similar to the feed-forward layer in the transformer block. 

The scalings are a matrix $\Lambda \in \mathbb{R}^{s \times l \times n}$. Therein, $s$, $l$, and $n$, respectively, denote the sequence length, the number of layers, and the number of adapters. For each layer, the corresponding scalings are extracted: A matrix $\lambda \in \mathbb{R}^{s \times n}$. The scalings $\lambda$ are then applied to each LoRA adapter by broadcast multiplying the corresponding X-LoRA scaling $\lambda_i \in \mathbb{R}^{1\times 1 \times s}$ to the input for the decomposition matrices $A$ and $B$.

With $W_0 \in \mathbb{R}^{d \times k}$ and $B_i \in \mathbb{R}^{d \times r_i}$ and $A_i \in \mathbb{R}^{r_i \times d}$ with the rank $r_i \ll {\rm min} (d,k)$ that are unique to each adapter $i$, the forward pass of an X-LoRA layer is expressed as the following equation:
\begin{equation}
    h = W_0x + \sum_{i=1}^{n} B_iA_i(x \cdot  \lambda_i) \alpha_i
\end{equation}
Visualizations of the scalings are depicted in the main paper. These were either visualized using bar plots or using the \texttt{contourf} function in Matplotlib~\cite{MatplotlibDocumentation}. 

\subsection{X-LoRA implementation algorithm and code}
The implementation chosen here is driven by ease-of-use and the ability to easily adapt the approach to a variety of models, specifically those available in the Hugging Face ecosystem.
The X-LoRA algorithm operates by:
\begin{enumerate}
    \item Freezing the base model and adapters, and then tuning performance.
    \item Mixing LoRA adapters according to the predictions of the X-LoRA scaling head.
\end{enumerate}

\begin{figure}
    \centering
    \includegraphics[width=1\linewidth]{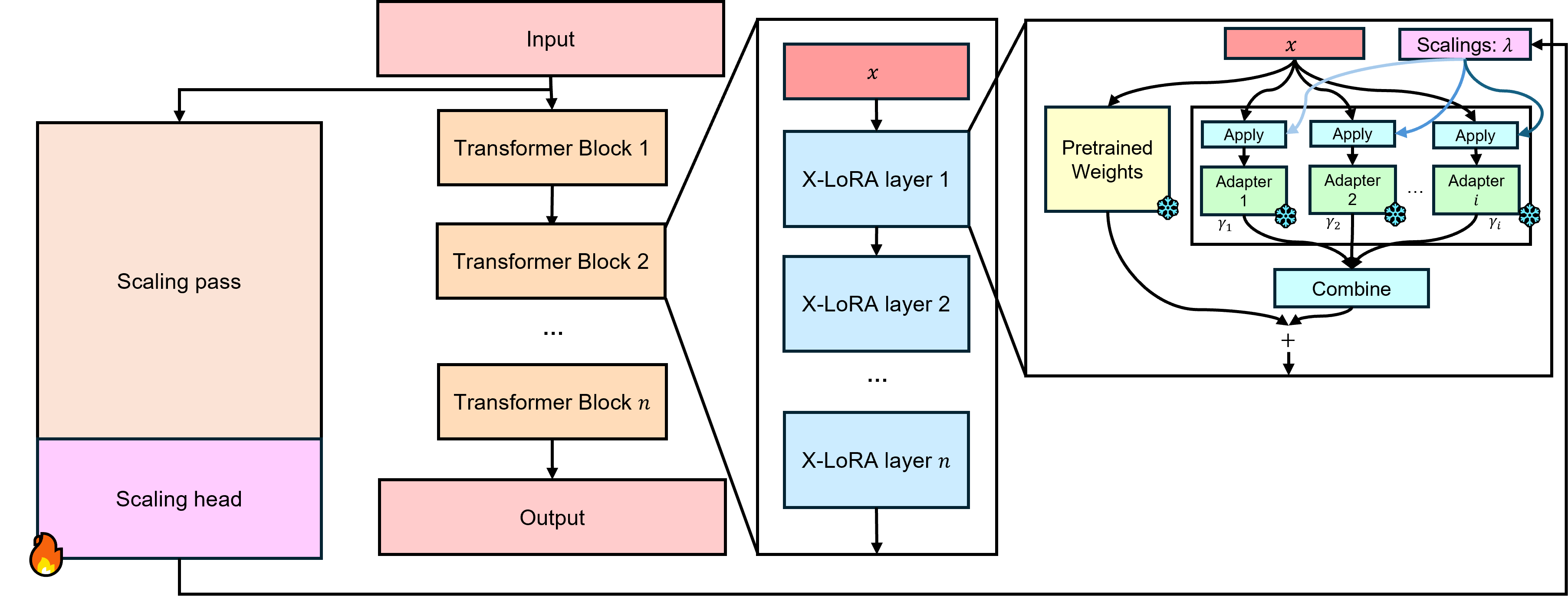}
    \caption{Detailed overview of the X-LoRA architecture, including the method of scaling inputs. As shown, the scalings are applied to the inputs of each frozen adapter. These are then additively combined. Finally, the output of the pretrained weights are added to form the final output of the X-LoRA layer. }
    \label{fig:general_arch}
\end{figure}

All code is written in Python and PyTorch~\cite{Paszke2019PyTorchLibrary}.

\subsubsection{Dual forward pass strategy for self-aware inference}
Because X-LoRA requires the calculation of the hidden states to predict the scalings, we implement a dual forward pass approach that utilizes a first pass for the model to consider the task, to decide how to reconfigure itself, followed by the inference step. This implements a type of `self-awareness' that endows X-LoRA with the capability to use additional compute for reflection rather than immediate response. 
We name the first pass the scaling pass and the second pass as the forward pass. We set the scalings $\Lambda$ during the scaling pass to zero. We  also experimented with setting the scalings $\Lambda$ to ${n}^{-1}$ and find similar results, indicating stable performance (in the X-LoRA package, this parameter can be set in the configuration). To easily be able to apply X-LoRA to a variety of models, we implement a forward pass hook. This strategy plays an essential role in the relationship between the forward and scaling passes and is the key feature which allows application to all models. 

\begin{enumerate}
    \item $\mathbf{Scaling}$ $\mathbf{pass}$: Implemented by executing the base model with adapters set to constant values. The resulting hidden states are used by the X-LoRA scaling head to generate $\Lambda$.
    \item $\mathbf{Forward}$ $\mathbf{pass}$: Apply $\Lambda$ to mix the adapters during the forward pass of the model. The resulting logits are returned as the output of the X-LoRA model.
\end{enumerate}

As a note of caution, using the same key-value cache for both the scalings pass and forward pass can potentially interfere with proper inputs to the scaling head. This is because the key-value cache will persist and would accumulate across the two forward passes. As such, we do not use the key-value cache during training or inference and implemented a proper handling of this in \texttt{Mistral.rs}.

\subsubsection{Freezing weights and opening segments of the model for training}
The base model and adapters are frozen ({\it i.e.}, their weights are not being trained) and as such the X-LoRA scaling head is the only trainable part of an X-LoRA model. It remains an option, in the code presented, to open up all adapters together with the X-LoRA scaling head. One could even opt to train the entire model - X-LoRA scaling head, adapter and the base foundation model jointly (or with different learning rates). Alternatively, future iterations of the concepts proposed here could be used to develop scaling functions between two or more foundation models to implement a traceable manifestation of SLERP-type modeling. This can be explored in future work. 

\subsubsection{X-LoRA layers}
An X-LoRA layer is a layer which scales and ultimately combines the outputs of several adapters. Each X-LoRA layer wraps the original LoRA layer and therefore has access to the specific adapters for that layer.

To effectively apply X-LoRA to any model, the X-LoRA algorithm swaps the original LoRA layers by creating X-LoRA layers and then injects their forward mechanics into the LoRA layer. This leverages aspects of Python's object model to enable swapping.

In summary, each X-LoRA layer executes the following steps during the forward pass (see Figure \ref{fig:general_arch}):
\begin{enumerate}
    \item $\mathbf{Scaling}$: Multiply the input signal of each LoRA adapter by the scaling value $\lambda_i$.
    \item $\mathbf{Forward}$: Run the forward pass for the wrapped LoRA layer.
\end{enumerate}

\subsection{Efficient Inference with \texttt{Mistral.rs} implemented in Rust}
To improve inference speed, we developed \texttt{Mistral.rs} in Rust, a LLM serving platform that implements X-LoRA in Rust~\cite{matsakis2014rust} along with several optimizations to improve performance. Specifically, the following methods are implemented, listed below in no particular order:
\begin{itemize}
    \item LoRA adapter weight stacking
    
        If all $A_i$ and $B_i$ matrices have the same shape for each LoRA adapter, we can stack the LoRA adapter weight matrices $A_i$ and $B_i$ along the first dimension. 
        
        Mathematically, given $W_0 \in \mathbb{R}^{d \times k}$, for adapter matrices $A_i \in \mathbb{R}^{r \times d}$ and $B_i \in \mathbb{R}^{d \times r}$ with $r_i \ll {\rm min} (d,k)$, we concatenate the matrices so that $A\in \mathbb{R}^{i \times r \times d}$ and $B\in \mathbb{R}^{i \times d \times r}$. The $\alpha_i$ parameters are also applied during initialization to the stacked matrices. 

        To apply scalings, we permute the dimensions such that $\lambda \in \mathbb{R} ^ {i \times 1 \times 1}$. Then, they are applied through a broadcast multiplication.
    
    \item Non-granular scalings

        To reduce the amount of times that the scaling pass is run, we implement an option of non-granular scalings. This caches the scalings after $n$ completion runs, as the KV cache ensures that the sequence length will remain at one for the remainder of the request. By using this approach, it decreases latency significantly as it reduces the number of invocations of the base model.
        
    \item Fused Compute Unified Device Architecture (CUDA) kernels

        To optimize the forward pass, we implement fused CUDA kernels. Fused kernels enable complex operations, such as Rotary Embedding (\cite{su2021roformer}) which would otherwise be executed sequentially, to be combined into a single kernel than can be executed in parallel on a GPU. We base our kernels on the vLLM implementation (\cite{kwon2023efficient}).

    \item Quantization

        The inference engine implements quantization, which allows one to use a quantized base model. In initial experiments, we find that quantization up to 8 bits works well even with a model that was trained with \texttt{bfloat16}, for instance.  
    
\end{itemize}

\subsection{Datasets}

Each of the adapters is trained with special-purpose datasets, as summarized in Table~\ref{tab:table_loradata} along with references to sources and original literature. 

\begin{table}
\tiny 
\caption{Summary of datasets used to train the individual adapters of the X-LoRA models reported here, along with references and sources. The X-LoRA model (also referred to as X-LoRA-Zephyr) features 9 adapters, whereas the X-LoRA-Gemma model 4 features adapters.}
\begin{center}
    \begin{tabular}{ |p{4cm} | p{8cm} | p{1cm}| p{1cm}|}
    \hline
   \textbf{Adapter}   &  \textbf{Dataset/source used, reference} & \textbf{Used in X-LoRA (Zephyr based)} &  \textbf{Used in X-LoRA-Gemma}   \\ \hline 
   Bioinspired materials   &   Dataset based on \cite{Luu2023BioinspiredLLM:Materials} expanded with new distilled data & \checkmark & \checkmark  \\ \hline
    Chain-of-thought (CoT) and reasoning  &  Dataset based on the FLAN dataset\cite{Wei2021FinetunedLearners}, as curated at~\cite{QingyiSi/Alpaca-CoTMain} (combination of 9 CoT datasets) & \checkmark  & \\\hline
    Chemistry&  Based on the CAMEL: Communicative Agents for “Mind” Exploration of Large Scale Language Model Society~\cite{Li2023CAMEL:Society} dataset for chemistry, CAMEL-AI Chemistry\cite{Camel-ai/chemistryFace} & \checkmark  & \\\hline
    Mathematics& CAMEL-AI Mathematics~\cite{Camel-ai/mathFace}   & \checkmark  & \\\hline
    Physics&   CAMEL-AI Physics~\cite{Camel-ai/physicsFace}  & \checkmark  &     \\\hline
    Biology&   CAMEL-AI Biology~\cite{Camel-ai/biologyFace}  & \checkmark  &    \\\hline
    Mechanics and materials&   Training data used in the MechGPT model as presented in \cite{Buehler2023MechGPTModalities}, and available at~\cite{Lamm-mit/MechanicsMaterialsFace} & \checkmark  &  \checkmark  \\ \hline
    Platypus: Logic and reasoning & Platypus\cite{Lee2023Platypus:LLMs} dataset as reported at~\cite{Garage-bAInd/Open-PlatypusFace}  & \checkmark  &       \\\hline
    Protein mechanics tasks (generative sequence-to-property and inverse capabilities focused on unfolding/physical properties) &   Dataset based on molecular modeling results as reported in \cite{Ni2023ForceGen:Model,Ghafarollahi2024ProtAgents:Learning}; forces and energies are normalized using the same factors as in the earlier reference that reported the original dataset. & \checkmark  &  \checkmark   \\
    \hline
    Quantum Machine 9 (QM9)& Dataset featuring a variety of computed properties for 134,000 molecules (expressed in SMILES code~\cite{Weininger1988SMILESRules,Weininger1989SMILES.Notation}), based on quantum mechanical calculations. These properties are essential for understanding the behavior and characteristics of molecules in various contexts~\cite{Ruddigkeit2012EnumerationGDB-17,Ramakrishnan2015ElectronicSpace}. & & \checkmark \\
    \hline
    \hline
    \end{tabular}
    \label{tab:table_loradata}
\end{center}
\end{table}

\subsection{Training strategy}
Training is conducted in stages. We use the  Zephyr-7B-$\beta$ model~\cite{HuggingFaceH4/zephyr-7b-betaFace} that was built on top of the Mistral-7B model\cite{Jiang2023Mistral7B} and train a series of adapters using the datasets described in the previous section. We use the Zephyr-7B-$\beta$ tokenizer. 

The first eight adapters are trained using question-answer pairs as provided by the datasets (see, Table~\ref{tab:table_loradata}). 
The protein mechanics tasks are trained similarly, but using specific forward and inverse instruction sets. The bidirectional instruction tasks are:
\begin{quote}
\tiny {\texttt{CalculateForce< G E E C D C G S P ....> [0.310]\\
CalculateEnergy< G E E C D C G S P ....> [0.220]\\ 
CalculateForceHistory< G E E C D C G S P ....> [0.004,0.034,0.125,0.142,0.159,0.102,0.079,0.073,0.131, ...]\\
GenerateForce<0.310>\,[ G E E C D C G S P ....]\\
GenerateEnergy<0.220>\,[ G E E C D C G S P ....]\\
GenerateForceHistory<0.004,0.034,0.125,0.142,0.159,0.102,0.079,0.073,0.131, ...>\,[ G E E C D C G S P ....]
}}
\end{quote}
These represent a total of three tasks (calculate the maximum unfolding force of a protein, the unfolding energy, and the force-deformation history). Each of the three forward tasks is complemented by three inverse tasks of the same nature, but yielding a protein sequence as a response (hence, a total of six tasks).  

The rank of each LoRA adapter is $r=16$ with $\alpha=16$. Each adapter is trained with the following five target  modules: \texttt{v\_proj}, \texttt{k\_proj}, \texttt{q\_proj},  \texttt{gate\_proj}, and \texttt{o\_proj}, with dropout 0.05. A LoRA adapter is usually trained with low ranks, such as 4 to 32. This is because it has been shown that increasing the rank of LoRA adapters does not typically improve performance, but it does increase computational requirements since a smaller rank results in a more compact model with fewer parameters~\cite{hu2021lora}. The choices for $r$, $\alpha$, and other parameters are made based on common values used in the literature that have been shown to work well. Moreover, earlier work by the author of this paper has experimented with various choices and found the values used here to be a reasonable choice~\cite{Luu2023BioinspiredLLM:Materials}.

The X-LoRA scaling head is trained with around 2,000 samples drawn as a fraction from the original dataset used to train the adapters (we also complemented the training data with a additional GPT-4 generated samples of multiple choice questions). We use a \texttt{paged\_adamw\_8bit} optimizer with gradient clipping of 0.3, a learning rate $2\times 10^{-4}$ with warmup, and four gradient accumulation steps, with a batch size of one. We train the X-LoRA model for around 10,000 steps. The loss is around 0.28 when we conclude training. 

Since the basic Mistral architecture used here as the foundation model has 32 layers, and since we use adapters in 5 of each of the transformer layers, there is a total of 160 LoRA layers, each of which is supplanted by an X-LoRA layer. The X-LoRA scaling head predicts scaling values for each of these 160 LoRA layers (for nine LoRA experts, this results in a total of $160\times 9=1440$ values of $\lambda_i$ computed for each token). We use single linear layer with ReLU activation functions and a dropout with $p=0.2$, with around 5M parameters. The code implements flexible structures so that more complex deep classifier heads can be constructed and explored in other applications. 

Both LoRA adapters and the X-LoRA scaling head are trained using next-token prediction and cross-entropy loss (we calculate the probability that the model assigns the correct next token in a sequence given the previous tokens; where the goal during training is to maximize the probability of the correct next token as denoted in the training set). We use token shifting, where the sequence is  shifted by one token to create the target sequence that the model should predict. For example, if the input sequence is ``Proteins are fundamental to", the target sequence for training would be ``are fundamental to biology". Therefore, the model tries to predict the next token in the sequence.
As shown in Figure~\ref{fig:hierarchical_concept} during X-LoRA scaling head training, only the scaling head is trainable and all other parameters are frozen. The native Zephyr tokenizer and prompt template is used.

\subsection{X-LoRA-Gemma model}
The X-LoRA-Gemma model is developed in a similar way as the X-LoRA model discussed above, but based on the Gemma-7B-it model~\cite{Google/gemma-7b-itFace}, and using four adapters (see main text for details). The four adapters are trained on the mechanics and materials, protein mechanics, bioinspired materials, as well as an additional quantum-mechanics based molecular properties QM9 (see Table~\ref{tab:table_QM9}) dataset~\cite{Ruddigkeit2012EnumerationGDB-17,Ramakrishnan2015ElectronicSpace}.
The rank of each LoRA adapter is $r=16$ with $\alpha=16$. Each adapter is trained with the following seven target  modules: \texttt{q\_proj}, \texttt{o\_proj}, \texttt{k\_proj}, \texttt{v\_proj} \texttt{gate\_proj}, and \texttt{up\_proj} and \texttt{down\_proj}, with dropout 0.05. The rank of each LoRA adapter is  chosen to be $r=16$ with $\alpha=16$. With four adapters and 7 target models, since there is a total of 28 layers in the Gemma-7B base model, there is a total of $28\times 28=784$ values of $\lambda_i$ computed for each token.

The first two adapters (bioinspired and mechanics of materials) used in the construction of this model are trained using question-answer pairs as provided by the datasets (see, Table~\ref{tab:table_loradata}). 
The protein mechanics tasks are trained in the same way as described in the preceding section, with bidirectional forward and inverse instruction sets. In addition, we train an adapter based on the QM9 dataset, with these two tasks:  
\begin{quote}
\tiny {\texttt{CalculateMolecularProperties<CC(C)N1CC1C=O> [0.098,0.358,0.581,0.395,0.309,0.330,0.570,0.649,0.519,0.519,0.519,0.518]\\
GenerateMolecularProperties<0.098,0.358,0.581,0.395,0.309,0.330,0.570,0.649,0.519,0.519,0.519,0.518> [CC(C)N1CC1C=O]\\ 
...
}}
\end{quote}
These represent a total of two tasks (calculate molecular properties, the forward task, and the inverse task to generate a molecule with target properties). All 12 properties featured in the QM9 dataset are predicted or designed for, respectively (see, Table~\ref{tab:table_QM9}). 
This model features a total of four adapters. 

The X-LoRA scaling head for this model has three layers, whereas the intermediate layer has a size of $3072\times {FF}_{\rm fac}=12,288$, where ${FF}_{\rm fac}=4$ is a feed-forward multiplier. We use linear layers with ReLU activation functions and a dropout of $p=0.2$, with around 47M parameters.
The X-LoRA-Gemma scaling head is trained with around 17,000 samples. These were drawn as a fraction from the training dataset used to train adapters; we note that even though the X-LoRA-Gemma model only contained four adapters, we used samples from across the entire training set of the model described above with the addition of samples from the QM9 dataset. We did this to experiment with the possibility that the X-LoRA model may be able to learn additional capabilities at the scaling head training stage. Since the loss shows good convergence to 0.58 at the end of training, together with excellent performance even in tasks for which the model was not explicitly trained for (e.g. chemistry, as shown in the molecule design example discussed in the paper), we believe that this can generally yield good results.
We used a \texttt{paged\_adamw\_8bit} optimizer with gradient clipping of 0.3, a learning rate $1\times 10^{-4}$ with warmup, and four gradient accumulation steps, with a batch size of one. We train the X-LoRA-Gemma model for around 62,000 steps. 
The native Gemma tokenizer and prompt template is used.

\subsection{Adversarial agentic modeling}
We implement an adversarial agentic strategy by instantiating two X-LoRA agents. One X-LoRA agent focuses on question asking. It asks the initial question and subsequently keeps developing questions to identify challenges and weaknesses in the responses. The other agent responds to the queries and continually provides answers. We implement this strategy in \{Guidance\}~\cite{Guidance-ai/guidance:Models.}.

For the example relating music and mechanics provided in the text, the features of the two agents are described as follows. First, the physicist and question asker:
\begin{quote}
\small\texttt{You are a critical physicist. You are taking part in a discussion, from the perspective of physics.
Keep your answers brief, and always challenge statements in a provocative way.
\newline\newline
As a creative individual, you inject ideas from other fields and push the boundaries.}
\end{quote}

Second, a philosopher (in some of the examples this role is revised to reflect a biology expert):
\begin{quote}
\small\texttt{You are a philosopher with knowledge in biology, chemistry and mathematics. 
You are taking part in a discussion.
\newline\newline
Keep your answers brief, but accurate, and creative. You come up with excellent ideas and new directions of thought, always logical.}
\end{quote}

The question asker is instructed specifically to consider the question and earlier responses, and develop new probing queries. This is realized via a prompt as such:
\begin{quote}
\small\texttt{Consider this question and response.
\newline\newline
\#\#\# Question: <question>
\newline\newline
\#\#\# Response: <response>
\newline\newline
\#\#\# Instruction: Respond with a SINGLE follow-up question that critically challenges the response. 
\newline\newline
DO NOT answer the question or comment on it yet. 
\newline\newline
The single question is: 
}
\end{quote}

Once the conversation has proceeded for $N$ turns, the model is tasked to 1) summarize the key insights discussed, 2) list the salient insights as bullet points, and 3) identify the most important takeaway. 

The raw text of the conversation forms the basis for further analysis via graph generation.

\subsection{Knowledge graph generation}
We use  Zephyr-7B-$\beta$ to extract triplets from text, following the strategy reported in~\cite{Rahulnyk/knowledge_graph:QnA} with additional features based on the Llama Index graph generation algorithm~\cite{Liu_LlamaIndex_2022}. Graphs are visualized using NetworX\cite{Networkx/networkx:Python} and Pyvis~\cite{WestHealth/pyvis:Graphs.}. One of the key instructions given is, following the approach suggested in~\cite{Liu_LlamaIndex_2022}:
\begin{quote}
\small\texttt{Your task is to extract the ontology of terms mentioned in the given context. These terms should represent the key concepts as per the context. 
\newline\newline
Format your output as a list of JSON. Each element of the list contains a pair of terms and the relation between them, like the following: [...]
}
\end{quote}
We instruct to build triplets of nodes and edge features as follows, following the approach used in\cite{Liu_LlamaIndex_2022}:
\begin{quote}
\small\texttt{\begin{itemize}
            \item "node\_1": "A concept from extracted ontology"
            \item "node\_2": "A related concept from extracted ontology"
            \item "edge": "Relationship between the two concepts, node\_1 and node\_2, succinctly described"
        \end{itemize}
        }
\end{quote}
We also provide example of ontological analysis to the model, such as:
\begin{quote}
\small\texttt{
       Context: ```Spider silk is a strong natural fiber used to catch prey in a web.```
       }
\end{quote}
We then give example triplets, such as 
\begin{quote}
\small\texttt{\begin{itemize}
            \item "node\_1": "spider silk"
            \item "node\_2": "fiber"
            \item "edge": "is"
        \end{itemize}
        }
\end{quote}
The resulting graph data is assembled and then grouped in to communities using the Girvan-Newman algorithm~\cite{Girvan2002CommunityNetworks}.

\subsection{Visualization of molecular structures}

We use PyMOL~\cite{Schrodinger2015The1.8} to visualize and analyze the predicted protein structures. Various scripts and functions within this software are used to identify certain features of the proteins, such a secondary structure, hydrophobic/hydrophilic regions, disulfide bonds, and hydrogen bonds.

\section*{Data Availability Statement}
The codes and data that support the findings of this study are openly available in \url{https://github.com/EricLBuehler/xlora} and \url{https://huggingface.co/lamm-mit/x-lora}. Additional data sources used for this study are referenced (see, specifically Table~\ref{tab:table_loradata}). 

\section*{Acknowledgments}
This was supported in part by MIT's Generative AI Initiative and Google. 
 
\section*{Author Contributions}
ELB and MJB conceived the concept, plan of study, developed the model and research, and wrote the paper. ELB developed the algorithms, codes and GitHub repository. MJB conducted the scientific investigations, carried out the protein modeling and analysis, molecular design and analysis, and revised and finalized the paper.  

\section*{Conflicts of interest}
The authors declare no conflicts of interest.

\bibliographystyle{apsrev} 
\bibliography{references,references-Mendeley}

\begin{thebibliography}{69}
\expandafter\ifx\csname natexlab\endcsname\relax\def\natexlab#1{#1}\fi
\expandafter\ifx\csname bibnamefont\endcsname\relax
  \def\bibnamefont#1{#1}\fi
\expandafter\ifx\csname bibfnamefont\endcsname\relax
  \def\bibfnamefont#1{#1}\fi
\expandafter\ifx\csname citenamefont\endcsname\relax
  \def\citenamefont#1{#1}\fi
\expandafter\ifx\csname url\endcsname\relax
  \def\url#1{\texttt{#1}}\fi
\expandafter\ifx\csname urlprefix\endcsname\relax\def\urlprefix{URL }\fi
\providecommand{\bibinfo}[2]{#2}
\providecommand{\eprint}[2][]{\url{#2}}

\bibitem[{\citenamefont{Vaswani et~al.}(2017)\citenamefont{Vaswani, Shazeer, Parmar, Uszkoreit, Jones, Gomez, Kaiser, and Polosukhin}}]{Vaswani2017AttentionNeedc}
\bibinfo{author}{\bibfnamefont{A.}~\bibnamefont{Vaswani}}, \bibinfo{author}{\bibfnamefont{N.}~\bibnamefont{Shazeer}}, \bibinfo{author}{\bibfnamefont{N.}~\bibnamefont{Parmar}}, \bibinfo{author}{\bibfnamefont{J.}~\bibnamefont{Uszkoreit}}, \bibinfo{author}{\bibfnamefont{L.}~\bibnamefont{Jones}}, \bibinfo{author}{\bibfnamefont{A.~N.} \bibnamefont{Gomez}}, \bibinfo{author}{\bibfnamefont{L.}~\bibnamefont{Kaiser}}, \bibnamefont{and} \bibinfo{author}{\bibfnamefont{I.}~\bibnamefont{Polosukhin}}, \emph{\bibinfo{title}{{Attention is All you Need}}} (\bibinfo{year}{2017}), \urlprefix\url{https://papers.nips.cc/paper/7181-attention-is-all-you-need}.

\bibitem[{\citenamefont{Touvron et~al.}(2023)\citenamefont{Touvron, Martin, Stone, Albert, Almahairi, Babaei, Bashlykov, Batra, Bhargava, Bhosale et~al.}}]{Touvron2023LlamaModels}
\bibinfo{author}{\bibfnamefont{H.}~\bibnamefont{Touvron}}, \bibinfo{author}{\bibfnamefont{L.}~\bibnamefont{Martin}}, \bibinfo{author}{\bibfnamefont{K.}~\bibnamefont{Stone}}, \bibinfo{author}{\bibfnamefont{P.}~\bibnamefont{Albert}}, \bibinfo{author}{\bibfnamefont{A.}~\bibnamefont{Almahairi}}, \bibinfo{author}{\bibfnamefont{Y.}~\bibnamefont{Babaei}}, \bibinfo{author}{\bibfnamefont{N.}~\bibnamefont{Bashlykov}}, \bibinfo{author}{\bibfnamefont{S.}~\bibnamefont{Batra}}, \bibinfo{author}{\bibfnamefont{P.}~\bibnamefont{Bhargava}}, \bibinfo{author}{\bibfnamefont{S.}~\bibnamefont{Bhosale}}, \bibnamefont{et~al.} (\bibinfo{year}{2023}), \urlprefix\url{https://arxiv.org/abs/2307.09288v2}.

\bibitem[{\citenamefont{{OpenAI}}(2023)}]{OpenAI2023GPT-4Report}
\bibinfo{author}{\bibnamefont{{OpenAI}}} (\bibinfo{year}{2023}), \urlprefix\url{http://arxiv.org/abs/2303.08774}.

\bibitem[{\citenamefont{Chowdhery et~al.}(2022)\citenamefont{Chowdhery, Narang, Devlin, Bosma, Mishra, Roberts, Barham, Chung, Sutton, Gehrmann et~al.}}]{Chowdhery2022PaLM:Pathways}
\bibinfo{author}{\bibfnamefont{A.}~\bibnamefont{Chowdhery}}, \bibinfo{author}{\bibfnamefont{S.}~\bibnamefont{Narang}}, \bibinfo{author}{\bibfnamefont{J.}~\bibnamefont{Devlin}}, \bibinfo{author}{\bibfnamefont{M.}~\bibnamefont{Bosma}}, \bibinfo{author}{\bibfnamefont{G.}~\bibnamefont{Mishra}}, \bibinfo{author}{\bibfnamefont{A.}~\bibnamefont{Roberts}}, \bibinfo{author}{\bibfnamefont{P.}~\bibnamefont{Barham}}, \bibinfo{author}{\bibfnamefont{H.~W.} \bibnamefont{Chung}}, \bibinfo{author}{\bibfnamefont{C.}~\bibnamefont{Sutton}}, \bibinfo{author}{\bibfnamefont{S.}~\bibnamefont{Gehrmann}}, \bibnamefont{et~al.} (\bibinfo{year}{2022}), \urlprefix\url{https://arxiv.org/abs/2204.02311v3}.

\bibitem[{\citenamefont{Jiang et~al.}(2023)\citenamefont{Jiang, Sablayrolles, Mensch, Bamford, Chaplot, Casas, Bressand, Lengyel, Lample, Saulnier et~al.}}]{Jiang2023Mistral7B}
\bibinfo{author}{\bibfnamefont{A.~Q.} \bibnamefont{Jiang}}, \bibinfo{author}{\bibfnamefont{A.}~\bibnamefont{Sablayrolles}}, \bibinfo{author}{\bibfnamefont{A.}~\bibnamefont{Mensch}}, \bibinfo{author}{\bibfnamefont{C.}~\bibnamefont{Bamford}}, \bibinfo{author}{\bibfnamefont{D.~S.} \bibnamefont{Chaplot}}, \bibinfo{author}{\bibfnamefont{D.~d.~l.} \bibnamefont{Casas}}, \bibinfo{author}{\bibfnamefont{F.}~\bibnamefont{Bressand}}, \bibinfo{author}{\bibfnamefont{G.}~\bibnamefont{Lengyel}}, \bibinfo{author}{\bibfnamefont{G.}~\bibnamefont{Lample}}, \bibinfo{author}{\bibfnamefont{L.}~\bibnamefont{Saulnier}}, \bibnamefont{et~al.} (\bibinfo{year}{2023}), \urlprefix\url{https://arxiv.org/abs/2310.06825v1}.

\bibitem[{\citenamefont{Gunasekar et~al.}(2023)\citenamefont{Gunasekar, Zhang, Aneja, C{\'{e}}sar, Mendes, Giorno, Gopi, Javaheripi, Kauffmann, De et~al.}}]{Gunasekar2023TextbooksNeed}
\bibinfo{author}{\bibfnamefont{S.}~\bibnamefont{Gunasekar}}, \bibinfo{author}{\bibfnamefont{Y.}~\bibnamefont{Zhang}}, \bibinfo{author}{\bibfnamefont{J.}~\bibnamefont{Aneja}}, \bibinfo{author}{\bibfnamefont{C.}~\bibnamefont{C{\'{e}}sar}}, \bibinfo{author}{\bibfnamefont{T.}~\bibnamefont{Mendes}}, \bibinfo{author}{\bibfnamefont{A.~D.} \bibnamefont{Giorno}}, \bibinfo{author}{\bibfnamefont{S.}~\bibnamefont{Gopi}}, \bibinfo{author}{\bibfnamefont{M.}~\bibnamefont{Javaheripi}}, \bibinfo{author}{\bibfnamefont{P.}~\bibnamefont{Kauffmann}}, \bibinfo{author}{\bibfnamefont{G.}~\bibnamefont{De}}, \bibnamefont{et~al.} (\bibinfo{year}{2023}), \urlprefix\url{https://arxiv.org/abs/2306.11644v2}.

\bibitem[{\citenamefont{Bubeck et~al.}(2023)\citenamefont{Bubeck, Chandrasekaran, Eldan, Gehrke, Horvitz, Kamar, Lee, Lee, Li, Lundberg et~al.}}]{Bubeck2023SparksGPT-4}
\bibinfo{author}{\bibfnamefont{S.}~\bibnamefont{Bubeck}}, \bibinfo{author}{\bibfnamefont{V.}~\bibnamefont{Chandrasekaran}}, \bibinfo{author}{\bibfnamefont{R.}~\bibnamefont{Eldan}}, \bibinfo{author}{\bibfnamefont{J.}~\bibnamefont{Gehrke}}, \bibinfo{author}{\bibfnamefont{E.}~\bibnamefont{Horvitz}}, \bibinfo{author}{\bibfnamefont{E.}~\bibnamefont{Kamar}}, \bibinfo{author}{\bibfnamefont{P.}~\bibnamefont{Lee}}, \bibinfo{author}{\bibfnamefont{Y.~T.} \bibnamefont{Lee}}, \bibinfo{author}{\bibfnamefont{Y.}~\bibnamefont{Li}}, \bibinfo{author}{\bibfnamefont{S.}~\bibnamefont{Lundberg}}, \bibnamefont{et~al.} (\bibinfo{year}{2023}), \urlprefix\url{https://arxiv.org/abs/2303.12712v1}.

\bibitem[{\citenamefont{Buehler}(2023{\natexlab{a}})}]{Buehler2023MechGPTModalities}
\bibinfo{author}{\bibfnamefont{M.~J.} \bibnamefont{Buehler}}, \bibinfo{journal}{Appl. Mech. Rev.}  (\bibinfo{year}{2023}{\natexlab{a}}), \urlprefix\url{https://doi.org/10.1115/1.4063843}.

\bibitem[{\citenamefont{Nejjar et~al.}(2023)\citenamefont{Nejjar, Luca, †1, Stiehle, and Weber}}]{Nejjar2023LLMsAnalysis}
\bibinfo{author}{\bibfnamefont{M.}~\bibnamefont{Nejjar}}, \bibinfo{author}{\bibfnamefont{.}~\bibnamefont{Luca}}, \bibinfo{author}{\bibfnamefont{Z.}~\bibnamefont{†1}}, \bibinfo{author}{\bibfnamefont{F.}~\bibnamefont{Stiehle}}, \bibnamefont{and} \bibinfo{author}{\bibfnamefont{I.}~\bibnamefont{Weber}} (\bibinfo{year}{2023}), \urlprefix\url{https://arxiv.org/abs/2311.16733v3}.

\bibitem[{\citenamefont{Buehler}(2023{\natexlab{b}})}]{Buehler2023GenerativeDesign}
\bibinfo{author}{\bibfnamefont{M.~J.} \bibnamefont{Buehler}}, \bibinfo{journal}{ACS Engineering Au}  (\bibinfo{year}{2023}{\natexlab{b}}), \urlprefix\url{10.1021/acsengineeringau.3c00058}.

\bibitem[{\citenamefont{Luu and Buehler}(2023)}]{Luu2023BioinspiredLLM:Materials}
\bibinfo{author}{\bibfnamefont{R.~K.} \bibnamefont{Luu}} \bibnamefont{and} \bibinfo{author}{\bibfnamefont{M.~J.} \bibnamefont{Buehler}}, \bibinfo{journal}{Adv. Science.}  (\bibinfo{year}{2023}), \urlprefix\url{https://doi.org/10.1002/advs.202306724}.

\bibitem[{\citenamefont{Luu et~al.}(2023)\citenamefont{Luu, Wysokowski, and Buehler}}]{Luu2023GenerativeSolvents}
\bibinfo{author}{\bibfnamefont{R.~K.} \bibnamefont{Luu}}, \bibinfo{author}{\bibfnamefont{M.}~\bibnamefont{Wysokowski}}, \bibnamefont{and} \bibinfo{author}{\bibfnamefont{M.~J.} \bibnamefont{Buehler}}, \bibinfo{journal}{Applied Physics Letters} \textbf{\bibinfo{volume}{122}} (\bibinfo{year}{2023}), \urlprefix\url{http://arxiv.org/abs/2304.12400 http://dx.doi.org/10.1063/5.0155890}.

\bibitem[{\citenamefont{Buehler}(2023{\natexlab{c}})}]{Buehler2023MeLMProblemsb}
\bibinfo{author}{\bibfnamefont{M.~J.} \bibnamefont{Buehler}} (\bibinfo{year}{2023}{\natexlab{c}}), \urlprefix\url{https://arxiv.org/abs/2306.17525v1}.

\bibitem[{\citenamefont{Ge et~al.}(2023)\citenamefont{Ge, Hua, Mei, Ji, Tan, Xu, Li, and Zhang}}]{Ge2023OpenAGI:Experts}
\bibinfo{author}{\bibfnamefont{Y.}~\bibnamefont{Ge}}, \bibinfo{author}{\bibfnamefont{W.}~\bibnamefont{Hua}}, \bibinfo{author}{\bibfnamefont{K.}~\bibnamefont{Mei}}, \bibinfo{author}{\bibfnamefont{J.}~\bibnamefont{Ji}}, \bibinfo{author}{\bibfnamefont{J.}~\bibnamefont{Tan}}, \bibinfo{author}{\bibfnamefont{S.}~\bibnamefont{Xu}}, \bibinfo{author}{\bibfnamefont{Z.}~\bibnamefont{Li}}, \bibnamefont{and} \bibinfo{author}{\bibfnamefont{Y.}~\bibnamefont{Zhang}} (\bibinfo{year}{2023}), \urlprefix\url{http://arxiv.org/abs/2304.04370}.

\bibitem[{\citenamefont{Hu et~al.}(2021)\citenamefont{Hu, Shen, Wallis, Allen-Zhu, Li, Wang, Wang, and Chen}}]{hu2021lora}
\bibinfo{author}{\bibfnamefont{E.~J.} \bibnamefont{Hu}}, \bibinfo{author}{\bibfnamefont{Y.}~\bibnamefont{Shen}}, \bibinfo{author}{\bibfnamefont{P.}~\bibnamefont{Wallis}}, \bibinfo{author}{\bibfnamefont{Z.}~\bibnamefont{Allen-Zhu}}, \bibinfo{author}{\bibfnamefont{Y.}~\bibnamefont{Li}}, \bibinfo{author}{\bibfnamefont{S.}~\bibnamefont{Wang}}, \bibinfo{author}{\bibfnamefont{L.}~\bibnamefont{Wang}}, \bibnamefont{and} \bibinfo{author}{\bibfnamefont{W.}~\bibnamefont{Chen}}, \emph{\bibinfo{title}{Lora: Low-rank adaptation of large language models}} (\bibinfo{year}{2021}), \eprint{2106.09685}.

\bibitem[{\citenamefont{Kim et~al.}(2023)\citenamefont{Kim, Park, Kim, Lee, Song, Kim, Kim, Kim, Lee, Kim et~al.}}]{Kim2023SOLARUp-Scaling}
\bibinfo{author}{\bibfnamefont{D.}~\bibnamefont{Kim}}, \bibinfo{author}{\bibfnamefont{C.}~\bibnamefont{Park}}, \bibinfo{author}{\bibfnamefont{S.}~\bibnamefont{Kim}}, \bibinfo{author}{\bibfnamefont{W.}~\bibnamefont{Lee}}, \bibinfo{author}{\bibfnamefont{W.}~\bibnamefont{Song}}, \bibinfo{author}{\bibfnamefont{Y.}~\bibnamefont{Kim}}, \bibinfo{author}{\bibfnamefont{H.}~\bibnamefont{Kim}}, \bibinfo{author}{\bibfnamefont{Y.}~\bibnamefont{Kim}}, \bibinfo{author}{\bibfnamefont{H.}~\bibnamefont{Lee}}, \bibinfo{author}{\bibfnamefont{J.}~\bibnamefont{Kim}}, \bibnamefont{et~al.} (\bibinfo{year}{2023}), \urlprefix\url{https://arxiv.org/abs/2312.15166v1}.

\bibitem[{Arc()}]{Arcee-ai/mergekit:Models.}
\emph{\bibinfo{title}{{arcee-ai/mergekit: Tools for merging pretrained large language models.}}}, \urlprefix\url{https://github.com/arcee-ai/mergekit}.

\bibitem[{\citenamefont{Jacobs et~al.}(1991)\citenamefont{Jacobs, Jordan, Nowlan, and Hinton}}]{Jacobs1991AdaptiveExperts}
\bibinfo{author}{\bibfnamefont{R.~A.} \bibnamefont{Jacobs}}, \bibinfo{author}{\bibfnamefont{M.~I.} \bibnamefont{Jordan}}, \bibinfo{author}{\bibfnamefont{S.~J.} \bibnamefont{Nowlan}}, \bibnamefont{and} \bibinfo{author}{\bibfnamefont{G.~E.} \bibnamefont{Hinton}}, \bibinfo{journal}{Neural Computation} \textbf{\bibinfo{volume}{3}}, \bibinfo{pages}{79} (\bibinfo{year}{1991}), ISSN \bibinfo{issn}{0899-7667}, \urlprefix\url{https://dx.doi.org/10.1162/neco.1991.3.1.79}.

\bibitem[{\citenamefont{Hampshire and Waibel}(1992)}]{Hampshire1992TheRecognition}
\bibinfo{author}{\bibfnamefont{J.~B.} \bibnamefont{Hampshire}} \bibnamefont{and} \bibinfo{author}{\bibfnamefont{A.}~\bibnamefont{Waibel}}, \bibinfo{journal}{IEEE Transactions on Pattern Analysis and Machine Intelligence} \textbf{\bibinfo{volume}{14}}, \bibinfo{pages}{751} (\bibinfo{year}{1992}), ISSN \bibinfo{issn}{01628828}.

\bibitem[{\citenamefont{Jordan and Jacobs}(1994)}]{Jordan1994HierarchicalAlgorithm}
\bibinfo{author}{\bibfnamefont{M.~I.} \bibnamefont{Jordan}} \bibnamefont{and} \bibinfo{author}{\bibfnamefont{R.~A.} \bibnamefont{Jacobs}}, \bibinfo{journal}{Neural Computation} \textbf{\bibinfo{volume}{6}}, \bibinfo{pages}{181} (\bibinfo{year}{1994}), ISSN \bibinfo{issn}{0899-7667}, \urlprefix\url{https://dx.doi.org/10.1162/neco.1994.6.2.181}.

\bibitem[{\citenamefont{Jiang et~al.}(2024)\citenamefont{Jiang, Sablayrolles, Roux, Mensch, Savary, Bamford, Chaplot, Casas, Hanna, Bressand et~al.}}]{Jiang2024MixtralExperts}
\bibinfo{author}{\bibfnamefont{A.~Q.} \bibnamefont{Jiang}}, \bibinfo{author}{\bibfnamefont{A.}~\bibnamefont{Sablayrolles}}, \bibinfo{author}{\bibfnamefont{A.}~\bibnamefont{Roux}}, \bibinfo{author}{\bibfnamefont{A.}~\bibnamefont{Mensch}}, \bibinfo{author}{\bibfnamefont{B.}~\bibnamefont{Savary}}, \bibinfo{author}{\bibfnamefont{C.}~\bibnamefont{Bamford}}, \bibinfo{author}{\bibfnamefont{D.~S.} \bibnamefont{Chaplot}}, \bibinfo{author}{\bibfnamefont{D.~d.~l.} \bibnamefont{Casas}}, \bibinfo{author}{\bibfnamefont{E.~B.} \bibnamefont{Hanna}}, \bibinfo{author}{\bibfnamefont{F.}~\bibnamefont{Bressand}}, \bibnamefont{et~al.} (\bibinfo{year}{2024}), \urlprefix\url{https://arxiv.org/abs/2401.04088v1}.

\bibitem[{\citenamefont{Hu and Buehler}(2023)}]{Hu2023DeepScience}
\bibinfo{author}{\bibfnamefont{Y.}~\bibnamefont{Hu}} \bibnamefont{and} \bibinfo{author}{\bibfnamefont{M.~J.} \bibnamefont{Buehler}}, \bibinfo{journal}{APL Machine Learning} \textbf{\bibinfo{volume}{1}}, \bibinfo{pages}{010901} (\bibinfo{year}{2023}), ISSN \bibinfo{issn}{2770-9019}, \urlprefix\url{https://aip.scitation.org/doi/abs/10.1063/5.0134317}.

\bibitem[{\citenamefont{Buehler}(2023{\natexlab{d}})}]{Buehler2023AMetamaterials}
\bibinfo{author}{\bibfnamefont{M.~J.} \bibnamefont{Buehler}}, \bibinfo{journal}{Modelling and Simulation in Materials Science and Engineering}  (\bibinfo{year}{2023}{\natexlab{d}}), ISSN \bibinfo{issn}{0965-0393}, \urlprefix\url{https://iopscience.iop.org/article/10.1088/1361-651X/accfb5 https://iopscience.iop.org/article/10.1088/1361-651X/accfb5/meta}.

\bibitem[{\citenamefont{Buehler and Guo}(2022)}]{Buehler2022ProteinNetworks}
\bibinfo{author}{\bibfnamefont{M.~J.} \bibnamefont{Buehler}} \bibnamefont{and} \bibinfo{author}{\bibfnamefont{K.}~\bibnamefont{Guo}} (\bibinfo{year}{2022}), \urlprefix\url{https://zenodo.org/record/6346661}.

\bibitem[{Hug()}]{HuggingFaceH4/zephyr-7b-betaFace}
\emph{\bibinfo{title}{{HuggingFaceH4/zephyr-7b-beta {\textperiodcentered} Hugging Face}}}, \urlprefix\url{https://huggingface.co/HuggingFaceH4/zephyr-7b-beta}.

\bibitem[{\citenamefont{Buehler}(2008)}]{Buehler2008AtomisticFailure}
\bibinfo{author}{\bibfnamefont{M.}~\bibnamefont{Buehler}}, \emph{\bibinfo{title}{{Atomistic modeling of materials failure}}} (\bibinfo{year}{2008}), ISBN \bibinfo{isbn}{9780387764252}.

\bibitem[{DAL()}]{DALLECard}
\emph{\bibinfo{title}{{DALL{\textperiodcentered}E 3 system card}}}, \urlprefix\url{https://openai.com/research/dall-e-3-system-card}.

\bibitem[{\citenamefont{Lu et~al.}(2023)\citenamefont{Lu, Kaplan, and Buehler}}]{Lu2023GenerativePropertiesb}
\bibinfo{author}{\bibfnamefont{W.}~\bibnamefont{Lu}}, \bibinfo{author}{\bibfnamefont{D.~L.} \bibnamefont{Kaplan}}, \bibnamefont{and} \bibinfo{author}{\bibfnamefont{M.~J.} \bibnamefont{Buehler}}, \bibinfo{journal}{Advanced Functional Materials} p. \bibinfo{pages}{2311324} (\bibinfo{year}{2023}), ISSN \bibinfo{issn}{1616-3028}, \urlprefix\url{https://onlinelibrary.wiley.com/doi/full/10.1002/adfm.202311324 https://onlinelibrary.wiley.com/doi/abs/10.1002/adfm.202311324 https://onlinelibrary.wiley.com/doi/10.1002/adfm.202311324}.

\bibitem[{\citenamefont{Ni et~al.}(2023)\citenamefont{Ni, Kaplan, and Buehler}}]{Ni2023ForceGen:Model}
\bibinfo{author}{\bibfnamefont{B.}~\bibnamefont{Ni}}, \bibinfo{author}{\bibfnamefont{D.~L.} \bibnamefont{Kaplan}}, \bibnamefont{and} \bibinfo{author}{\bibfnamefont{M.~J.} \bibnamefont{Buehler}} (\bibinfo{year}{2023}), \urlprefix\url{https://arxiv.org/abs/2310.10605v3}.

\bibitem[{\citenamefont{Lu et~al.}(1998)\citenamefont{Lu, Isralewitz, Krammer, Vogel, and Schulten}}]{Lu1998UnfoldingSimulation.}
\bibinfo{author}{\bibfnamefont{H.}~\bibnamefont{Lu}}, \bibinfo{author}{\bibfnamefont{B.}~\bibnamefont{Isralewitz}}, \bibinfo{author}{\bibfnamefont{A.}~\bibnamefont{Krammer}}, \bibinfo{author}{\bibfnamefont{V.}~\bibnamefont{Vogel}}, \bibnamefont{and} \bibinfo{author}{\bibfnamefont{K.}~\bibnamefont{Schulten}}, \bibinfo{journal}{Biophysical Journal} \textbf{\bibinfo{volume}{75}}, \bibinfo{pages}{662} (\bibinfo{year}{1998}), ISSN \bibinfo{issn}{00063495}, \urlprefix\url{/pmc/articles/PMC1299741/?report=abstract https://www.ncbi.nlm.nih.gov/pmc/articles/PMC1299741/}.

\bibitem[{\citenamefont{Jumper et~al.}(2021)\citenamefont{Jumper, Evans, Pritzel, Green, Figurnov, Ronneberger, Tunyasuvunakool, Bates, {\v{Z}}{\'{i}}dek, Potapenko et~al.}}]{Jumper2021HighlyAlphaFoldb}
\bibinfo{author}{\bibfnamefont{J.}~\bibnamefont{Jumper}}, \bibinfo{author}{\bibfnamefont{R.}~\bibnamefont{Evans}}, \bibinfo{author}{\bibfnamefont{A.}~\bibnamefont{Pritzel}}, \bibinfo{author}{\bibfnamefont{T.}~\bibnamefont{Green}}, \bibinfo{author}{\bibfnamefont{M.}~\bibnamefont{Figurnov}}, \bibinfo{author}{\bibfnamefont{O.}~\bibnamefont{Ronneberger}}, \bibinfo{author}{\bibfnamefont{K.}~\bibnamefont{Tunyasuvunakool}}, \bibinfo{author}{\bibfnamefont{R.}~\bibnamefont{Bates}}, \bibinfo{author}{\bibfnamefont{A.}~\bibnamefont{{\v{Z}}{\'{i}}dek}}, \bibinfo{author}{\bibfnamefont{A.}~\bibnamefont{Potapenko}}, \bibnamefont{et~al.}, \bibinfo{journal}{Nature} \textbf{\bibinfo{volume}{596}}, \bibinfo{pages}{583} (\bibinfo{year}{2021}), ISSN \bibinfo{issn}{14764687}.

\bibitem[{\citenamefont{Altschul et~al.}(1990)\citenamefont{Altschul, Gish, Miller, Myers, and Lipman}}]{Altschul1990BasicTool}
\bibinfo{author}{\bibfnamefont{S.~F.} \bibnamefont{Altschul}}, \bibinfo{author}{\bibfnamefont{W.}~\bibnamefont{Gish}}, \bibinfo{author}{\bibfnamefont{W.}~\bibnamefont{Miller}}, \bibinfo{author}{\bibfnamefont{E.~W.} \bibnamefont{Myers}}, \bibnamefont{and} \bibinfo{author}{\bibfnamefont{D.~J.} \bibnamefont{Lipman}}, \bibinfo{journal}{Journal of Molecular Biology} \textbf{\bibinfo{volume}{215}}, \bibinfo{pages}{403} (\bibinfo{year}{1990}), ISSN \bibinfo{issn}{00222836}, \urlprefix\url{http://www.ncbi.nlm.nih.gov/pubmed/2231712 https://linkinghub.elsevier.com/retrieve/pii/S0022283605803602}.

\bibitem[{\citenamefont{Mirdita et~al.}(2022)\citenamefont{Mirdita, Sch{\"{u}}tze, Moriwaki, Heo, Ovchinnikov, and Steinegger}}]{Mirdita2022ColabFold:All}
\bibinfo{author}{\bibfnamefont{M.}~\bibnamefont{Mirdita}}, \bibinfo{author}{\bibfnamefont{K.}~\bibnamefont{Sch{\"{u}}tze}}, \bibinfo{author}{\bibfnamefont{Y.}~\bibnamefont{Moriwaki}}, \bibinfo{author}{\bibfnamefont{L.}~\bibnamefont{Heo}}, \bibinfo{author}{\bibfnamefont{S.}~\bibnamefont{Ovchinnikov}}, \bibnamefont{and} \bibinfo{author}{\bibfnamefont{M.}~\bibnamefont{Steinegger}}, \bibinfo{journal}{Nature Methods 2022 19:6} \textbf{\bibinfo{volume}{19}}, \bibinfo{pages}{679} (\bibinfo{year}{2022}), ISSN \bibinfo{issn}{1548-7105}, \urlprefix\url{https://www.nature.com/articles/s41592-022-01488-1}.

\bibitem[{\citenamefont{Giesa et~al.}(2012)\citenamefont{Giesa, Spivak, and Buehler}}]{Giesa2012CategoryDesign}
\bibinfo{author}{\bibfnamefont{T.}~\bibnamefont{Giesa}}, \bibinfo{author}{\bibfnamefont{D.}~\bibnamefont{Spivak}}, \bibnamefont{and} \bibinfo{author}{\bibfnamefont{M.}~\bibnamefont{Buehler}}, \bibinfo{journal}{Advanced Engineering Materials} \textbf{\bibinfo{volume}{14}} (\bibinfo{year}{2012}), ISSN \bibinfo{issn}{14381656 15272648}.

\bibitem[{\citenamefont{Spivak and Reoccurring}(2011)}]{Spivak2011ReoccurringAnalogies}
\bibinfo{author}{\bibfnamefont{D.~I.} \bibnamefont{Spivak}} \bibnamefont{and} \bibinfo{author}{\bibfnamefont{M.~J.~B.} \bibnamefont{Reoccurring}}, pp. \bibinfo{pages}{0--13} (\bibinfo{year}{2011}).

\bibitem[{\citenamefont{Girvan and Newman}(2002)}]{Girvan2002CommunityNetworks}
\bibinfo{author}{\bibfnamefont{M.}~\bibnamefont{Girvan}} \bibnamefont{and} \bibinfo{author}{\bibfnamefont{M.~E.} \bibnamefont{Newman}}, \bibinfo{journal}{Proceedings of the National Academy of Sciences of the United States of America} \textbf{\bibinfo{volume}{99}}, \bibinfo{pages}{7821} (\bibinfo{year}{2002}), ISSN \bibinfo{issn}{00278424}, \urlprefix\url{https://www.pnas.org/doi/abs/10.1073/pnas.122653799}.

\bibitem[{Goo()}]{Google/gemma-7b-itFace}
\emph{\bibinfo{title}{{google/gemma-7b-it {\textperiodcentered} Hugging Face}}}, \urlprefix\url{https://huggingface.co/google/gemma-7b-it}.

\bibitem[{\citenamefont{Ruddigkeit et~al.}(2012)\citenamefont{Ruddigkeit, Van~Deursen, Blum, and Reymond}}]{Ruddigkeit2012EnumerationGDB-17}
\bibinfo{author}{\bibfnamefont{L.}~\bibnamefont{Ruddigkeit}}, \bibinfo{author}{\bibfnamefont{R.}~\bibnamefont{Van~Deursen}}, \bibinfo{author}{\bibfnamefont{L.~C.} \bibnamefont{Blum}}, \bibnamefont{and} \bibinfo{author}{\bibfnamefont{J.~L.} \bibnamefont{Reymond}}, \bibinfo{journal}{Journal of Chemical Information and Modeling} \textbf{\bibinfo{volume}{52}}, \bibinfo{pages}{2864} (\bibinfo{year}{2012}), ISSN \bibinfo{issn}{1549960X}, \urlprefix\url{https://pubs.acs.org/doi/full/10.1021/ci300415d}.

\bibitem[{\citenamefont{Ramakrishnan et~al.}(2015)\citenamefont{Ramakrishnan, Hartmann, Tapavicza, and Von~Lilienfeld}}]{Ramakrishnan2015ElectronicSpace}
\bibinfo{author}{\bibfnamefont{R.}~\bibnamefont{Ramakrishnan}}, \bibinfo{author}{\bibfnamefont{M.}~\bibnamefont{Hartmann}}, \bibinfo{author}{\bibfnamefont{E.}~\bibnamefont{Tapavicza}}, \bibnamefont{and} \bibinfo{author}{\bibfnamefont{O.~A.} \bibnamefont{Von~Lilienfeld}}, \bibinfo{journal}{Journal of Chemical Physics} \textbf{\bibinfo{volume}{143}}, \bibinfo{pages}{84111} (\bibinfo{year}{2015}), ISSN \bibinfo{issn}{00219606}, \urlprefix\url{/aip/jcp/article/143/8/084111/73278/Electronic-spectra-from-TDDFT-and-machine-learning}.

\bibitem[{\citenamefont{Rapp{\'{e}} et~al.}(1992)\citenamefont{Rapp{\'{e}}, Casewit, Colwell, Goddard, and Skiff}}]{Rappe1992UFFSimulations}
\bibinfo{author}{\bibfnamefont{A.~K.} \bibnamefont{Rapp{\'{e}}}}, \bibinfo{author}{\bibfnamefont{C.~J.} \bibnamefont{Casewit}}, \bibinfo{author}{\bibfnamefont{K.~S.} \bibnamefont{Colwell}}, \bibinfo{author}{\bibfnamefont{W.~A.} \bibnamefont{Goddard}}, \bibnamefont{and} \bibinfo{author}{\bibfnamefont{W.~M.} \bibnamefont{Skiff}}, \bibinfo{journal}{Journal of the American Chemical Society} \textbf{\bibinfo{volume}{114}}, \bibinfo{pages}{10024} (\bibinfo{year}{1992}), ISSN \bibinfo{issn}{15205126}, \urlprefix\url{https://pubs.acs.org/doi/pdf/10.1021/ja00051a040}.

\bibitem[{\citenamefont{Gasteiger and Marsili}(1980)}]{Gasteiger1980IterativeCharges}
\bibinfo{author}{\bibfnamefont{J.}~\bibnamefont{Gasteiger}} \bibnamefont{and} \bibinfo{author}{\bibfnamefont{M.}~\bibnamefont{Marsili}}, \bibinfo{journal}{Tetrahedron} \textbf{\bibinfo{volume}{36}}, \bibinfo{pages}{3219} (\bibinfo{year}{1980}), ISSN \bibinfo{issn}{0040-4020}.

\bibitem[{\citenamefont{McQuarrie and Simon}(1997)}]{McQuarrieSimonPhysicalChemistry}
\bibinfo{author}{\bibfnamefont{D.~A.} \bibnamefont{McQuarrie}} \bibnamefont{and} \bibinfo{author}{\bibfnamefont{J.~D.} \bibnamefont{Simon}}, \emph{\bibinfo{title}{Physical Chemistry: A Molecular Approach}} (\bibinfo{publisher}{University Science Books}, \bibinfo{year}{1997}), ISBN \bibinfo{isbn}{978-0935702996}.

\bibitem[{\citenamefont{Carey and Sundberg}(2007)}]{CareySundbergAdvancedOrganicChemistry}
\bibinfo{author}{\bibfnamefont{F.~A.} \bibnamefont{Carey}} \bibnamefont{and} \bibinfo{author}{\bibfnamefont{R.~J.} \bibnamefont{Sundberg}}, \emph{\bibinfo{title}{Advanced Organic Chemistry: Part A: Structure and Mechanisms}} (\bibinfo{publisher}{Springer}, \bibinfo{year}{2007}), \bibinfo{edition}{5th} ed., ISBN \bibinfo{isbn}{978-0387683461}.

\bibitem[{\citenamefont{Varghese and Grinstaff}(2022)}]{Varghese2022BeyondApplications}
\bibinfo{author}{\bibfnamefont{M.}~\bibnamefont{Varghese}} \bibnamefont{and} \bibinfo{author}{\bibfnamefont{M.~W.} \bibnamefont{Grinstaff}}, \bibinfo{journal}{Chemical Society Reviews} \textbf{\bibinfo{volume}{51}}, \bibinfo{pages}{8258} (\bibinfo{year}{2022}), ISSN \bibinfo{issn}{1460-4744}, \urlprefix\url{https://pubs.rsc.org/en/content/articlehtml/2022/cs/d1cs00930c https://pubs.rsc.org/en/content/articlelanding/2022/cs/d1cs00930c}.

\bibitem[{\citenamefont{Hendrycks et~al.}(2020)\citenamefont{Hendrycks, Burns, Basart, Zou, Mazeika, Song, and Steinhardt}}]{Hendrycks2020MeasuringUnderstanding}
\bibinfo{author}{\bibfnamefont{D.}~\bibnamefont{Hendrycks}}, \bibinfo{author}{\bibfnamefont{C.}~\bibnamefont{Burns}}, \bibinfo{author}{\bibfnamefont{S.}~\bibnamefont{Basart}}, \bibinfo{author}{\bibfnamefont{A.}~\bibnamefont{Zou}}, \bibinfo{author}{\bibfnamefont{M.}~\bibnamefont{Mazeika}}, \bibinfo{author}{\bibfnamefont{D.}~\bibnamefont{Song}}, \bibnamefont{and} \bibinfo{author}{\bibfnamefont{J.}~\bibnamefont{Steinhardt}}, \bibinfo{journal}{ICLR 2021 - 9th International Conference on Learning Representations}  (\bibinfo{year}{2020}), \urlprefix\url{https://arxiv.org/abs/2009.03300v3}.

\bibitem[{\citenamefont{Matsakis and Klock~II}(2014)}]{matsakis2014rust}
\bibinfo{author}{\bibfnamefont{N.~D.} \bibnamefont{Matsakis}} \bibnamefont{and} \bibinfo{author}{\bibfnamefont{F.~S.} \bibnamefont{Klock~II}}, in \emph{\bibinfo{booktitle}{ACM SIGAda Ada Letters}} (\bibinfo{organization}{ACM}, \bibinfo{year}{2014}), vol.~\bibinfo{volume}{34}, pp. \bibinfo{pages}{103--104}.

\bibitem[{\citenamefont{Ghafarollahi and Buehler}(2024)}]{Ghafarollahi2024ProtAgents:Learning}
\bibinfo{author}{\bibfnamefont{A.}~\bibnamefont{Ghafarollahi}} \bibnamefont{and} \bibinfo{author}{\bibfnamefont{M.~J.} \bibnamefont{Buehler}} (\bibinfo{year}{2024}), \urlprefix\url{https://arxiv.org/abs/2402.04268v1}.

\bibitem[{Mat()}]{MatplotlibDocumentation}
\emph{\bibinfo{title}{{Matplotlib documentation — Matplotlib 3.5.1 documentation}}}, \urlprefix\url{https://matplotlib.org/stable/index.html}.

\bibitem[{\citenamefont{Paszke et~al.}(2019)\citenamefont{Paszke, Gross, Bradbury, Lin, Devito, Massa, Steiner, Killeen, and Yang}}]{Paszke2019PyTorchLibrary}
\bibinfo{author}{\bibfnamefont{A.}~\bibnamefont{Paszke}}, \bibinfo{author}{\bibfnamefont{S.}~\bibnamefont{Gross}}, \bibinfo{author}{\bibfnamefont{J.}~\bibnamefont{Bradbury}}, \bibinfo{author}{\bibfnamefont{Z.}~\bibnamefont{Lin}}, \bibinfo{author}{\bibfnamefont{Z.}~\bibnamefont{Devito}}, \bibinfo{author}{\bibfnamefont{F.}~\bibnamefont{Massa}}, \bibinfo{author}{\bibfnamefont{B.}~\bibnamefont{Steiner}}, \bibinfo{author}{\bibfnamefont{T.}~\bibnamefont{Killeen}}, \bibnamefont{and} \bibinfo{author}{\bibfnamefont{E.}~\bibnamefont{Yang}} (\bibinfo{year}{2019}).

\bibitem[{\citenamefont{Su et~al.}(2021)\citenamefont{Su, Lu, Pan, Wen, and Liu}}]{su2021roformer}
\bibinfo{author}{\bibfnamefont{J.}~\bibnamefont{Su}}, \bibinfo{author}{\bibfnamefont{Y.}~\bibnamefont{Lu}}, \bibinfo{author}{\bibfnamefont{S.}~\bibnamefont{Pan}}, \bibinfo{author}{\bibfnamefont{B.}~\bibnamefont{Wen}}, \bibnamefont{and} \bibinfo{author}{\bibfnamefont{Y.}~\bibnamefont{Liu}}, \emph{\bibinfo{title}{Roformer: Enhanced transformer with rotary position embedding}} (\bibinfo{year}{2021}), \eprint{2104.09864}.

\bibitem[{\citenamefont{Kwon et~al.}(2023)\citenamefont{Kwon, Li, Zhuang, Sheng, Zheng, Yu, Gonzalez, Zhang, and Stoica}}]{kwon2023efficient}
\bibinfo{author}{\bibfnamefont{W.}~\bibnamefont{Kwon}}, \bibinfo{author}{\bibfnamefont{Z.}~\bibnamefont{Li}}, \bibinfo{author}{\bibfnamefont{S.}~\bibnamefont{Zhuang}}, \bibinfo{author}{\bibfnamefont{Y.}~\bibnamefont{Sheng}}, \bibinfo{author}{\bibfnamefont{L.}~\bibnamefont{Zheng}}, \bibinfo{author}{\bibfnamefont{C.~H.} \bibnamefont{Yu}}, \bibinfo{author}{\bibfnamefont{J.~E.} \bibnamefont{Gonzalez}}, \bibinfo{author}{\bibfnamefont{H.}~\bibnamefont{Zhang}}, \bibnamefont{and} \bibinfo{author}{\bibfnamefont{I.}~\bibnamefont{Stoica}}, in \emph{\bibinfo{booktitle}{Proceedings of the ACM SIGOPS 29th Symposium on Operating Systems Principles}} (\bibinfo{year}{2023}).

\bibitem[{\citenamefont{Wei et~al.}(2021)\citenamefont{Wei, Bosma, Zhao, Guu, Yu, Lester, Du, Dai, and Le}}]{Wei2021FinetunedLearners}
\bibinfo{author}{\bibfnamefont{J.}~\bibnamefont{Wei}}, \bibinfo{author}{\bibfnamefont{M.}~\bibnamefont{Bosma}}, \bibinfo{author}{\bibfnamefont{V.~Y.} \bibnamefont{Zhao}}, \bibinfo{author}{\bibfnamefont{K.}~\bibnamefont{Guu}}, \bibinfo{author}{\bibfnamefont{A.~W.} \bibnamefont{Yu}}, \bibinfo{author}{\bibfnamefont{B.}~\bibnamefont{Lester}}, \bibinfo{author}{\bibfnamefont{N.}~\bibnamefont{Du}}, \bibinfo{author}{\bibfnamefont{A.~M.} \bibnamefont{Dai}}, \bibnamefont{and} \bibinfo{author}{\bibfnamefont{Q.~V.} \bibnamefont{Le}}, \bibinfo{journal}{ICLR 2022 - 10th International Conference on Learning Representations}  (\bibinfo{year}{2021}), \urlprefix\url{https://arxiv.org/abs/2109.01652v5}.

\bibitem[{Qin()}]{QingyiSi/Alpaca-CoTMain}
\emph{\bibinfo{title}{{QingyiSi/Alpaca-CoT at main}}}, \urlprefix\url{https://huggingface.co/datasets/QingyiSi/Alpaca-CoT/tree/main/Chain-of-Thought}.

\bibitem[{\citenamefont{Li et~al.}(2023)\citenamefont{Li, Hasan, Al, Hammoud, Itani, Khizbullin, and Ghanem}}]{Li2023CAMEL:Society}
\bibinfo{author}{\bibfnamefont{G.}~\bibnamefont{Li}}, \bibinfo{author}{\bibfnamefont{.}~\bibnamefont{Hasan}}, \bibinfo{author}{\bibfnamefont{A.}~\bibnamefont{Al}}, \bibinfo{author}{\bibfnamefont{K.}~\bibnamefont{Hammoud}}, \bibinfo{author}{\bibfnamefont{H.}~\bibnamefont{Itani}}, \bibinfo{author}{\bibfnamefont{D.}~\bibnamefont{Khizbullin}}, \bibnamefont{and} \bibinfo{author}{\bibfnamefont{B.}~\bibnamefont{Ghanem}} (\bibinfo{year}{2023}), \urlprefix\url{https://arxiv.org/abs/2303.17760v2}.

\bibitem[{Cam({\natexlab{a}})}]{Camel-ai/chemistryFace}
\emph{\bibinfo{title}{{camel-ai/chemistry {\textperiodcentered} Datasets at Hugging Face}}}, \urlprefix\url{https://huggingface.co/datasets/camel-ai/chemistry}.

\bibitem[{Cam({\natexlab{b}})}]{Camel-ai/mathFace}
\emph{\bibinfo{title}{{camel-ai/math {\textperiodcentered} Datasets at Hugging Face}}}, \urlprefix\url{https://huggingface.co/datasets/camel-ai/math}.

\bibitem[{Cam({\natexlab{c}})}]{Camel-ai/physicsFace}
\emph{\bibinfo{title}{{camel-ai/physics {\textperiodcentered} Datasets at Hugging Face}}}, \urlprefix\url{https://huggingface.co/datasets/camel-ai/physics}.

\bibitem[{Cam({\natexlab{d}})}]{Camel-ai/biologyFace}
\emph{\bibinfo{title}{{camel-ai/biology {\textperiodcentered} Datasets at Hugging Face}}}, \urlprefix\url{https://huggingface.co/datasets/camel-ai/biology}.

\bibitem[{Lam()}]{Lamm-mit/MechanicsMaterialsFace}
\emph{\bibinfo{title}{{lamm-mit/MechanicsMaterials {\textperiodcentered} Datasets at Hugging Face}}}, \urlprefix\url{https://huggingface.co/datasets/lamm-mit/MechanicsMaterials}.

\bibitem[{\citenamefont{Lee et~al.}(2023)\citenamefont{Lee, Hunter, and Ruiz}}]{Lee2023Platypus:LLMs}
\bibinfo{author}{\bibfnamefont{A.~N.} \bibnamefont{Lee}}, \bibinfo{author}{\bibfnamefont{C.~J.} \bibnamefont{Hunter}}, \bibnamefont{and} \bibinfo{author}{\bibfnamefont{N.}~\bibnamefont{Ruiz}} (\bibinfo{year}{2023}), \urlprefix\url{https://arxiv.org/abs/2308.07317v1}.

\bibitem[{Gar()}]{Garage-bAInd/Open-PlatypusFace}
\emph{\bibinfo{title}{{garage-bAInd/Open-Platypus {\textperiodcentered} Datasets at Hugging Face}}}, \urlprefix\url{https://huggingface.co/datasets/garage-bAInd/Open-Platypus}.

\bibitem[{\citenamefont{Weininger}(1988)}]{Weininger1988SMILESRules}
\bibinfo{author}{\bibfnamefont{D.}~\bibnamefont{Weininger}}, \bibinfo{journal}{Journal of Chemical Information and Computer Sciences} \textbf{\bibinfo{volume}{28}}, \bibinfo{pages}{31} (\bibinfo{year}{1988}), ISSN \bibinfo{issn}{00952338}, \urlprefix\url{https://pubs.acs.org/doi/abs/10.1021/ci00057a005}.

\bibitem[{\citenamefont{Weininger et~al.}(1989)\citenamefont{Weininger, Weininger, and Weininger}}]{Weininger1989SMILES.Notation}
\bibinfo{author}{\bibfnamefont{D.}~\bibnamefont{Weininger}}, \bibinfo{author}{\bibfnamefont{A.}~\bibnamefont{Weininger}}, \bibnamefont{and} \bibinfo{author}{\bibfnamefont{J.~L.} \bibnamefont{Weininger}}, \bibinfo{journal}{Journal of Chemical Information and Computer Sciences} \textbf{\bibinfo{volume}{29}}, \bibinfo{pages}{97} (\bibinfo{year}{1989}), ISSN \bibinfo{issn}{00952338}, \urlprefix\url{https://pubs.acs.org/doi/abs/10.1021/ci00062a008}.

\bibitem[{Gui()}]{Guidance-ai/guidance:Models.}
\emph{\bibinfo{title}{{guidance-ai/guidance: A guidance language for controlling large language models.}}}, \urlprefix\url{https://github.com/guidance-ai/guidance}.

\bibitem[{Rah()}]{Rahulnyk/knowledge_graph:QnA}
\emph{\bibinfo{title}{{rahulnyk/knowledge{\_}graph: Convert any text to a graph of knowledge. This can be used for Graph Augmented Generation or Knowledge Graph based QnA}}}, \urlprefix\url{https://github.com/rahulnyk/knowledge_graph}.

\bibitem[{\citenamefont{Liu}(2022)}]{Liu_LlamaIndex_2022}
\bibinfo{author}{\bibfnamefont{J.}~\bibnamefont{Liu}}, \emph{\bibinfo{title}{{LlamaIndex}}} (\bibinfo{year}{2022}), \urlprefix\url{https://github.com/jerryjliu/llama_index}.

\bibitem[{Net()}]{Networkx/networkx:Python}
\emph{\bibinfo{title}{{networkx/networkx: Network Analysis in Python}}}, \urlprefix\url{https://github.com/networkx/networkx}.

\bibitem[{Wes()}]{WestHealth/pyvis:Graphs.}
\emph{\bibinfo{title}{{WestHealth/pyvis: Python package for creating and visualizing interactive network graphs.}}}, \urlprefix\url{https://github.com/WestHealth/pyvis/tree/master}.

\bibitem[{\citenamefont{Schr{\"{o}}dinger}(2015)}]{Schrodinger2015The1.8}
\bibinfo{author}{\bibfnamefont{L.~L.~C.} \bibnamefont{Schr{\"{o}}dinger}}, \emph{\bibinfo{title}{{The PyMOL molecular graphics system, version 1.8}}} (\bibinfo{year}{2015}).

\end{thebibliography}

\end{document}